\documentclass[preprint,groupedaddress,showpacs,amsmath,amssymb]{revtex4} 
\usepackage[english]{babel} 
\usepackage[latin1]{inputenc} 
\usepackage[T1]{fontenc}
\usepackage{latexsym} 
\usepackage{epsfig} 
\usepackage{rotating}
\usepackage{graphicx}
\usepackage{lscape}
\usepackage{array}
\usepackage{color}

\newcommand\disp{\displaystyle}

\newcommand\refi[1]
{Fig.\ref{#1}}

\newcommand\reft[1]
{Table \ref{#1}}

\newcommand\refs[1]
{(cf. section \ref{#1})}

\newcommand\refa[1]
{(cf. appendix \ref{#1})}

\newcommand\refaa[1]
{appendix \ref{#1}}

\newcommand\refm[1]
{(\ref{#1})}

\newcommand\cellcolor[2]
{\textcolor{#1}{#2}}

\newcommand\po[1]
{\paragraph{#1}

 ~

}

\newcommand\imagea[4]
{\begin{figure}[h!]
\begin{center}
\includegraphics[height=#4 cm]{#1}
\caption{(Color online) #2}
\label{#3}
\end{center}
\end{figure}}

\newcommand\doublimage[6]
{
\begin{figure}[h!]

\begin{center}

\begin{tabular}{c c}
\includegraphics[width=6cm]{#1} & \includegraphics[width=6cm]{#3} \\
#2 & #4 \\
\end{tabular}

\caption{(Color online) #5}
\label{#6}
\end{center}
\end{figure}

}

\newcommand\triplimage[8]
{

\begin{figure}[h!]
   \begin{minipage}[c]{.3\linewidth}
\begin{center}
      \includegraphics[width=4cm]{#1}

#2
\end{center}
   \end{minipage} \hfill
   \begin{minipage}[c]{.3\linewidth}
\begin{center}
      \includegraphics[width=4cm]{#3}

 #4
\end{center}
   \end{minipage}\hfill
   \begin{minipage}[c]{.3\linewidth}
\begin{center}
      \includegraphics[width=4cm]{#5} 

#6
\end{center}
   \end{minipage}

\caption{(Color online) #7}\label{#8}
\end{figure}

}

\newcommand\doublimagem[6]
{

\begin{figure}[h!]
   \begin{minipage}[c]{.46\linewidth}
\begin{center}
      \includegraphics[width=6cm]{#1}

#2
\end{center}
   \end{minipage} \hfill
   \begin{minipage}[c]{.46\linewidth}
\begin{center}
      \includegraphics[width=6cm]{#3}

 #4
\end{center}
   \end{minipage}\hfill

\caption{(Color online) #5}\label{#6}
\end{figure}

}

\begin{document}

\title{Robustness of optimal intermittent search strategies in dimension 1, 2 and 3}

\author{C. Loverdo}
\affiliation{Laboratoire de Physique Th{é}orique de la Mati{è}re Condens{é}e, UMR CNRS 7600,
Universit{é} Pierre et Marie Curie, 4 Place Jussieu, 75252 Paris, France}

\author{O. Bénichou}
%\email{moreau@lptl.jussieu.fr}
\affiliation{Laboratoire de Physique Th{é}orique de la Mati{è}re Condens{é}e, UMR CNRS 7600,
Universit{é} Pierre et Marie Curie, 4 Place Jussieu, 75252 Paris, France}

\author{M. Moreau}
\affiliation{Laboratoire de Physique Th{é}orique de la Mati{è}re Condens{é}e, UMR CNRS 7600,
Universit{é} Pierre et Marie Curie, 4 Place Jussieu, 75252 Paris, France}

\author{R. Voituriez}
\affiliation{Laboratoire de Physique Th{é}orique de la Mati{è}re Condens{é}e, UMR CNRS 7600,
Universit{é} Pierre et Marie Curie, 4 Place Jussieu, 75252 Paris, France}

\date{\today}

\begin{abstract}
Search problems at various scales  involve a searcher, be it a molecule before reaction or a 
foraging animal,  which performs an intermittent motion.
%In search problems, like animals searching for a prey, or reactivity within biological cells,
%there are cases where detection and fast motion are not compatible. 
Here we analyze a generic model based on such type of intermittent motion, in which the searcher   
alternates phases of slow motion allowing detection,
and phases of fast motion without detection. 
We present full and systematic results for different modeling hypotheses of the detection mechanism
in space dimension 1, 2 and 3. 
Our study completes and extends the results of 
our recent letter [Loverdo {\it et al.} Nature Physics {\bf 4}, 134 (2008)] 
and gives the necessary calculation details. 
In addition, a new modeling of the detection phase is presented. 
We show that   the mean target detection time can be minimized as a function of 
the mean duration of each phase in dimension 1, 2 and 3. 
Importantly, this optimal strategy  does not depend on the details of the modeling 
of the slow detection phase, 
which shows the robustness of our results.  
We believe that this systematic analysis can be used as a basis to study quantitatively 
various real search problems involving intermittent behaviors. 

\end{abstract}


\maketitle

\section{Introduction}

%\subsection{Importance of search problems and modelings}

Search problems, involving a "searcher" and a "target",  pop up in a wide range of domains. 
They have been subject of intense work of modeling in situations as various as castaway rescue~\cite{gardecote}, 
foraging animals~\cite{viswaNat,revisitingViswaNat,Viswanathan:2008zm,NewsViewsNature,Animaux1D,Benichou:2005uk,PRE2006,SpecialIssue2006} or proteins  reacting on a specific DNA sequence
\cite{PCCP,intermittentlevy,MirnySlutsky,DNA,Eliazar2007,Lomholt2005,NAR08,epladn,Broek:2008kl,Loverdo:2009a}. 
%In some cases, there is the alternation of two phases, 
%one of  slow motion with detection, 
%the other with fast motion but no detection. 
%Our model is based on this intermittence. 
Among the wide panel of search strategies, the so-called intermittent strategies -- combining  "slow" phases enabling detection of the targets and "rapid" phases 
during which the searcher is unable to detect the targets --  have been proved to be relevant at various scales. 
%before stating the general questions.

Indeed, at the macroscopic scale, 
numerous animal species have been reported to perform 
such kind of intermittent motion~\cite{Bell,animauxObrien} 
while searching either for food, shelter or mate.  In the case of an exploratory behavior, i.e. when the searcher has no previous knowledge or "mental map" of the location of targets, trajectories can be considered as random. Actually, the observed search trajectories  are often described as a  
sequence of ballistic segments  interrupted by much slower phases. 
These slow, and sometimes even immobile~\cite{animauxObrien}, phases  
are not always well characterized, but it seems clear that they are aimed at sensing the environment 
and trying to detect the targets~\cite{kramer}. 
On the other hand, during the fast moving phases, perception is generally degraded so 
that the detection is very unlikely. 
An example of such intermittent behavior is given by the \textit{C.elegans} worm, which  alternates 
between a fast and almost straight displacement (``roaming'')
and a much more sinuous and slower trajectory (``dwelling'')~\cite{CelFunda}.
During this last phase, 
the worm's head, bearing most of its sensory organs, moves 
and touches the surface nearby.

Intermittent strategies 	are actually also relevant at the 
microscopic scale, as exemplified by reaction kinetics in biological cells~\cite{natphys08,PCCP}. 
As the cellular environment is intrinsically out of equilibrium, the
transport of a given tracer particle which has to react  with  a target molecule cannot be described as mere thermal diffusion~: 
if the tracer particle can indeed diffuse freely in the medium, it also intermittently binds and unbinds
 to motor proteins, 
which perform an  active ballistic motion powered by ATP hydrolysis along cytoskeletal filaments~\cite{alberts, karatekid, natphys08}. Such intermittent trajectories of reactive particles 
are observed for example in the case of vesicles before they react with their target membrane proteins~\cite{karatekid}. In that case, targets are not accessible during the ballistic phases when the vesicle is bound to motors, but only during 
the free diffusive phase.

As illustrated in the previous examples the search time is often a limiting quantity whose optimization can be very beneficial for the system -- be it an animal or a single cell. In the case of  intermittent search strategies, the minimization of the search time can be qualitatively discussed~:
on the one hand, the fast but non-reactive phases can appear as a waste of time, since they do not give any chance of target detection. On the other hand, such fast phases can   provide an efficient way to relocate and explore space. This puts forward the following questions~: is it beneficial for the search to perform such fast but non reactive phases?
Is it possible, by properly tuning the kinetic parameters  of trajectories (such as the durations of each of the two phases) to minimize the search time?
These questions have been addressed quantitatively on specific examples in~\cite{Animaux1D,Benichou:2005uk,PRE2006,SpecialIssue2006,natphys08,Benichou:2007fv,Oshanin:2007a,Rojo:2009,Levitz08}, where it was shown that intermittent search strategies can be optimized. In this paper, we perform a systematic analytical study of intermittent search strategies in dimensions 1,2 and 3 and fully  characterize the optimal regimes. This study completes our previous works, and in particular our recent letter~\cite{natphys08}, by providing all calculations details and specifying the validity domains of our approach. It also presents
 new cases of potential relevance to model real search problems.
%Besides the comparison with other models, 
Overall, this systematic approach allows us to identify  robust features of intermittent search strategies. 
In particular, the slow  phase that enables detection 
is often hard to characterize experimentally. 
Here we propose and study 3 distinct  modelings  for this phase, which allows us 
to assess to which extent our results are robust and model dependent. %We add the description of the slow detection phase as a ballistic motion, 
%which allow us to compare our results to the results of Viswanatahn et al~\cite{viswaNat}. 
Our analysis covers in details  intermittent search problems in dimension 1, 2 and 3, and is aimed at giving a quantitative basis -- as complete as possible -- to model real search problems involving intermittent searchers.

We first define our model and give general notations 
that we will use in this article.  
Then we systematically examine each case, studying the search problem in dimension  1, 2 and 3, where
 for each dimension 
 different types of motion in the slow phase are considered. 
Each case is ended by a short summary, and 
 we highlight the main results for each dimension.  
Eventually we synthesize the results 
in the table \ref{recapgeneral} where all cases, their differences and similarities are gathered.
This table finally leads us  
 to draw general conclusions.

\section{Model and notations}

\subsection{Model}

The general framework of the model relies on intermittent trajectories, which have been put forward for example  in ~\cite{Animaux1D}. 
We consider a searcher that switches  between two phases.
The switching rate $\lambda_1$ (resp. $\lambda_2$) from phase 1 to phase 2 (resp. from phase 2 to phase 1)  is  time-independent, which assumes that the searcher has no memory and  implies an  exponential distribution of durations of each phase $i$ 
of mean $\tau_i=1/\lambda_i$. 

Phase 1 denotes the phase of slow motion, 
 during which the target can be detected if it lies within a distance from the searcher which is smaller than a given detection radius  $a$.
$a$ is the maximum distance within which the searcher can get information about target location. 
We propose 3 different modelings of this phase, 
in order to cover various real life situations.
\begin{itemize}
\item In the first modeling of phase 1, hereafter referred to as  the ``static mode'',  the searcher is immobile, and detects the target with probability per unit time $k$ if it lies at a distance  less than $a$.
\item In the second modeling, called the ``diffusive mode'', the searcher performs a continuous diffusive motion, with  diffusion coefficient $D$,
and finds immediately the target if it lies at a distance  less than $a$. 
\item In the last modeling,  called the ``ballistic mode'', the searcher  moves ballistically in a random direction with constant speed $v_l$ and reacts immediately with  the target if it lies at a distance  less than $a$. 
\end{itemize}
Some comments on these different modelings 
of the slow phase 1 are to be made.  The first two modes have already been introduced~\cite{natphys08,PRE2006}, while the analysis of the ballistic mode has never been performed. These 3 modes  schematically cover experimental observations of the behavior of animals searching for food~\cite{Bell,animauxObrien}, where the slow phases of detection are often described as either static, random or with slow velocity.
Several real situations 
are likely to involve a combination of two modes.  For instance the motion of a reactive particle in a cell not bound to motors can be described by a combination of the diffusive and static modes. For the sake of simplicity, here  we treat these modes independently, and our approach can therefore be considered as a limit of more realistic models. 
We note that a similar version of the ballistic mode of detection has been discussed by Viswanathan et al.\cite{viswaNat, james}.
They introduced a model searcher performing  
 randomly oriented ballistic movements (and of power law distributed duration),  with detection capability all along the trajectory. For this 1 state searcher, the search time for a target (which is assumed to disappear after the first encounter) is minimized when the searcher performs a purely ballistic motion and never reorientates. In fact, the ballistic mode version of our model extends this model of 1 state ballistic searcher, by allowing the searcher to switch to a mode of faster motion, but with no perception. As it will be discussed, adding this possibility of intermittence enables a further minimization of the search time.
Finally, combining  these three schematic modes covers a wide range of possible motions, from subdiffusive (even static), 
diffusive, to superdiffusive (even ballistic). 

The phase 2 denotes the fast phase during which the target cannot be found. 
In this phase, the searcher performs  a ballistic motion at constant speed $V$ and random direction, redrawn each time the searcher enters a phase 2, independently of previous phases. 
In real examples correlations between successive phases could exist. 
If correlations are very high, it is close to a 1-dimensional problem 
with all the phases 2 in the same direction, a different problem already treated in \cite{Animaux1D}. 
We consider here the limit of low correlation, that is of a searcher with no memory skills.

We assume that the searcher evolves  in d-dimensional spherical domain of radius $b$, with reflective boundaries and with one centered immobile target
(bounds can be obtained in the case of mobile targets   \cite{Moreau:2003xe,Pascal_suite}). 
 As the searcher does not initially know the target location, 
 we start the walk from a random point of the d-dimensional sphere,
and average the mean target detection time over the initial position.
 This geometry models the case of a single target in a finite domain, and  also provides a good approximation
of an infinite space with regularly spaced targets. Such regular array of targets  corresponds to a mean-field approximation of random distributions of targets, which can be more realistic in some experimental situations. We note that 
% In the real systems, targets localisations vary.  
%Targets distributions are also sometimes described\cite{Bell}
%as random or patched. 
%In this last case, 
%the patch 
%could be treated as a bigger target, 
%with a change of behavior when the first target is encountered, 
%so as to exploit the patch~\cite{BenhamouPatch}. 
in the 1  dimensional case,
 we have shown that a Poisson distribution of targets can lead to significantly different results 
 from the regular distribution\cite{Europhys2006,desordonnegros}. We expect this difference to be less  in dimension 2 and 3, and  
we limit ourselves in this paper to the mean field treatment  for the  sake of simplicity.

\subsection{Notations}

$t_i(\vec{r})$ denotes the mean first passage time on the target, 
for a  searcher {\it starting} in the phase $i$ from point $\vec{r}$, 
where phase $i=1$ is the slow motion phase with detection and 
phase $i=2$ is the fast motion phase without target detection. 
Note that in dimension 1, the space coordinate will be denoted by $x$, 
and in the case of a  ballistic mode for phase 1, the upper index in $t_i^\pm $ stands for  
ballistic motion  with direction $\pm x$. 
The general method which will be used at length in this paper, 
consists in deriving and solving  backward equations for $t_i(\vec{r})$~\cite{gardiner}. 
These linear equations involve derivatives with respect to the starting position $\vec{r}$.

Assuming that the searcher starts in phase 1, the mean  detection time for a target is then defined as~: 
\begin{equation}
t_m=\frac{1}{V(\Omega_d)}\int_{\Omega_d} t_1(\vec{r})d\vec{r}
\end{equation}
with $\Omega_d$ the d-dimensional sphere of radius $b$ and $V(\Omega_d)$ its volume. Unless specified, we will consider the low target density limit $a\ll b$.

Our general aim is to minimize $t_m$ as a function of the mean durations $\tau_1,\tau_2$ of each phase, and in particular to determine under which conditions an intermittent strategy (with finite $\tau_2$) is faster than a usual  1 state search in the phase 1 only, which is given by the limit $\tau_1\to\infty$.  In the static mode, intermittence is necessary for the searcher to move, 
and is therefore always favorable. 
In the diffusive mode, we will compare 
the mean search time with intermittence $t_m$ to the mean search time for a 1 state diffusive searcher  $t_{diff}$, 
and define the gain as $gain=t_{diff}/t_m$.
Similarly in the ballistic mode, we will compare $t_m$ to the mean search time for a 1 state ballistic searcher  $t_{bal}$ and 
define the gain  as $gain=t_{bal}/t_m$. 

Throughout the paper, the upper index 
``opt'' is used to denote the value of a parameter or variable at the minimum of $t_m$ .

%\part{Detailed results}

\section{Dimension 1}

Besides the fact that it involves more tractable calculations, the 1-dimensional case can also be interesting to model real search problems. 
At the microscopic scale,  tubular structures of cells such as axons or dendrites in neurons can be considered as 1-dimensional~\cite{alberts}. 
The active transport of reactive particles, which alternate diffusion phases and ballistic phases when bound to molecular motors, can be schematically captured by our model with diffusive mode~\cite{natphys08}.
At the macroscopic scale, one could cite animals like ants~\cite{fourcassieWallFollowingAnts} which 
tend to follow tracks or one-dimensional boundaries.

\subsection{Static mode}

\imagea{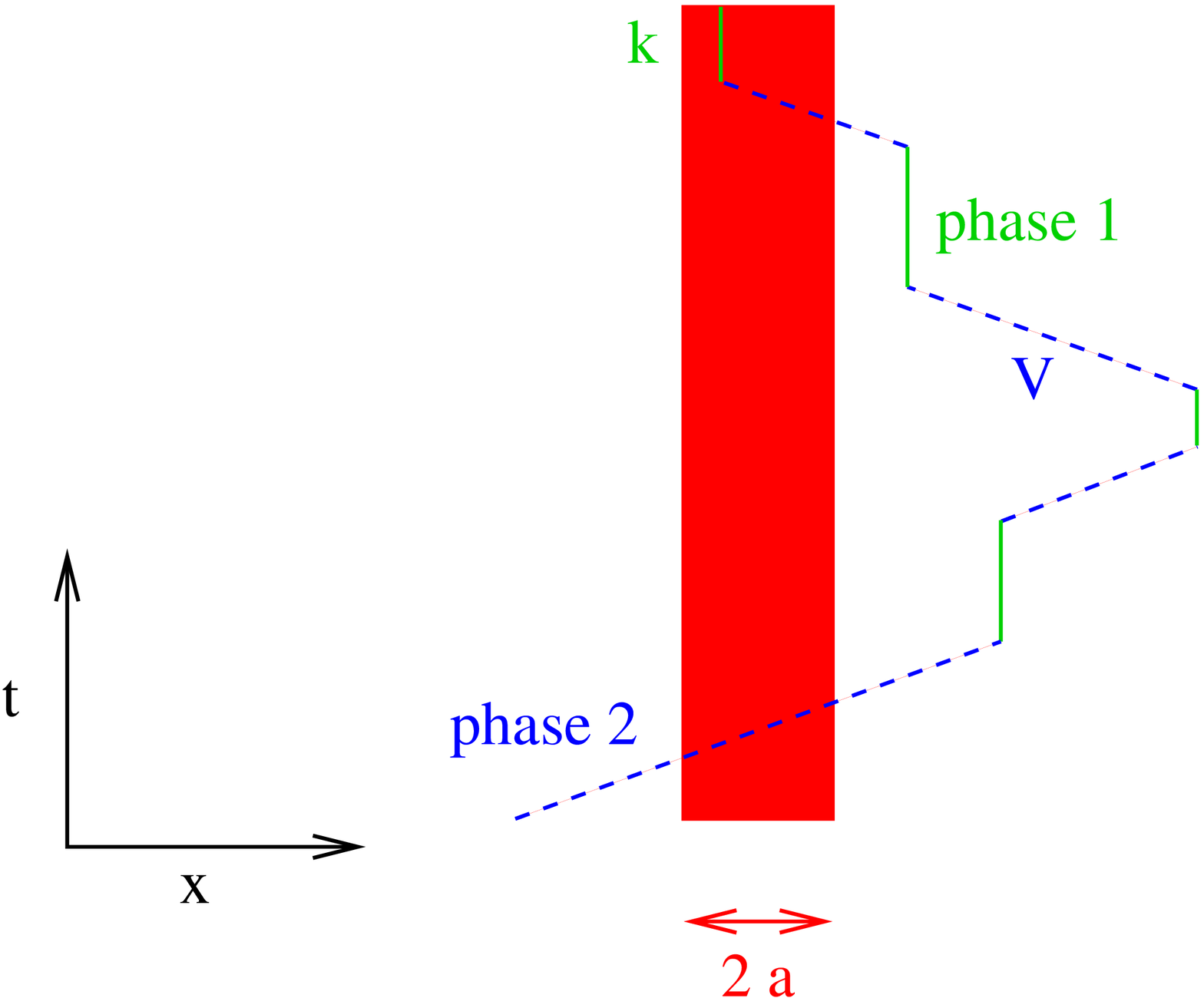}{Static mode in  dimension 1}{1k}{5}

In this section we assume that the detection phase is modeled by  the static mode. Hence the searcher does not move during the reactive phase 1, 
and has a fixed reaction  rate $k$ per unit time  with the target if it lies within its detection radius  $a$. 
It is the limit of a very slow searcher in  the reactive phase.

\subsubsection{Equations}
Outside the target (for $x>a$, we have the following backward equations for the mean first-passage time~: 
\begin{equation}
 V \frac{d t_2^+}{dx}+\frac{1}{\tau_2}(t_1-t_2^+)=-1
\end{equation}
\begin{equation}
 - V \frac{d t_2^-}{dx}+\frac{1}{\tau_2}(t_1-t_2^-)=-1
\end{equation}
\begin{equation}
 \frac{1}{\tau_1}\left( \frac{t_2^++t_2^-}{2} -t_1 \right)=-1
\end{equation}
Inside the target ($x\le a$), the first two equations are identical, but the third one is written~: 
\begin{equation}
  \frac{1}{\tau_1} \frac{t_2^++t_2^-}{2} -\left(\frac{1}{\tau_1}+k  \right)t_1=-1.
\end{equation}
We introduce $t_2=\frac{t_2^++t_2^-}{2}$ and $t_2^d=\frac{t_2^+-t_2^-}{2}$. 
Then outside the target we have the following equations~: 
\begin{equation}
  V \frac{d t_2}{dx}-\frac{1}{\tau_2}t_2^d=0
\end{equation}
\begin{equation}
  V^2 \tau_2 \frac{d^2 t_2}{dx^2}+\frac{1}{\tau_2}(t_1-t_2)=0
\end{equation}
\begin{equation}
 \frac{1}{\tau_1}(t_2-t_1)=-1.
\end{equation}
Inside the target the first two  equations are identical, but the last one writes~: 
\begin{equation}
  \frac{1}{\tau_1}t_2-\left(\frac{1}{\tau_1}+k  \right)t_1=-1.
\end{equation}
Due to the symmetry $x \leftrightarrow -x$, we can restrict the study to the part $x \in [0,a]$ and the part $x\in [a,b]$. 
This symmetry also implies ~: 
\begin{equation}
\left. \frac{dt_2^{in}}{dx}\right|_{x=0}=0
\end{equation}
\begin{equation}
\left. \frac{dt_2^{out}}{dx}\right|_{x=b}=0.
\end{equation}
In addition,  continuity at $x=a$ for $t_2^+$ and $t_2^-$ gives: 
\begin{equation}
t_2^{in}(x=a)=t_2^{out}(x=a)
\end{equation}
\begin{equation}
t_2^{d,in}(x=a)=t_2^{d,out}(x=a).
\end{equation}
This set of linear equations enables an explicit determination of  $t_1$, $t_2$, $t_2^d$  inside and outside the target.

\subsubsection{Results}
An exact analytical expression of the mean first passage time at the target is then given by ~: 
\begin{equation}\label{tm1Dvk}
 t_m = \frac{\tau_1+\tau_2}{b}\left( \frac{b}{k\tau_1} + \frac{(b-a)^3}{3V^2 \tau_2^2}+\frac{\beta (b-a)^2}{V\tau_2}coth\left(\frac{a}{V\tau_2 \beta} \right) + (b-a)\beta^2\right)
\end{equation}
where $\beta=\sqrt{(k\tau_1)^{-1} +1}$.

\begin{figure}[h!]
   \begin{minipage}[c]{.46\linewidth}
\begin{center}
      \includegraphics[width=6cm]{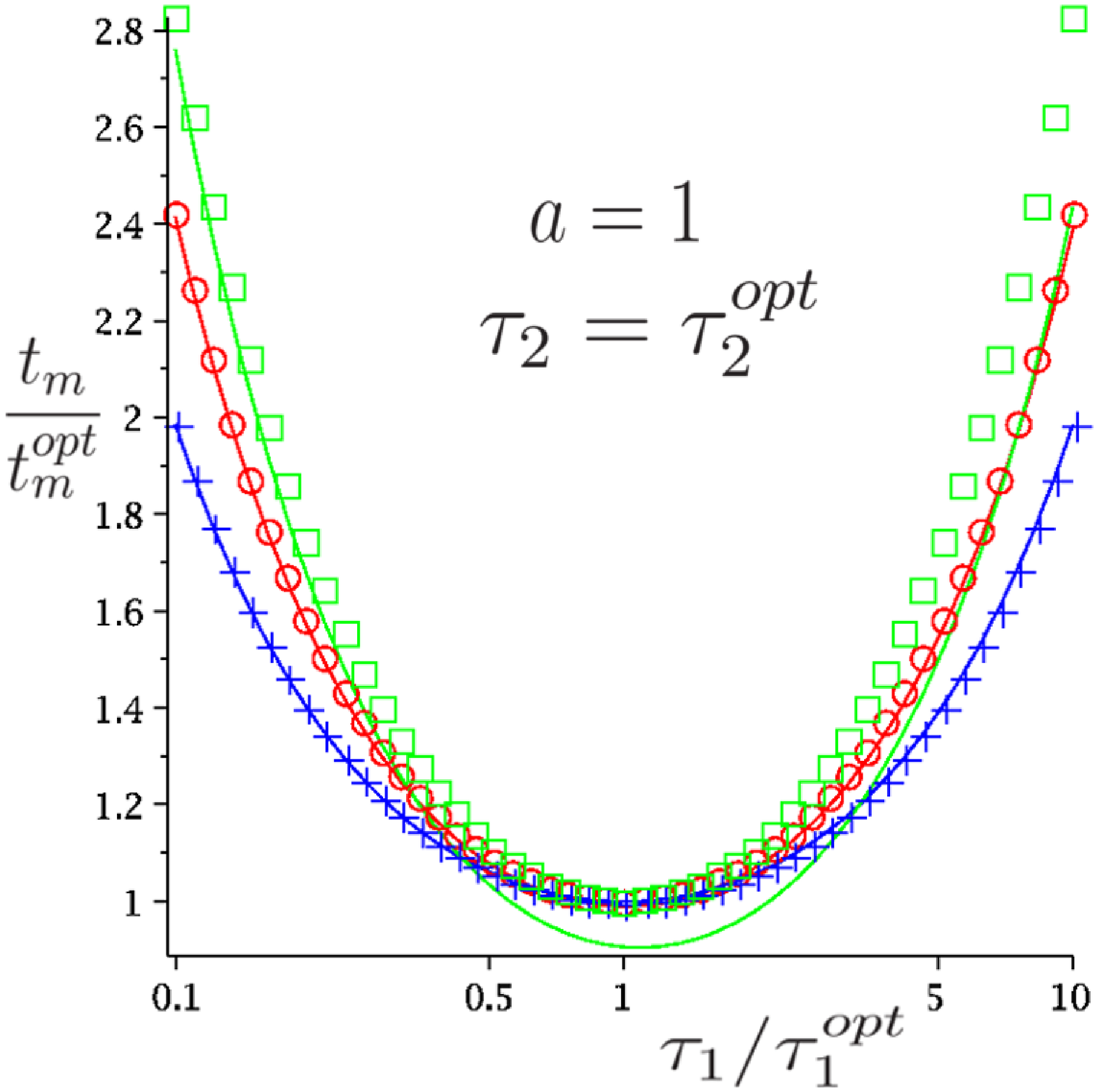}

\end{center}
   \end{minipage} \hfill
   \begin{minipage}[c]{.46\linewidth}
\begin{center}
      \includegraphics[width=6cm]{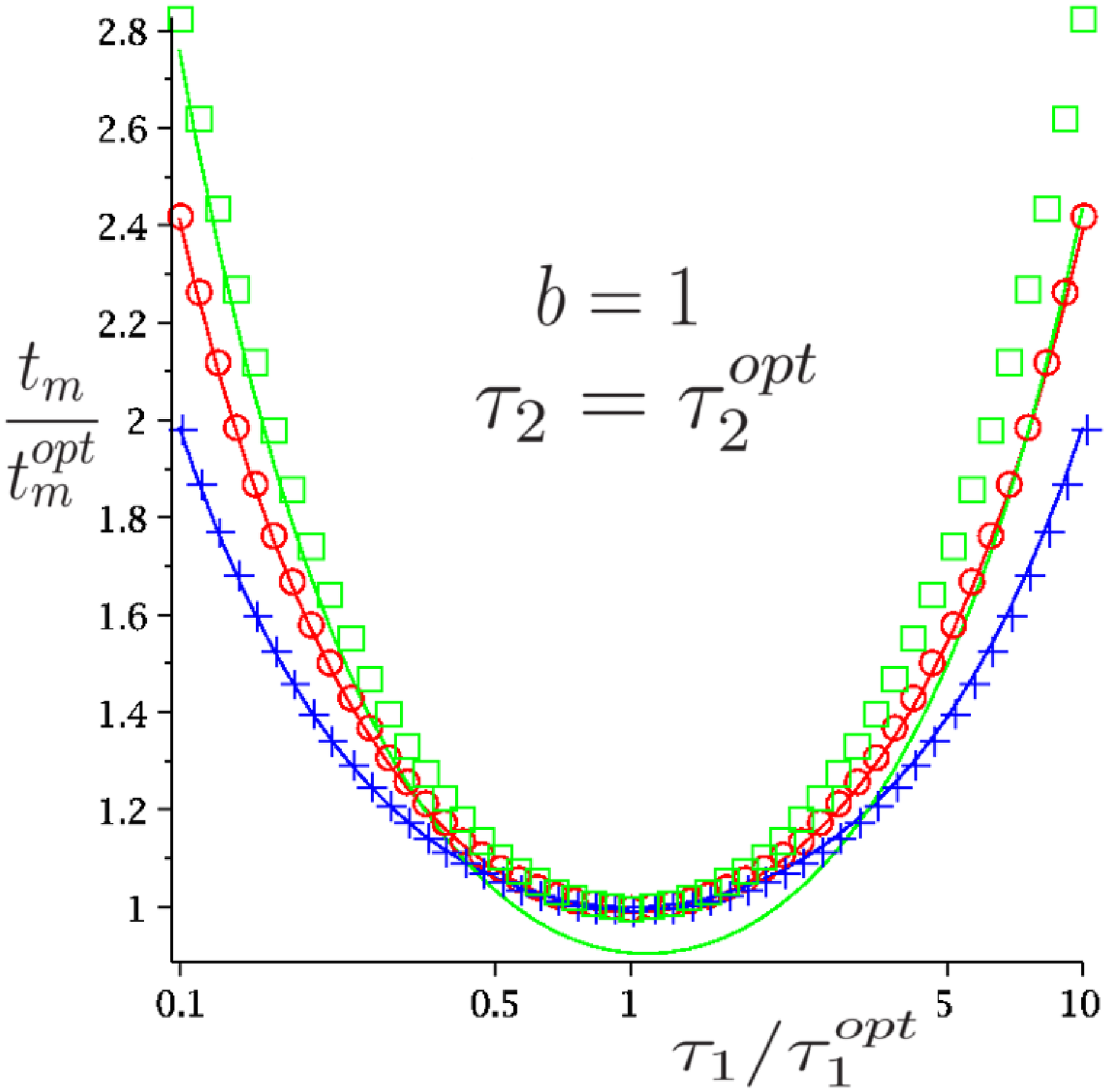}

\end{center}
   \end{minipage}\hfill

   \begin{minipage}[c]{.46\linewidth}
\begin{center}
      \includegraphics[width=6cm]{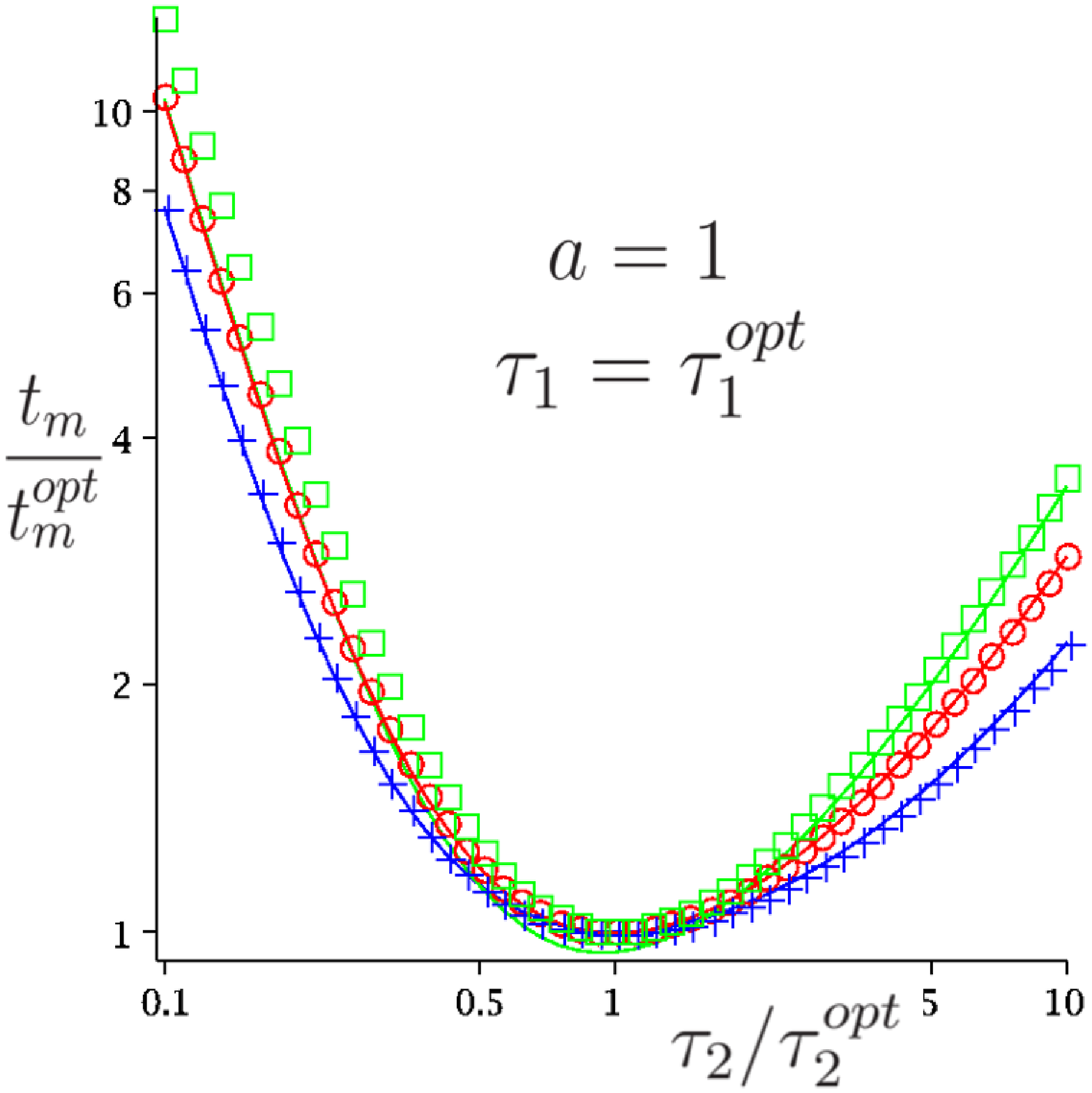}

\end{center}
   \end{minipage} \hfill
   \begin{minipage}[c]{.46\linewidth}
\begin{center}
      \includegraphics[width=6cm]{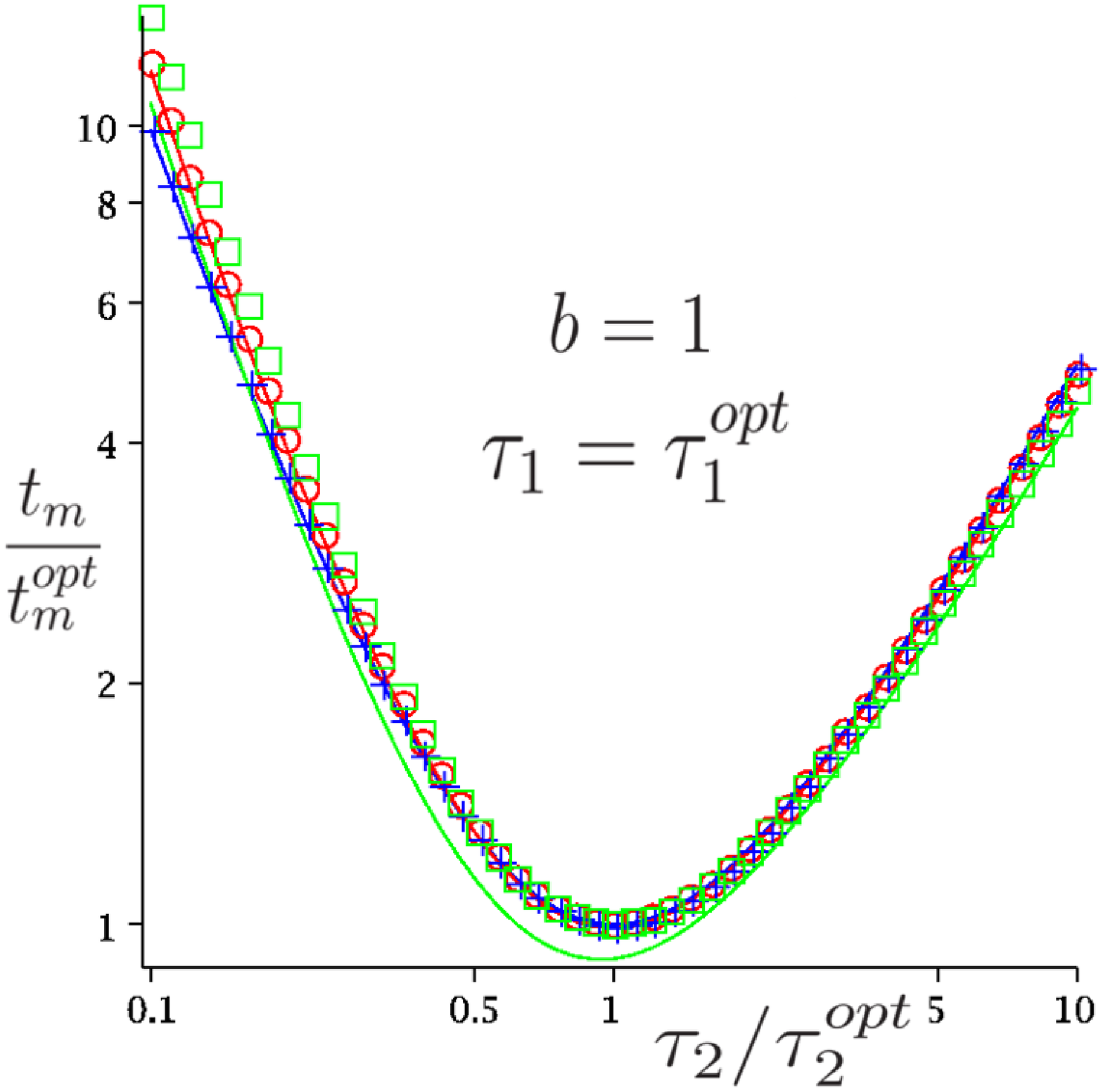}

\end{center}
   \end{minipage}\hfill

\caption{(Color online) Static mode in 1 dimension. Exact expression of $t_m$ \refm{tm1Dvk} (lines) compared to the approximation of $t_m$  \refm{tm1Dvksimp} (symbols), 
both rescaled by $t_m^{opt}$ \refm{topt1Dvk}. $\tau_1^{opt}$ from \refm{1Dvk_t1},  $\tau_2^{opt}$ from \refm{1Dvk_t2}. $V=1$, $k=1$. $b/a=10$ (green, squares), $b/a=100$ (red, circles), $b/a=1000$ (blue, crosses).}\label{1Dvk_approx}
\end{figure}

In order to determine the optimal strategy, we need to simplify this expression, 
by expanding  \refm{tm1Dvk} in the regime $b\gg a$~: 
\begin{equation}\label{tm1Dvkb}
 t_m = (\tau_1+\tau_2)\left( \frac{1}{k\tau_1} + \frac{b^2}{3V^2 \tau_2^2}+\frac{\beta b}{V\tau_2}coth\left(\frac{a}{V\tau_2 \beta} \right) + \beta^2\right)
\end{equation}
We make the further assumption   $\frac{a}{V\tau_2} \ll 1$ and obtain, using $\beta >1 $~: 
\begin{equation}
 t_m = (\tau_1+\tau_2)\left( \frac{1}{k\tau_1} + \frac{b^2}{3V^2 \tau_2^2}+\beta^2 \frac{ b}{a} + \beta^2\right)
\end{equation}
Since $\beta > 1$ and $\beta > 1/(k \tau_1)$, we obtain in the limit  $b\gg a$ ~: 
\begin{equation}\label{tm1Dvksimp}
 t_m = (\tau_1+\tau_2)\left( \frac{b^2}{3V^2 \tau_2^2}+\left(\frac{1}{k\tau_1}+1\right) \frac{ b}{a}\right)
\end{equation}
This simple expression gives a very good and convinient approximation of the mean first passage time at the target as shown in \refi{1Dvk_approx}. 

We use this approximation \refm{tm1Dvksimp} to 
find $\tau_1$ and $\tau_2$ values which minimize $t_m$~:
\begin{equation}\label{1Dvk_t1}
 \tau_1^{opt}=\sqrt{\frac{a}{Vk}}\left(\frac{b}{12a} \right)^{1/4}
\end{equation}
\begin{equation}\label{1Dvk_t2}
 \tau_2^{opt}=\frac{a}{V}\sqrt{\frac{b}{3a}}.
\end{equation}
It can be noticed than $\tau_2^{opt}$ does not depend on $k$. 
Then the expression of the minimal value of the search time $t_m$ \refm{tm1Dvksimp} with $\tau_1=\tau_1^{opt}$ and $\tau_2=\tau_2^{opt}$ is~:
\begin{equation}\label{topt1Dvk}
 t_m^{opt} = \frac{b}{ak}\left(\frac{b}{3a} \right)^{1/4}\left( \sqrt{\frac{2bk}{3V}}\left(\frac{3a}{b}\right)^{1/4}+1 \right)\left( \sqrt{\frac{2ka}{V}}+ \left(\frac{3a}{b}\right)^{1/4}\right).
\end{equation}

\subsubsection{Summary}\label{1dvkf}

For the static modeling of the detection phase in dimension 1, in the $b \gg a$ limit, the mean detection time is~:
\begin{equation}
 t_m = (\tau_1+\tau_2)\left( \frac{b^2}{3V^2 \tau_2^2}+\left(\frac{1}{k\tau_1}+1\right) \frac{ b}{a}\right).
\end{equation}
Intermittence is always favorable, 
and the optimal strategy is realized when $ \tau_1^{opt}=\sqrt{\frac{a}{Vk}}\left(\frac{b}{12a} \right)^{1/4}$ and $\tau_2^{opt}=\frac{a}{V}\sqrt{\frac{b}{3a}}$. Importantly,  the optimal duration of the relocation phase 
does not depend on $k$, i.e. on the description of the detection phase.

\subsection{Diffusive mode}

\imagea{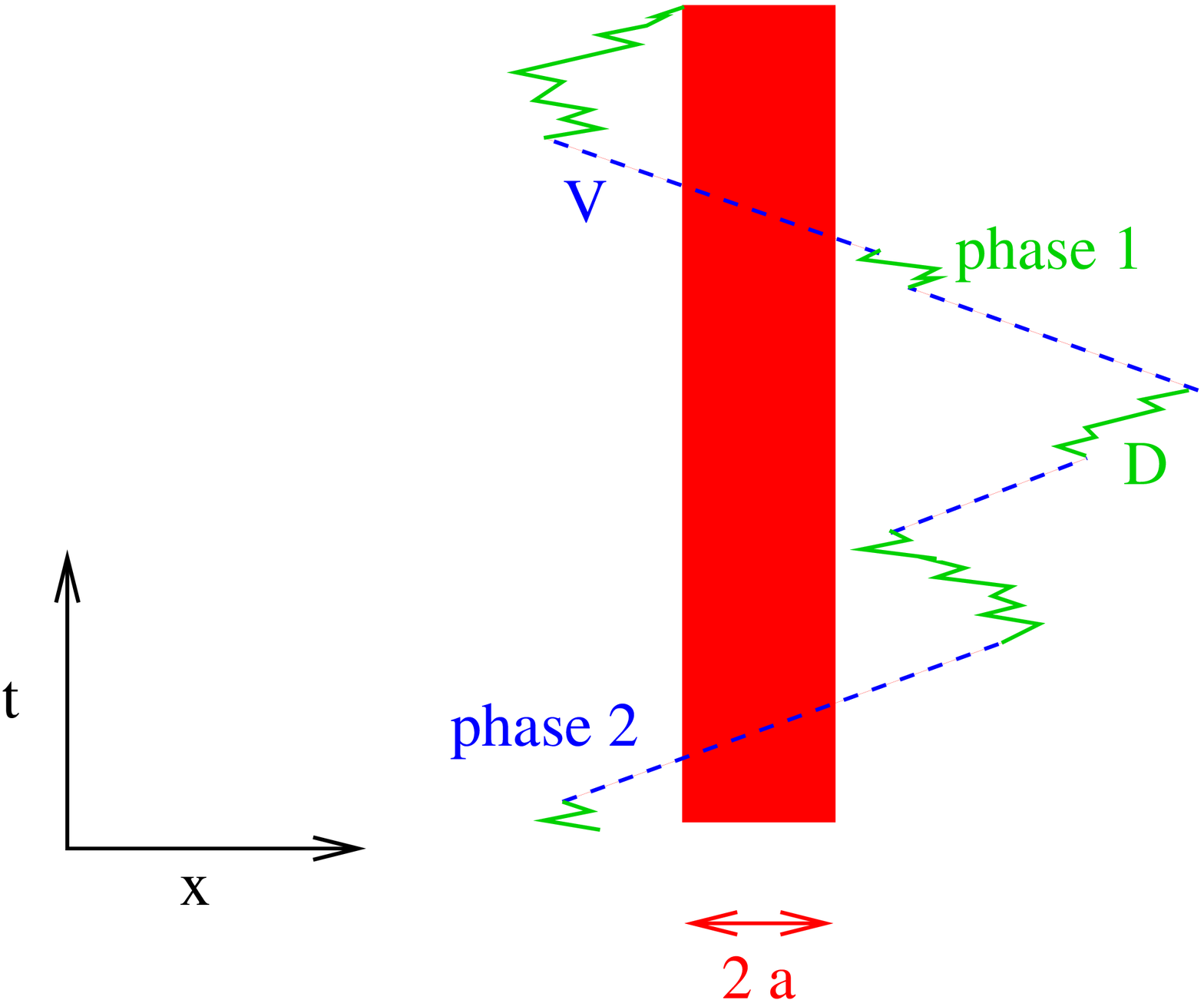}{Diffusive mode in one dimension}{1D}{5}

We now turn to the diffusive modeling of the detection phase. 
The detection phase 1 is now diffusive, with immediate detection
of the target if it is within a radius $a$ from the searcher. 

\subsubsection{Equations}

Along the same lines, the backward equations for the mean first-passage time read outside the target ($x>a$) ~:
\begin{equation}
 V \frac{d t_{2}^+}{dx} + \frac{1}{\tau_2} (t_1-t_2^+)=-1
\end{equation}
\begin{equation}
-V \frac{d t_{2}^-}{dx} + \frac{1}{\tau_2} (t_1-t_2^-)=-1
\end{equation}
\begin{equation}
 D \frac{d^2 t_{1}}{dx^2} + \frac{1}{\tau_1} \left(\frac{t_2^+}{2}+\frac{t_2^-}{2}-t_1 \right)=-1,
\end{equation}
and inside the target ($x\le a$)~: 
\begin{equation}
  V \frac{d t_{2}^+}{dx} - \frac{1}{\tau_2} t_2^+=-1
\end{equation}
\begin{equation}
-V \frac{d t_{2}^-}{dx} - \frac{1}{\tau_2} t_2^-=-1
\end{equation}
\begin{equation}
 t_{1}=0.
\end{equation}
We introduce the variables $t_2=\frac{ t_{2}^+ +t_{2}^-}{2}$
and $t_2^{d}=\frac{ t_{2}^+ -t_{2}^-}{2}$.
This leads to the following system outside the target ($x>a$)~:
\begin{equation}
V \frac{d t_2}{dx} = \frac{1}{\tau_2} t_2^{d}
\end{equation}
\begin{equation}
 V^2 \tau_2 \frac{d t_2}{dx} + \frac{1}{\tau_2} (t_1-t_2)=-1
\end{equation}
\begin{equation}
 D \frac{d^2 t_{1}}{dx^2} + \frac{1}{\tau_1} (t_2-t_1 )=-1,
\end{equation}
and inside the target ($x\le a$)~:
\begin{equation}
  V \frac{d t_{2,in}}{dx} = \frac{1}{\tau_2} t_{2,in}^{d}
\end{equation}
\begin{equation}
 V^2 \tau_2 \frac{d t_{2,in}}{dx} - \frac{1}{\tau_2} t_{2,in}=-1 
\end{equation}
\begin{equation}
 t_{1}=0.
\end{equation}
Interestingly, this system is exactly of the same type that what would be obtained with 2 diffusive phases, 
with $D_2^{eff}= V^2\tau_2$ in phase 2.  Boundary conditions result from   continuity and symmetry~:
\begin{equation}
 t_1(a)=0
\end{equation}
\begin{equation}
 t_2^+(a)=t_{2,in}^+(a)
\end{equation}
\begin{equation}
 t_2^-(a)=t_{2,in}^-(a)
\end{equation}
\begin{equation}
 \left.\frac{dt_2}{dx}\right|_{x=b}=0
\end{equation}
\begin{equation}
 \left.\frac{dt_1}{dx}\right|_{x=b}=0
\end{equation}
\begin{equation}
 \left. \frac{dt_{2,in}}{dx}\right|_{x=0}=0.
\end{equation}

\subsubsection{Results}\label{r1dvd1}

Standard but lengthy calculations lead to an exact expression of mean first detection time of the target $t_m$ given in \refaa{a1dvd1}.
%\subsubsection{Numerical study}\label{r1dvd2}
We first studied numerically the minimum of $t_m$  in \refaa{a1dvd2}, and identified 
 3 regimes. In the first regime ($b<\frac{D}{V}$) intermittence is not favorable. For $b >\frac{D}{V}$ intermittence is favorable  and two regimes ($ \frac{bD^2}{a^3 V^2}<1$ and $ \frac{bD^2}{a^3 V^2}>1$) should be distinguished . 
We now study analytically each of these regimes.

\subsubsection{Regime where intermittence is not favorable~: $b<\frac{D}{V}$}

If $b < \frac{D}{V}$, 
the time spent to explore the search space is smaller in the diffusive phase than in the ballistic phase.
Intermittence cannot be favorable in this regime, 
as confirmed by the numerical study in \refaa{a1dvd2}.

Without intermittence, 
the searcher only performs diffusive motion and the problem can be solved straightforwardly. The backward equations read  
$t_{diff}=0$ inside the target ($x\le a$), 
and outside the target ($x>a$)~: 
\begin{equation}
 D \frac{d^2 t_{diff}}{dx^2}=-1.
\end{equation}
Since  $t_{diff} (x=a)=0$ and $\left. \frac{dt_{diff}}{dx}\right|_{x=b} =0$, we get 
$t_{diff} (x) = \frac{1}{2D}((b-a)^2-(b-x)^2)$. 
The mean first passage time at the target then reads~: 
\begin{equation}\label{tdiff1D}
 t_{diff}= \frac{(b-a)^3}{3Db},
\end{equation}
which in  the limit $b\gg a$ leads to~:
\begin{equation}\label{tdiff1Ds}
t_{diff} \simeq \frac{b^2}{3D}.
\end{equation}

\subsubsection{Optimization in the first regime where  intermittence is favorable~:  $b<\frac{D}{V}$ and $ \frac{bD^2}{a^3 V^2} \gg 1$}\label{r1dvd3}

As explained in details in \refaa{a1dvd3},  we use the approximation of low target density ($b \gg a$), 
and we use assumptions on the dependence of   $\tau_1^{opt}$ and $\tau_2^{opt}$ on $b$ and $a$.  
These assumptions lead to the following  approximation of the mean first passage time~: 
\begin{equation}\label{tmn}
 t_m=  \left(\tau_1+\tau_2\right)b\left(\frac{b}{3V^2\tau_2^2}+\frac{1}{\sqrt{D\tau_1}} \right).
\end{equation}
We checked numerically that this expression gives a good approximation of $t_m$ 
in this regime, in particular around the optimum (\refi{tcasn}).
\doublimagem{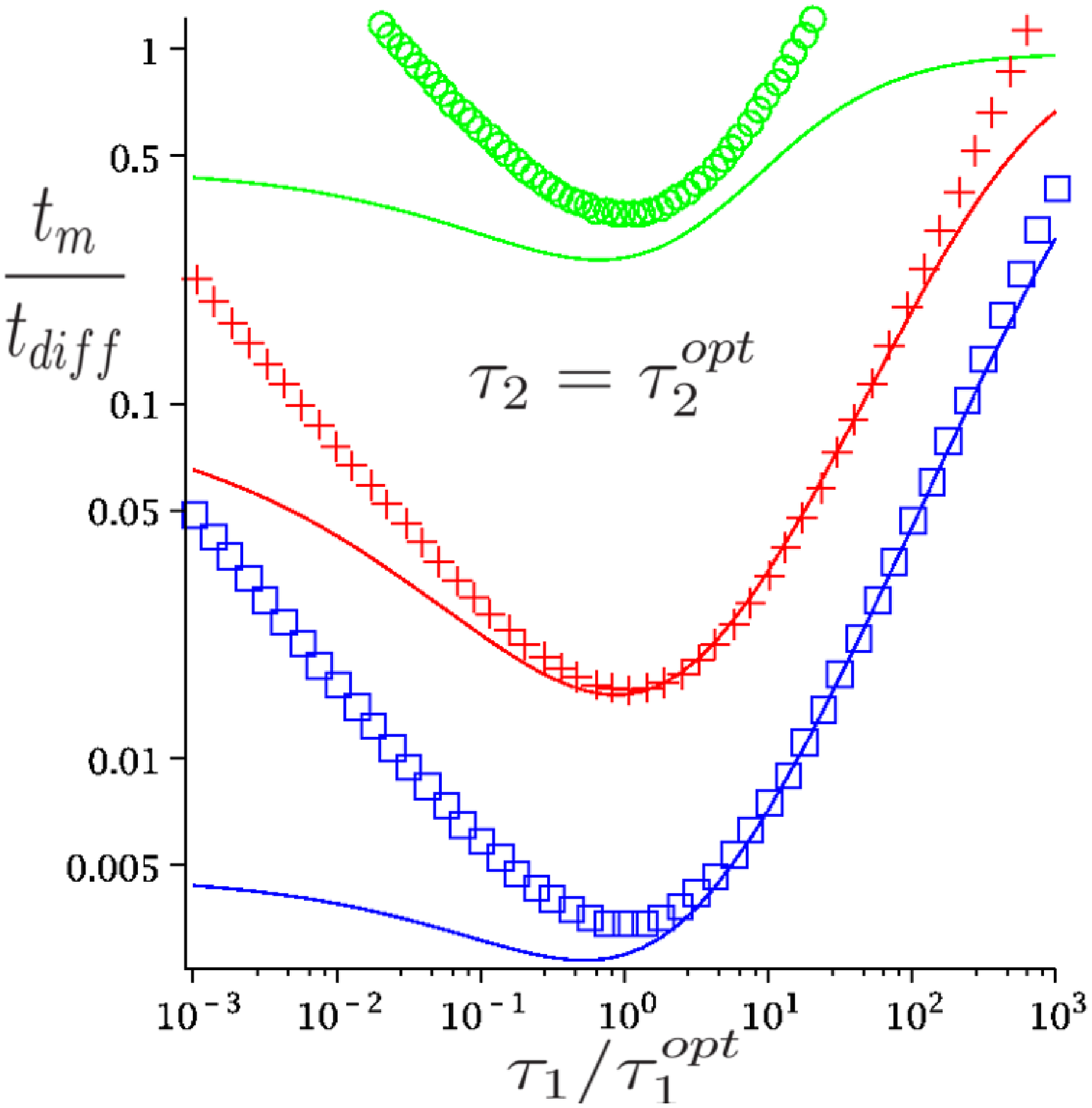}
{}
{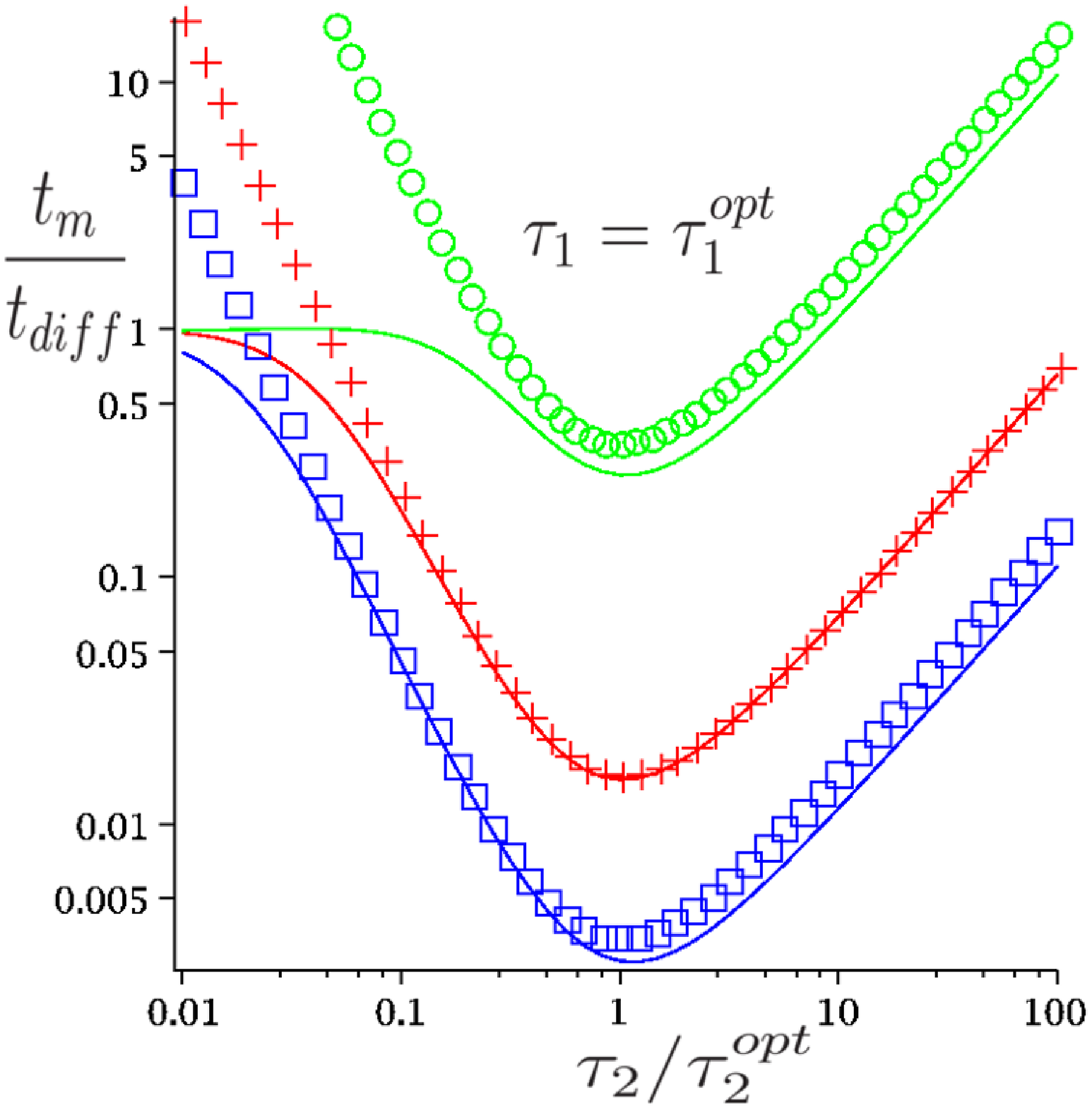}
{}
{Diffusive mode in 1 dimension. $\frac{t_m}{t_{diff}}$, $t_{diff}$ from \refm{tdiff1Ds}, and $t_m$  exact expression  \refm{tm1DvD} (line), approximation in the regime of favorable intermittence and $ \frac{bD^2}{a^3 V^2} \gg 1$ \refm{tmn} (symbols). $a=1$ and $b=100$ (green, circles), $a=1$, $b=10^4$ (red, crosses), $a=10$, $b=10^5$ (blue, squares). $D=1$, $V=1$. $\tau_1^{opt}$ is from expression \refm{l11dvd1}, $\tau_2^{opt}$ is from expression \refm{l21dvd1}}{tcasn}
The simplified $t_m$ expression \refm{tmn} is minimized for~:
\begin{equation}\label{l11dvd1}
 \tau_1^{opt}=\frac{1}{2} \sqrt [3]{\frac{2b^2D}{9V^4}}
\end{equation}
\begin{equation}\label{l21dvd1}
 \tau_2^{opt}=\sqrt [3]{\frac{2b^2D}{9V^4}}
\end{equation}
\begin{equation}\label{toptn}
 t_m^{opt} \simeq  \sqrt[3]{\frac{3^5}{2^4} \frac{b^4}{DV^2}}.
\end{equation}
This compares to the case without  intermittence \refm{tdiff1D} according to~:
\begin{equation}\label{gainn}
 gain^{opt}=\frac{t_{diff}}{t_m^{opt}}\simeq \sqrt[3]{\frac{2^4}{3^8}} \left(\frac{bV}{D} \right)^{\frac{2}{3} } \simeq 0.13 \left(\frac{bV}{D} \right)^{\frac{2}{3}} .
\end{equation}
These results are in agreement with numerical minimization of the exact $t_m$ (\reft{tabnum} in \refaa{a1dvd2}).

\subsubsection{Optimization in the second regime where  intermittence is favorable  :  $b<\frac{D}{V}$ and  $1 \gg \frac{bD^2}{a^3 V^2}$}\label{r1dvd4}

\doublimagem{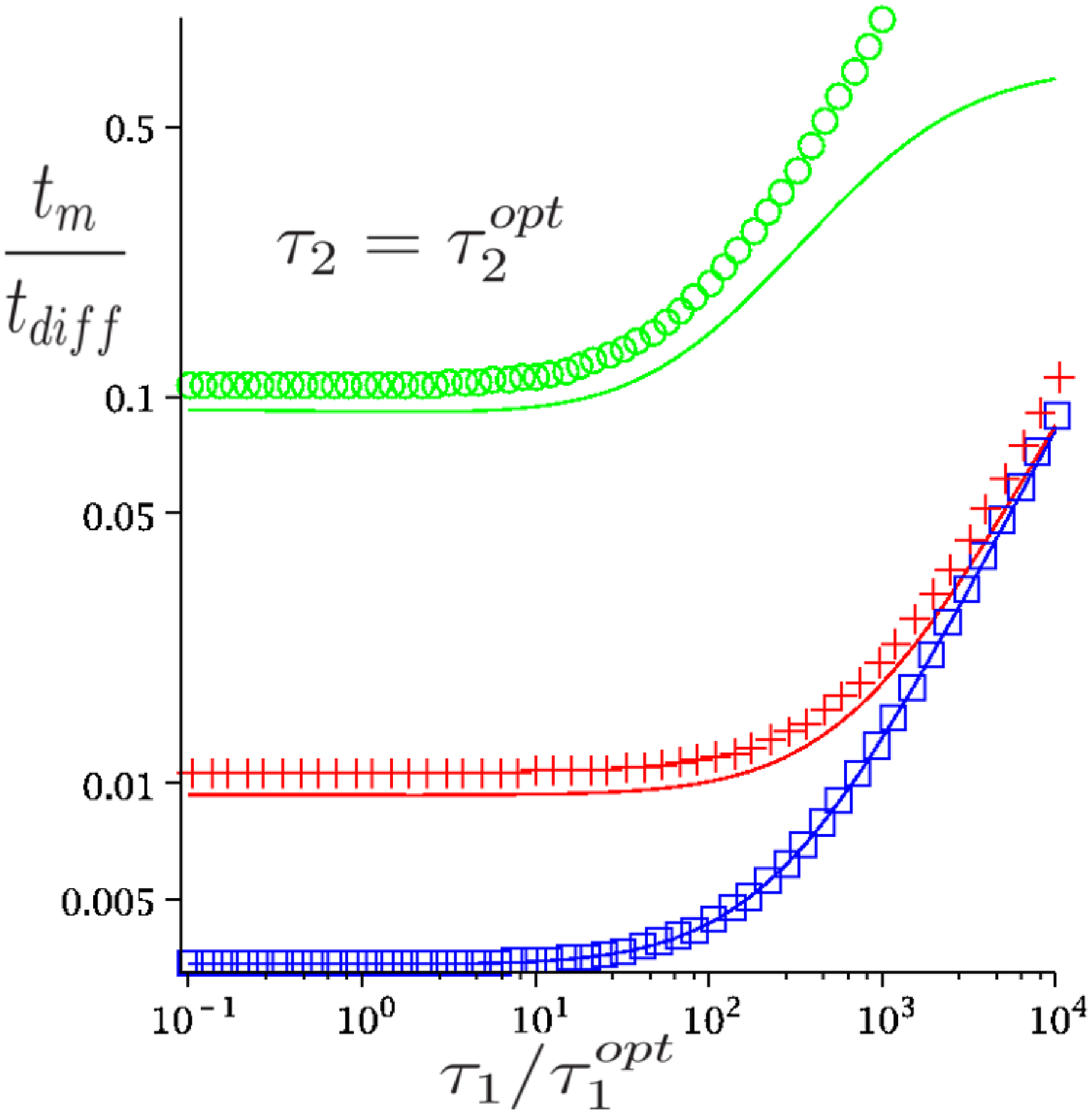}
{ }
{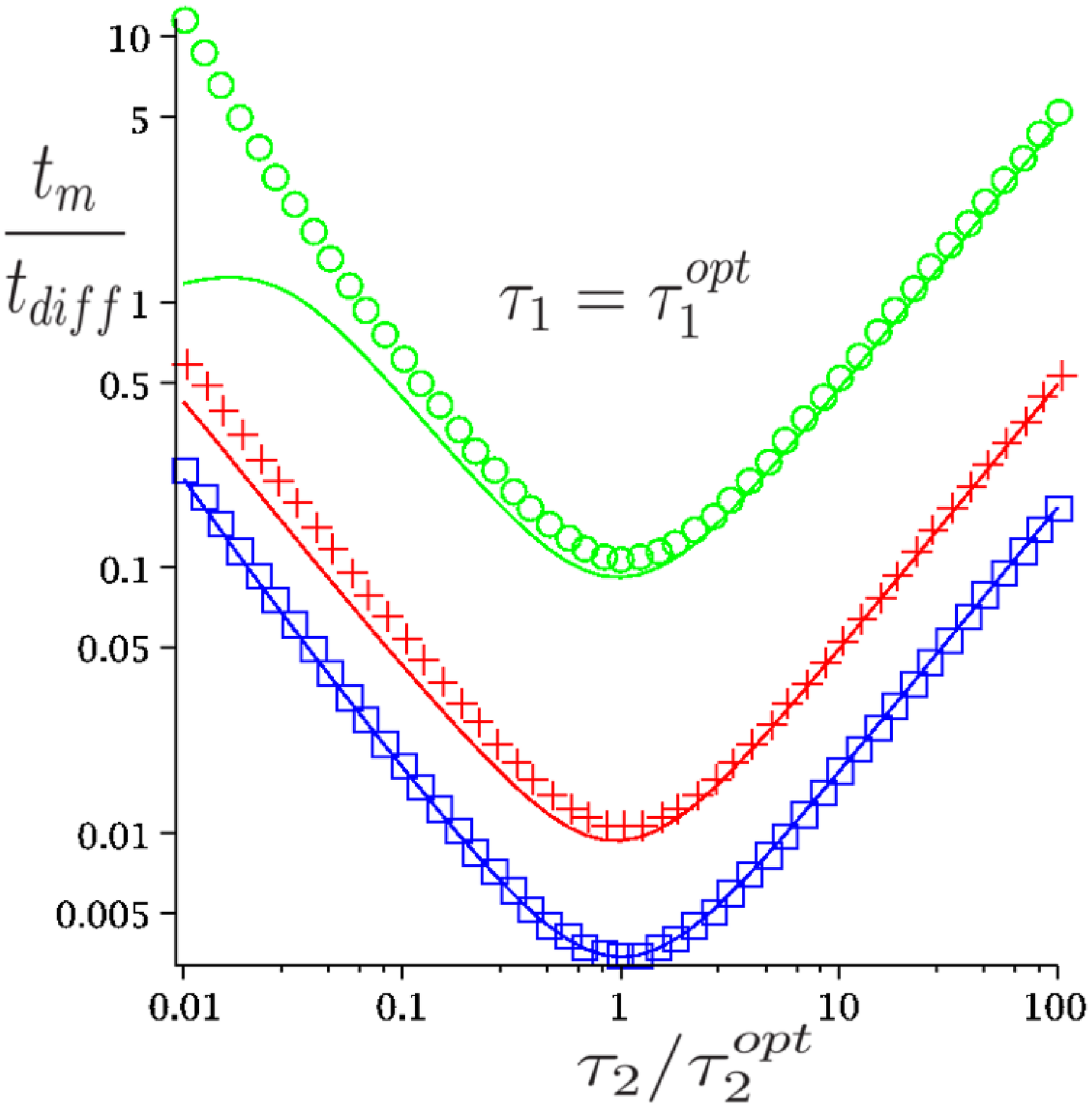}
{}
{Diffusive mode in 1 dimension. $\frac{t_m}{t_{diff}}$, $t_{diff}$ from \refm{tdiff1Ds}, and $t_m$  exact expression  \refm{tm1DvD} (line), approximation in the regime of favorable intermittence and $ \frac{bD^2}{a^3 V^2} \ll 1$ \refm{tsimp1DvD} (symbols). $a=10$ and $b=100$ (green, circles), $a=10$, $b=1000$ (red, crosses), $a=100$, $b=10^4$ (blue, squares). $D=1$, $V=1$. $\tau_1^{opt}$ is from expression \refm{l11dvd2}, $\tau_2^{opt}$ is from expression \refm{l21dvd2}}{tsimpfig}

We start from the exact expression of  $t_m$ \refm{tm1DvD}. As detailed in \refaa{a1dvd4}, 
we make assumptions on the dependence of $\tau_1^{opt}$ and $\tau_2^{opt}$ with $b$ and $a$, and use 
the assumptions that $b\gg a$ and $1 \gg \frac{bD^2}{a^3 V^2}$. It leads to~:
\begin{equation}\label{tsimp1DvD}
 t_m \simeq \frac{b}{a}(\tau_1+\tau_2)\left(\frac{a}{a+\sqrt{D\tau_1}}+\frac{ab}{3V^2\tau_2^2} \right).
\end{equation}
This expression gives a good approximation of $t_m$, at least around the optimum (\refi{tsimpfig}), which is characterized by:
\begin{equation}\label{l11dvd2}
 \tau_1^{opt}=\frac{D b}{48V^2 a}
\end{equation}
\begin{equation}\label{l21dvd2}
 \tau_2^{opt} = \frac{a}{V} \sqrt{\frac{b}{3a}}
\end{equation}
\begin{equation}
  t_m^{opt}\simeq \frac{2a}{v\sqrt{3}}\left(\frac{b}{a} \right)^{3/2}
\end{equation}
\begin{equation}\label{gain1dvd2}
 gain\simeq \frac{1}{2\sqrt{3}}\frac{aV}{D}\sqrt{\frac{b}{a}}.
\end{equation}
These results are in very good agreement with numerical data  (\reft{tabnum} in \refaa{a1dvd2}).
Note that the gain can be very large at low target density.

\subsubsection{Summary}

We calculated explicitly the mean first passage time $t_m$ 
in the case where the detection phase is modeled by  the diffusive mode.  
We minimized $t_m$ as a function of $\tau_1$ and $\tau_2$, the mean phases durations, 
with the assumption $a\ll b$. 
There are three regimes: 
\begin{itemize}
\item  when $b <\frac{D}{V}$, intermittence is not favorable. $\tau_1^{opt}\to \infty$, $\tau_2^{opt} \to 0$, $t_m^{opt} =t_{diff}\simeq \frac{b^2}{3D}$
\item  when $b > \frac{D}{V}$ and  $\frac{bD^2}{a^3 V^2}\gg 1$, intermittence is favorable, 
with $\tau_2^{opt}=2\tau_1^{opt}=\sqrt [3]{\frac{2b^2D}{9V^4}}$, and $t_m^{opt} \simeq  \sqrt[3]{\frac{3^5}{2^4} \frac{b^4}{DV^2}}$
\item when $b > \frac{D}{V}$ and  $\frac{bD^2}{a^3 V^2}\ll 1$,  intermittence is favorable, with 
$\tau_1^{opt}=\frac{D b}{48V^2 a}$, $\tau_2^{opt}=\frac{a}{V} \sqrt{\frac{b}{3a}}$,  $t_m^{opt}\simeq \frac{2a}{v\sqrt{3}}\left(\frac{b}{a} \right)^{3/2}$.
\end{itemize}
This last regime is of particular interest, since the value obtained for $\tau_2^{opt}$ 
is the same as in the static mode \refs{1dvkf}.

\subsection{Ballistic mode}

\imagea{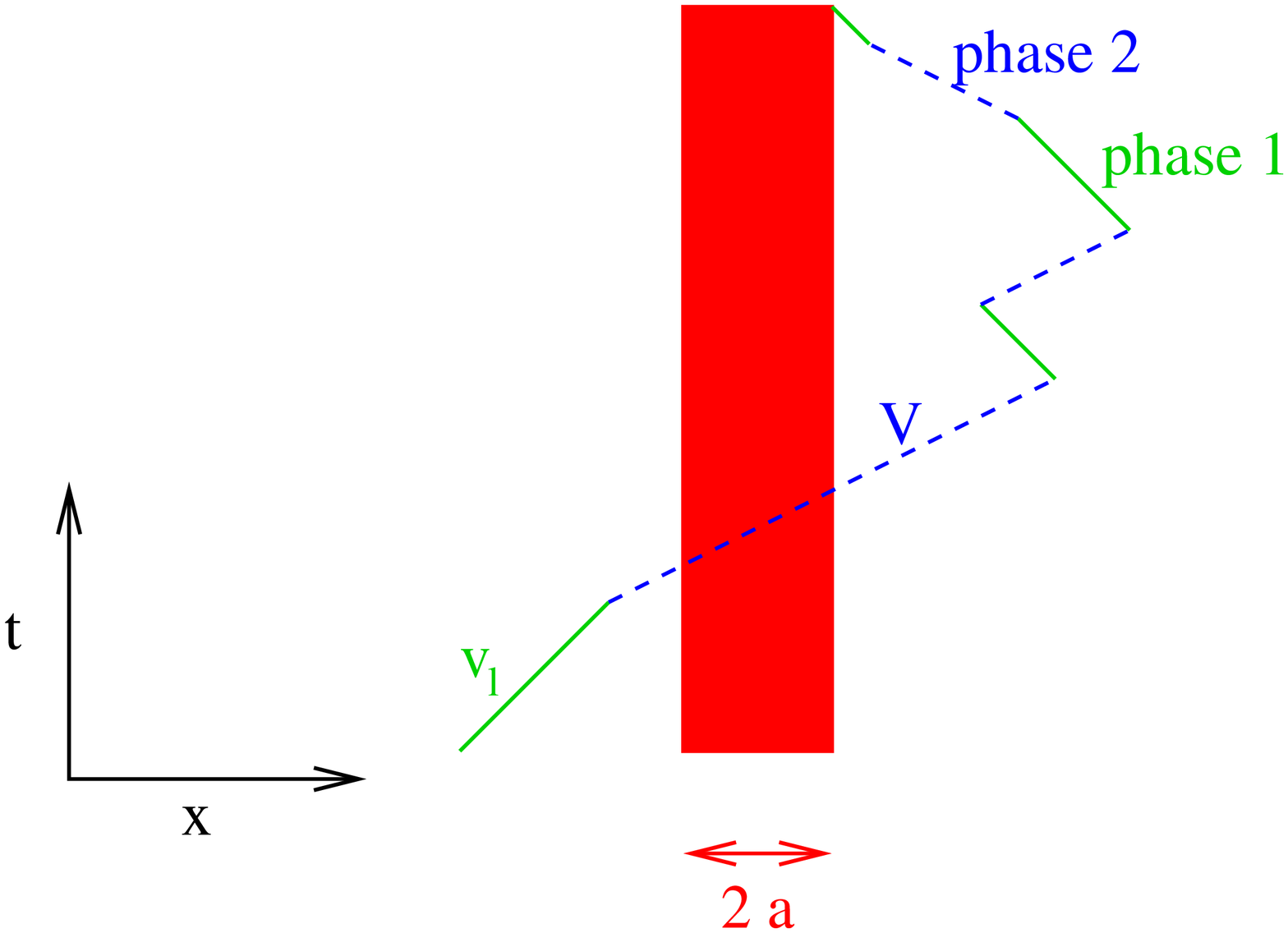}{Ballistic mode in one dimension}{1v}{5}

We now treat the case where the detection  phase 1 is modeled by the  ballistic mode. This model schematically accounts for the general observation that  speed often degrades perception abilities.  Our model corresponds to the extreme case where only two modes are available~: either the motion is slow and the target can be found, 
or the motion is fast and the target cannot be found. Note that this model  can be compared  with~\cite{viswaNat}.

\subsubsection{Equations}

The backward equations read outside the target ($x>a$)~:
\begin{equation}
v_l \frac{dt_1^+}{dx} +\frac{1}{\tau_1} \left( \frac{t_2^+}{2}+ \frac{t_2^-}{2} -t_1^+ \right) =-1
\end{equation}
\begin{equation}
-v_l \frac{dt_1^-}{dx} +\frac{1}{\tau_1} \left( \frac{t_2^+}{2}+ \frac{t_2^-}{2} -t_1^- \right) =-1
\end{equation}
\begin{equation}
V \frac{dt_2^+}{dx} +\frac{1}{\tau_2} \left( \frac{t_1^+}{2}+ \frac{t_1^-}{2} -t_2^+ \right) =-1
\end{equation}
\begin{equation}
-V \frac{dt_2^-}{dx} +\frac{1}{\tau_2} \left( \frac{t_1^+}{2}+ \frac{t_1^-}{2} -t_2^- \right) =-1.
\end{equation}
Defining  $t_i^d=\frac{t_i^+-t_i^-}{2}$ and $t_i=\frac{t_i^++t_i^-}{2}$,
we get the following equations (and similar expressions  with $v_l \to V$, $t_1 \to t_2$, $t_2 \to t_1$)~:
\begin{equation}
 v_l\frac{dt_1^d}{dx}+\frac{1}{\tau_1} (t_2-t_1)=-1
\end{equation}
\begin{equation}
 v_l\frac{dt_1}{dx}-\frac{1}{\tau_1} t_1^d =0, 
\end{equation}
which eventually  lead to the following system~:
\begin{equation}
v_l^2\tau_1\frac{d^2 t_1}{dx^2}+\frac{1}{\tau_1} (t_2-t_1)=-1
\end{equation}
\begin{equation}
 V^2\tau_2\frac{d^2 t_2}{dx^2}+\frac{1}{\tau_2} (t_1-t_2)=-1,
\end{equation}
together with~:
\begin{equation}
 t_1^d =v_l\tau_1\frac{dt_1}{dx} 
\end{equation}
\begin{equation}
 t_2^d =V\tau_2\frac{dt_2}{dx} .
\end{equation}
Inside the target ($x\le a$), one has 
 $t_1^{+,in}(x)=t_1^{-,in}(x)=0$, and~:
\begin{equation}
V \frac{dt_2^{+,in}}{dx} - \frac{1}{\tau_2} t_2^{+,in} =-1
\end{equation}
\begin{equation}
-V \frac{dt_2^{-,in}}{dx} - \frac{1}{\tau_2} t_2^{-,in} =-1.
\end{equation}
Finally, the  boundary conditions read~: 
\begin{equation}
\left.\frac{dt_2}{dx}\right|_{x=b}=0
\end{equation}
\begin{equation}
\left.\frac{dt_1}{dx}\right|_{x=b}=0
\end{equation}
\begin{equation}
t_2^+(a)=t_{2,in}^+(a)
\end{equation}
\begin{equation}
t_2^-(a)=t_{2,in}^-(a)
\end{equation}
\begin{equation}
\left.\frac{dt_{2,in}}{dx}\right|_{x=0}=0
\end{equation}
\begin{equation}
t_1^-(a)=0
\end{equation}

\subsubsection{Results}\label{r1dvv1}

\begin{figure}[h!]
   \begin{minipage}[c]{.46\linewidth}
\begin{center}
      \includegraphics[width=6cm]{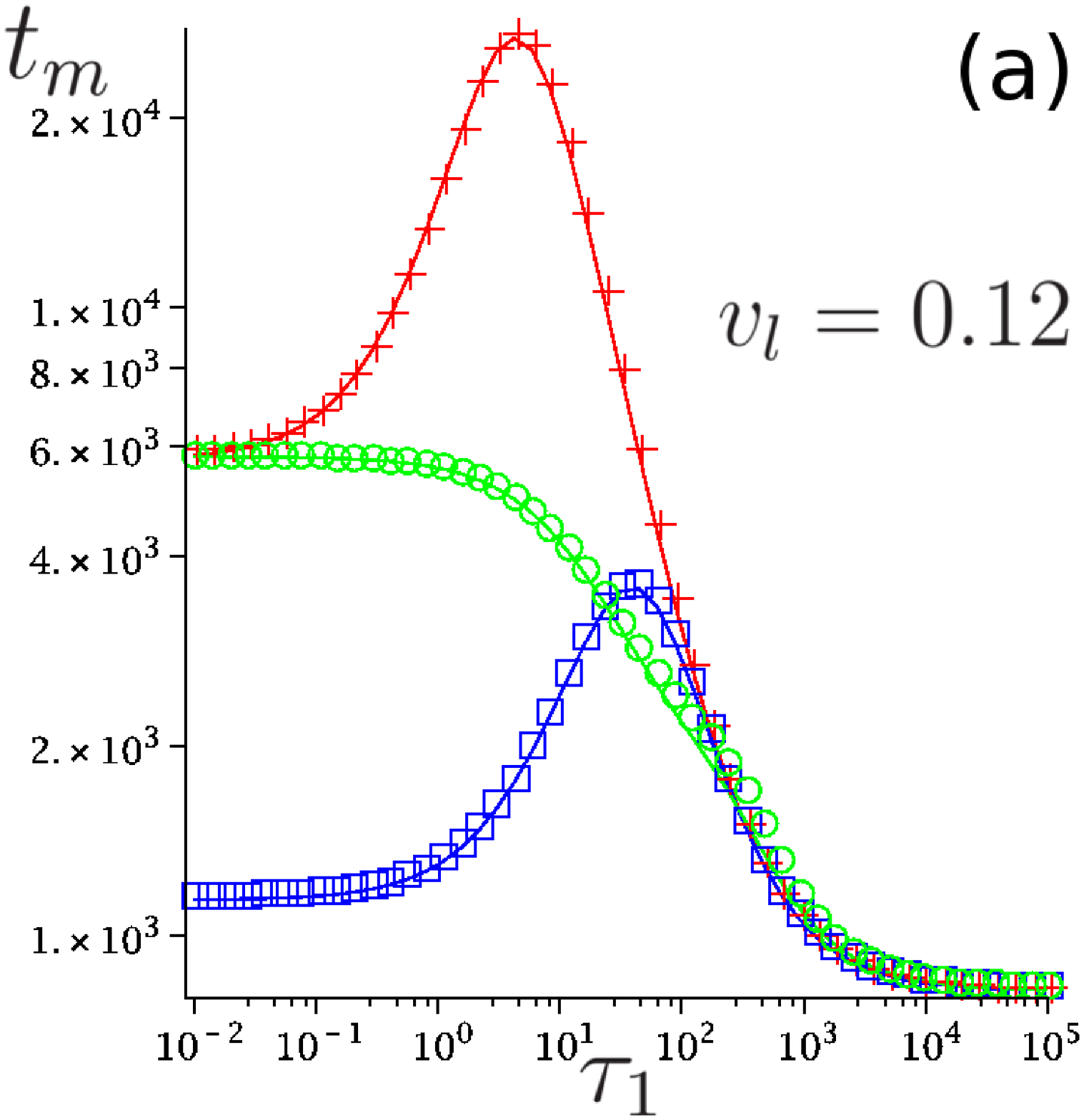}

\end{center}
   \end{minipage} \hfill
   \begin{minipage}[c]{.46\linewidth}
\begin{center}
      \includegraphics[width=6cm]{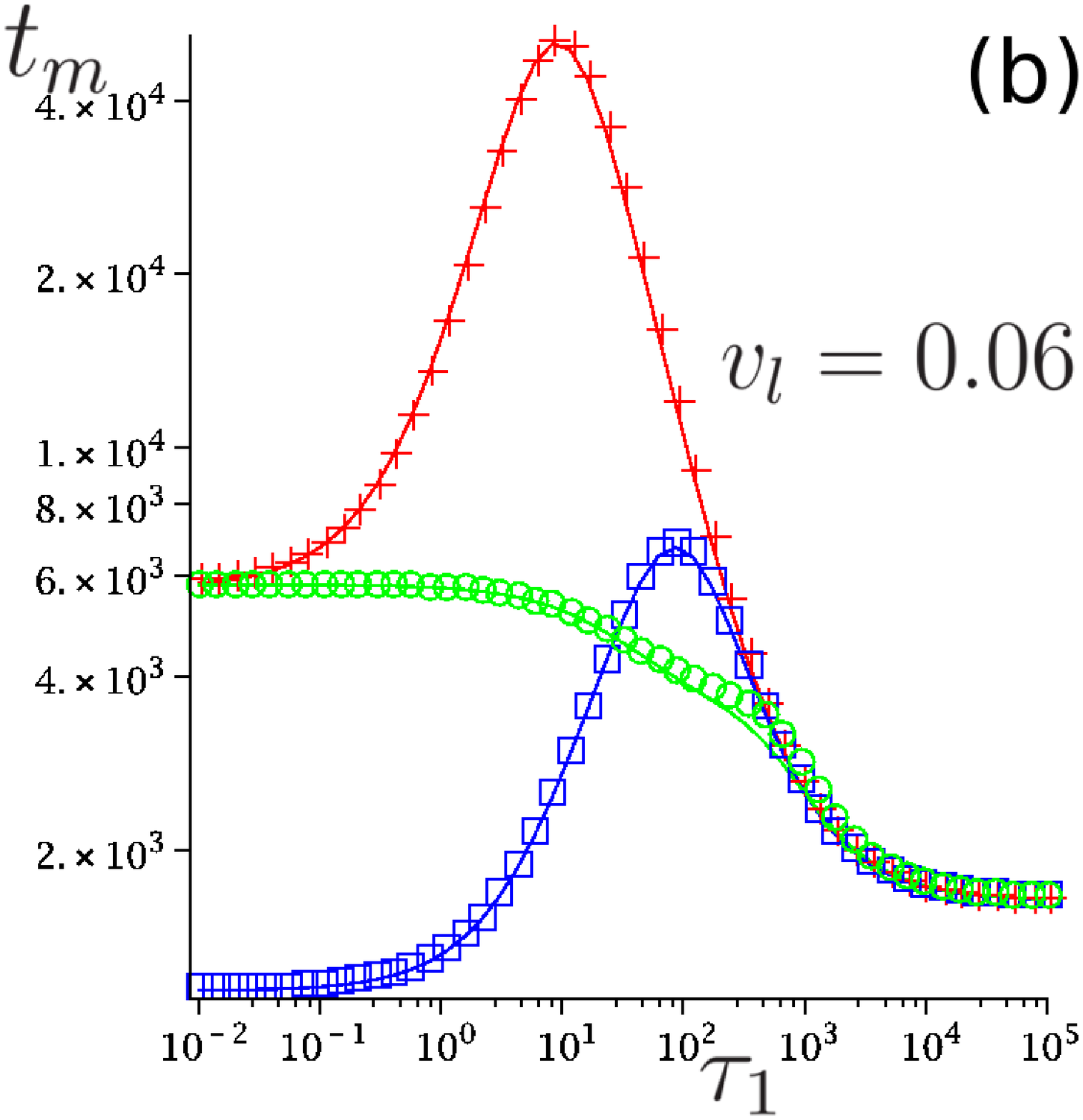}

\end{center}
   \end{minipage}\hfill
   \begin{minipage}[c]{.46\linewidth}
\begin{center}
      \includegraphics[width=6cm]{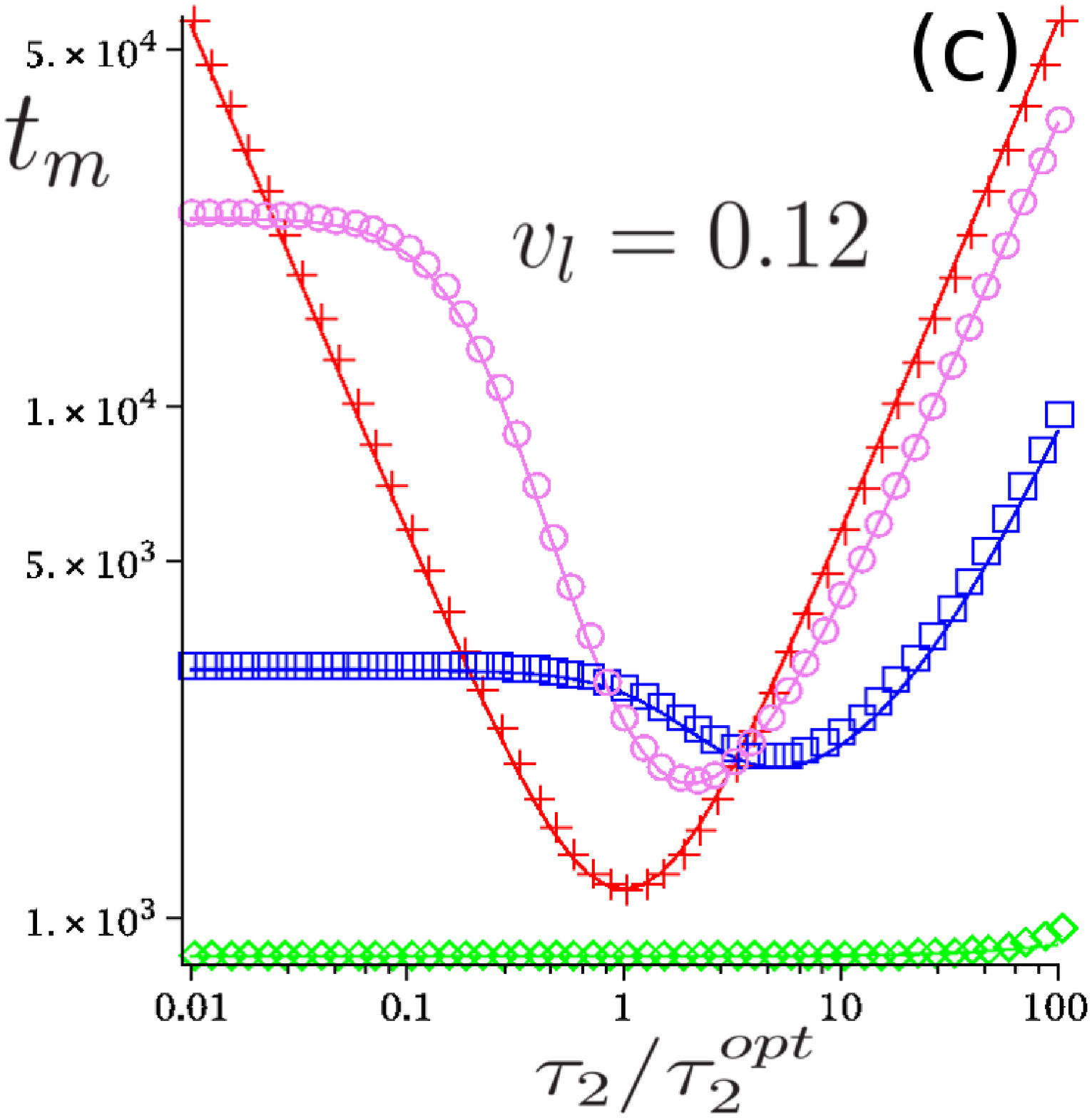}

\end{center}
   \end{minipage} \hfill
   \begin{minipage}[c]{.46\linewidth}
\begin{center}
      \includegraphics[width=6cm]{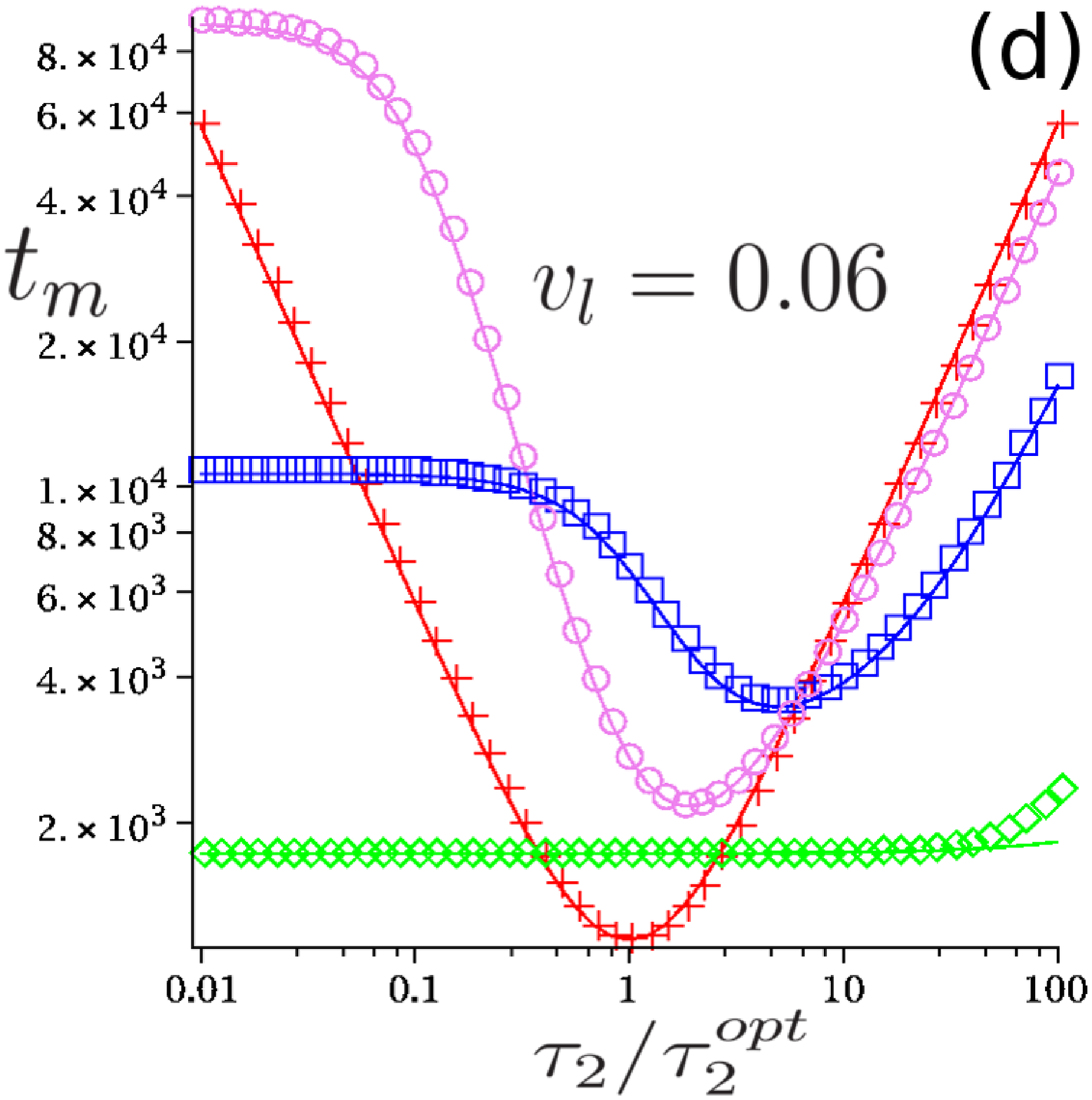}

\end{center}
   \end{minipage}\hfill
\caption{(Color online) Ballistic mode in 1 dimension. 
Comparison between low density approximation \refm{td} (symbols) 
and the exact expression of $t_m$ \refm{tm1v} (line). 
(a),~(b)~: $t_m$ as a function of $\tau_1$, 
with $\tau_2=0.1\tau_2^{opt}$ (red, crosses), $\tau_2=\tau_2^{opt}$ (blue, squares), $\tau_2=10 \tau_2^{opt}$ (green, circles).
(c),~(d)~: $t_m$ as a function of $\tau_2/\tau_2^{opt}$, 
with $\tau_1=0$ (red, crosses), $\tau_1=10$ (violet, circles), $\tau_1=100$ (blue, squares), $\tau_1=10000$ (green, diamonds).
(a),~(c)~: $v_l=0.12>v_l^c$~: intermittence is not favorable.
(b),~(d)~: $v_l=0.06<v_l^c$~: intermittence is favorable.
$\tau_2^{opt}$ is from the analytical prediction \refm{l2opt}. $a=1$, $V=1$, $b=100$.}\label{approxt}
\end{figure}

The exact expression of $t_m$ \refa{a1dvv1} is obtained through lengthy but standard calculations.
To simplify this expression, 
we consider the small density limit  $a/b \to 0$ and finally obtain
 the following very good approximation of $t_{m}$ (\refi{approxt})~:
\begin{equation}\label{td}
 t_{m}=\frac{(\tau_1+\tau_2) b}{\alpha^{3/2}}\left(\left( \frac{b}{3}+L_1 \right) \sqrt{\alpha}+ \Gamma L_2(\sqrt{\alpha}+L_2) \right)
\end{equation}
where~:
\begin{equation}
 \Gamma =\frac{(\sqrt{\alpha}-L_1)(L_1+L_2)+\sqrt{\alpha}  (L_2-L_1+\sqrt{\alpha})X  +X^2L_2 (L_2-L_1) }
{((L_1+\sqrt{\alpha}) X^2+(L_1-\sqrt{\alpha})) (\sqrt{\alpha}+L_2-L_1) }
\end{equation}
\begin{equation}
 X=e^{\frac{2 a}{L_2}}
\end{equation}
\begin{equation}
 \alpha=L_1^2+L_2^2
\end{equation}
\begin{equation}
 L_2=V\tau_2
\end{equation}
\begin{equation}
 L_1=v_l\tau_1
\end{equation}

\begin{table}[h!]
\begin{tabular}{|c|c|c|c|c|c|}
 \hline
& $v_l=1$& $v_l=0.1$& $v_l=0.01$& $v_l=0.001$&$\tau_2^{opt,th}$\refm{l2opt}\\
\hline
$b=5$& $\tau_1\to \infty$&\multicolumn{3}{|c|}{ $\tau_1 \to 0$, $\tau_2^{opt} = 0.86$}&$\tau_2^{opt,th}=0.91$\\
\hline
$b=50$&\multicolumn{2}{|c|}{$\tau_1\to \infty$}&\multicolumn{2}{|c|}{$\tau_1 \to 0$, $\tau_2^{opt} = 2.9$}&$\tau_2^{opt,th}=2.9$\\
\hline
$b=500$&\multicolumn{2}{|c|}{$\tau_1\to \infty$}&\multicolumn{2}{|c|}{$\tau_1 \to 0$, $\tau_2^{opt} = 9.1$}&$\tau_2^{opt,th}=9.1$\\
\hline
$b=5000$&\multicolumn{3}{|c|}{$\tau_1\to \infty$}&$\tau_1 \to 0$, $\tau_2^{opt} = 29$&$\tau_2^{opt,th}=29$\\
\hline
\end{tabular}
\caption{Ballistic mode in 1 dimension. Numerical minimization of the exact $t_m$ \refm{tm1v}. Values 
of $\tau_1$ and $\tau_2$ at the minimum. Comparison with theoretical $\tau_2$. $a=0.5$, $V=1$.}
\label{tableauvaleurs}
\end{table}

A numerical analysis indicates  (\refi{approxt}, \reft{tableauvaleurs})  that, depending on the parameters,  
there are two possible optimal strategies~:
\begin{itemize}
 \item $\tau_1\to \infty$ . Intermittence is not favorable. 
\item  $\tau_1 \to 0, \tau_2=\tau_2^{opt}$. Intermittence is favorable.
\end{itemize}
We now study analytically these regimes.

\subsubsection{Regime without intermittence~: $\tau_1\to \infty$}

In this regime, there is no intermittence. 
The searcher starts   either inside the target ($x$ in $[-a,a]$) 
and immediately finds the target, 
or  it starts at a position $x$ outside the target. 
We can therefore take $x \in [a,b]$. 
If the searcher goes in the $-x$ direction, 
it find its target after $T=(x-a)/v_l$.
If the searcher goes in the $+x$ direction, 
it finds its target after $T=((b-x)+(b-a))/v_l$. 
This leads to~:
\begin{equation}\label{tsans}
t_{bal}=\frac{1}{b}\int_{a}^{b}\frac{b-a}{v_l}dx=\frac{(b-a)^2}{ b v_l}.
\end{equation}

\subsubsection{Intermittent regime}

We take the limit $\tau_1 \to 0$ in the expression of $t_m$ \refm{td} and 
obtain~:
\begin{equation}\label{limt1}
\begin{split}
\lim_{\tau_1 \to 0} t_m =\frac{b}{V}\left( \frac{b}{3L_2}+\frac{e^{\frac{2a}{L_2}}+1}{e^{\frac{2a}{L_2}}-1}\right)
\end{split}
\end{equation}
Taking the derivative with respect to  $L_2$ yields~:
\begin{equation}
\frac{d}{dL_2}\left(\lim_{\tau_1 \to 0} t_m \right)\propto 12a e^{\frac{2a}{L_2}}+2b e^{\frac{2a}{L_2}}-b-b e^{\frac{4a}{L_2}}
\end{equation}
which has only one positive root~:
\begin{equation}
L_2^{opt}=\frac{2a}{\ln\left(1+6a/b+2\sqrt{3a/b+9a^2/b^2} \right)}.
\end{equation}
In the limit $b\gg a$ it leads to~: 
\begin{equation}\label{l2opt}
\tau_2^{opt}= \frac{a}{3V} \sqrt{\frac{b}{a}}, 
\end{equation}
which  is in agreement with numerical minimization of the exact 
mean detection time shown in the table \ref{tableauvaleurs}.

The mean first passage time at the target is minimized in the intermittent regime for  $\tau_1 \to 0$ and $\tau_2=\tau_2^{opt}$.
We replace $\tau_2$ by \refm{l2opt} in the expression \refm{limt1}, and take $b\gg a$ to finally obtain~:
\begin{equation}
 t^{opt}_m=\frac{2}{\sqrt{3}} \frac{b}{V} \sqrt{\frac{b}{a}}
\end{equation}
\begin{equation}
 gain=\frac{\sqrt{3}}{2}\frac{V}{v_l}  \sqrt{\frac{a}{b}}.
\end{equation}
This shows that 
the gain is larger than  1 for  $v_l<v_l^c=V\frac{\sqrt{3}}{2}\sqrt{\frac{a}{b}}$, which defines the regime where intermittence is favorable.
%It is in agreement with  \ref{tableauvaleurs}.

\subsubsection{Summary}

In the case where the phase 1 is modeled by  the ballistic mode in dimension 1, 
we calculated the exact mean first passage time $t_m$ at the target.
$t_m$ can be minimized  as a function of $\tau_1$ and $\tau_2$, yielding  
 two possible optimal strategies~: 
\begin{itemize}
\item for  $v_l>v_l^c=V\frac{\sqrt{3}}{2}\sqrt{\frac{a}{b}}$, intermittence is not favorable ~:  $\tau_1^{opt} \to \infty$,  $\tau_2^{opt} \to 0$
\item for  $v_l<v_l^c=V\frac{\sqrt{3}}{2}\sqrt{\frac{a}{b}}$,   intermittence is favorable, with $\tau_1^{opt} \to 0$ and  $\tau_2^{opt}=\frac{a}{3V} \sqrt{\frac{b}{a}} $ 
\end{itemize}

Note that the model studied in ~\cite{viswaNat} shows that when targets   are not revisitable,  the optimal strategy for a 1 state searcher is to perform a straight ballistic motion. This strategy corresponds to $\tau_1\to\infty$ in our model. 
Our results show that if a faster phase without detection is allowed, this 
 straight line strategy can be outperformed.

\subsection{Conclusion in dimension 1}

Intermittent search strategies  in dimension 1 share  similar features for the static, diffusive and ballistic detection modes. 
In particular, all modes show regimes where intermittence is favorable and lead to a minimization of the search time.
Strikingly, the optimal duration of the non-reactive relocation phase 2 is quite independent of the modeling of the 
reactive phase~: $\tau_2^{opt} = \frac{a}{3V} \sqrt{\frac{b}{a}}$ for the static mode, 
for  the ballistic mode (in the regime $v_l < v_l^c \simeq \frac{V}{2}\sqrt{\frac{3a}{b}}$), and for the diffusive mode (in the regime $b>\frac{D}{V}$ and $a\gg \frac{D}{V}\sqrt{\frac{b}{a}}$). This shows the robustness of the optimal value  $\tau_2^{opt}$.

\section{Dimension 2}

The 2-dimensional problem is particularly well suited to model animal behaviors.  
It is also relevant to  the microscopic scale, since it  mimics for example the case of  
cellular traffic on membranes~\cite{alberts}. 
% In this section, the results for the ballistic mode of detection  are new. 
The results for  the static and diffusive modes, already treated in ~\cite{PRE2006,SpecialIssue2006}, 
are summarized here for completeness. 
While in dimension  1 the mean search time can be calculated analytically, 
we introduce in dimension 2 (and later dimension 3) approximation schemes, 
which we check by numerical simulations. 
In these numerical simulations, diffusion was simulated  using variable step lengths, 
as in~\cite{methodDif}, and we used square domains instead of disks for numerical convenience 
(it  was checked numerically that results are not affected  as soon as $b\gg a$).

\subsection{Static mode}

\imagea{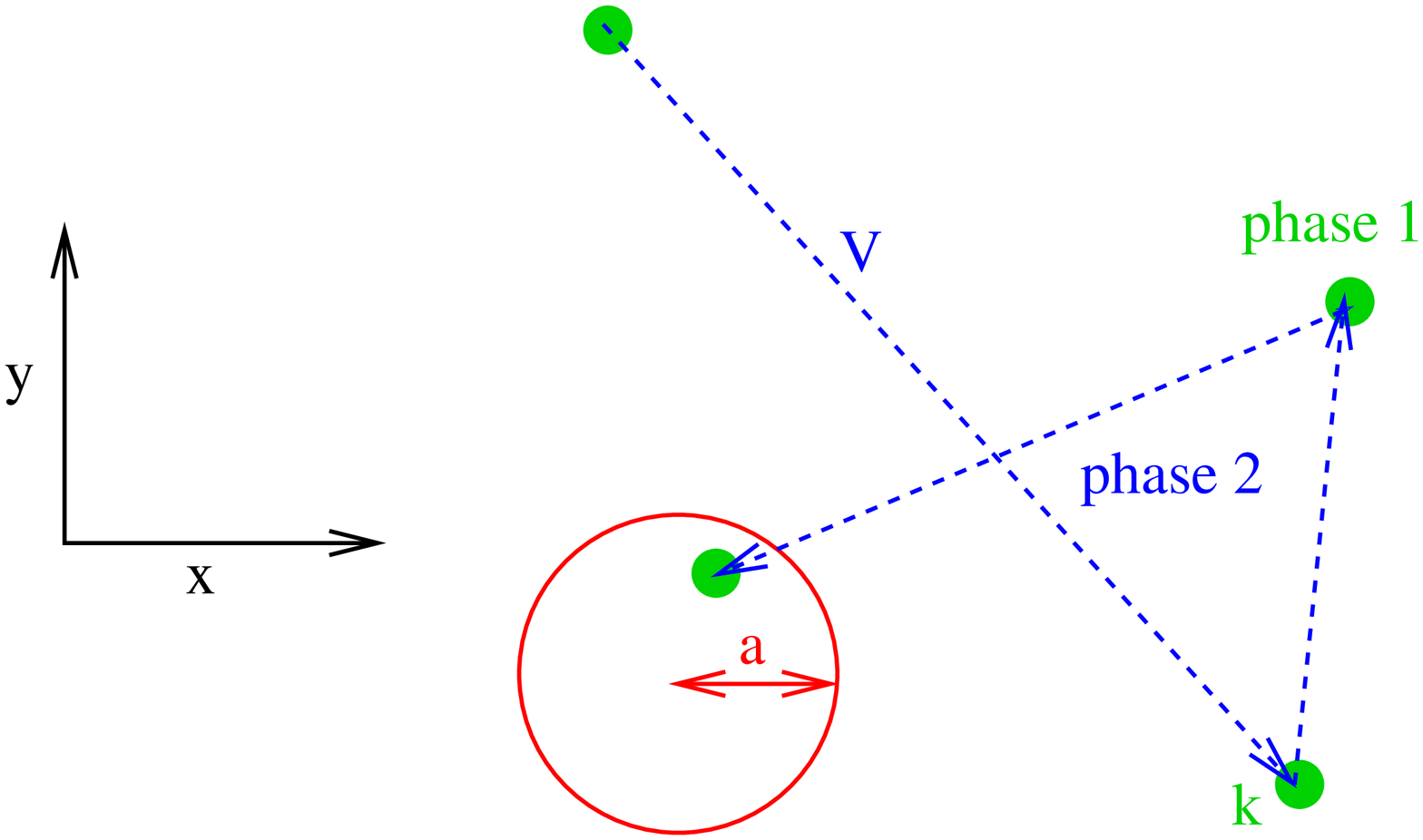}{Static mode in dimension 2}{2k}{5}

We study here the case where the detection phase is modeled by the static mode~: 
the searcher does not move during the detection phase 
and has a finite reaction rate with the target if it is within its detection radius $a$.  
%This problem has already been treated in~\cite{PRE2006,SpecialIssue2006}. 

\subsubsection{Equations}

 The mean first passage time (MFPT) at a target
satisfies the following backward equations~\cite{redner}:
\begin{equation}\label{back1k}
\frac{1}{2\pi\tau_1}\int_{0}^{2\pi}(t_2(\overrightarrow{r})-t_1(\overrightarrow{r}))d\theta_{\overrightarrow{V}}-k{\rm I}_a(\overrightarrow{r})t_1(\overrightarrow{r})=-1.
\end{equation}
\begin{equation}\label{back2k}
\overrightarrow{V}\cdot\nabla_{\bf r}t_2(\overrightarrow{r})-\frac{1}{\tau_2}(t_2(\overrightarrow{r})-t_1(\overrightarrow{r}))=-1
\end{equation}
The function ${\rm I}_a$ is defined by ${\rm I}_a(\overrightarrow{r})=1$ inside the target (if $|\overrightarrow{r}|\le a$) and  ${\rm I}_a(\overrightarrow{r})=0$ outside the target (if $|\overrightarrow{r}|> a$).
In the present form, these integro-differential equations (completed with boundary conditions)  do not seem to allow for an exact resolution with standard methods.
$t_2$ is the mean first passage time on the target, 
starting from $\overrightarrow{r}$ in phase 2, with speed $\overrightarrow{V}$, 
of angle $\theta_v$, 
and with projections on the axes $V_{x}$, $V_{y}$. $i$ and $j$ can take either $x$ or $y$ as a value. 
 We use the following decoupling assumption~:
\begin{equation}\label{deck}
\langle V_i V_j t_2\rangle_{\theta_{\bf V}}\simeq\langle V_i V_j\rangle_{\theta_{\bf V}}\langle t_2\rangle_{\theta_{\bf V}}
\end{equation}
and finally obtain the following approximation of the mean search time, which can be checked by numerical simulations~:  
\begin{equation}\label{searchtime2}
t_m = \frac{\tau_1+\tau_2}{2k\tau_1 y^2}\left\{\frac{1}{x}(1+k\tau_1)(y^2-x^2)^2\frac{{\rm I}_0(x)}{{{\rm I}_1(x)}}
+\frac{1}{4}\left[8y^2+(1+k\tau_1)\left(4y^4\ln(y/x)+(y^2-x^2)(x^2-3y^2+8)\right)\right]\right\}
\end{equation}
\begin{equation}
{\rm where}\;x=\sqrt{\frac{2k\tau_1}{1+k\tau_1}}\frac{a}{V\tau_2} \;{\rm and}\;y=\sqrt{\frac{2k\tau_1}{1+k\tau_1}}\frac{b}{V\tau_2}
\end{equation}
In that case, intermittence is trivially necessary to find the target: indeed, if the searcher does  not move, the MFPT is infinite.
In the regime  $b\gg a$, the optimization of the search time (\ref{searchtime2})
 leads to~: 
\begin{equation} \label{statique}
\tau_{1}^{opt}=\left(\frac{a}{Vk}\right)^{1/2}\left(\frac{2\ln(b/a)-1}{8}\right)^{1/4},\;
\end{equation}
\begin{equation}\label{statique2}
\tau_{2}^{opt}=\frac{a}{V}\left(\ln(b/a)-1/2\right)^{1/2},
\end{equation}
and the minimum search time is given in the large volume limit by~: 
\begin{eqnarray}
 &&t_m^{opt}={\frac {{b}^{2}}{{a}^{2}k}}- \frac{2^{1/4}}{\sqrt{Vka^3}}\,\frac{ ({a}^{2}-4b^2)\ln(b/a)+2b^2-a^2}{( 2\ln(b/a) -1 ) ^{3/4}}
 \nonumber\\
&&-\frac{\sqrt{2}}{48ab^2V}\,\frac{(96a^2-192b^4)\ln^2(b/a)+(192b^4-144a^2b^2)\ln(b/a)+46a^2b^2-47b^4+a^4}{( 2\ln(b/a) -1 ) ^{3/2}}
\end{eqnarray}

\subsubsection{Summary}

In the case of a  static detection mode in  dimension 2, we obtained  a simple approximate expression of the mean first passage time $ t_m$ at the target. With the static  detection mode, intermittence is always favorable and leads to a single  optimal intermittent strategy. The minimal search time is realized for 
 $\tau_{1}^{opt}=\left(\frac{a}{Vk}\right)^{1/2}\left(\frac{2\ln(b/a)-1}{8}\right)^{1/4}$ and $\tau_{2}^{opt}=\frac{a}{V}\left(\ln(b/a)-1/2\right)^{1/2}$.
Importantly,  the optimal duration of the relocation phase 
does not depend on $k$, i.e. on the description of the detection phase, as in dimension 1.

\subsection{Diffusive mode}

\imagea{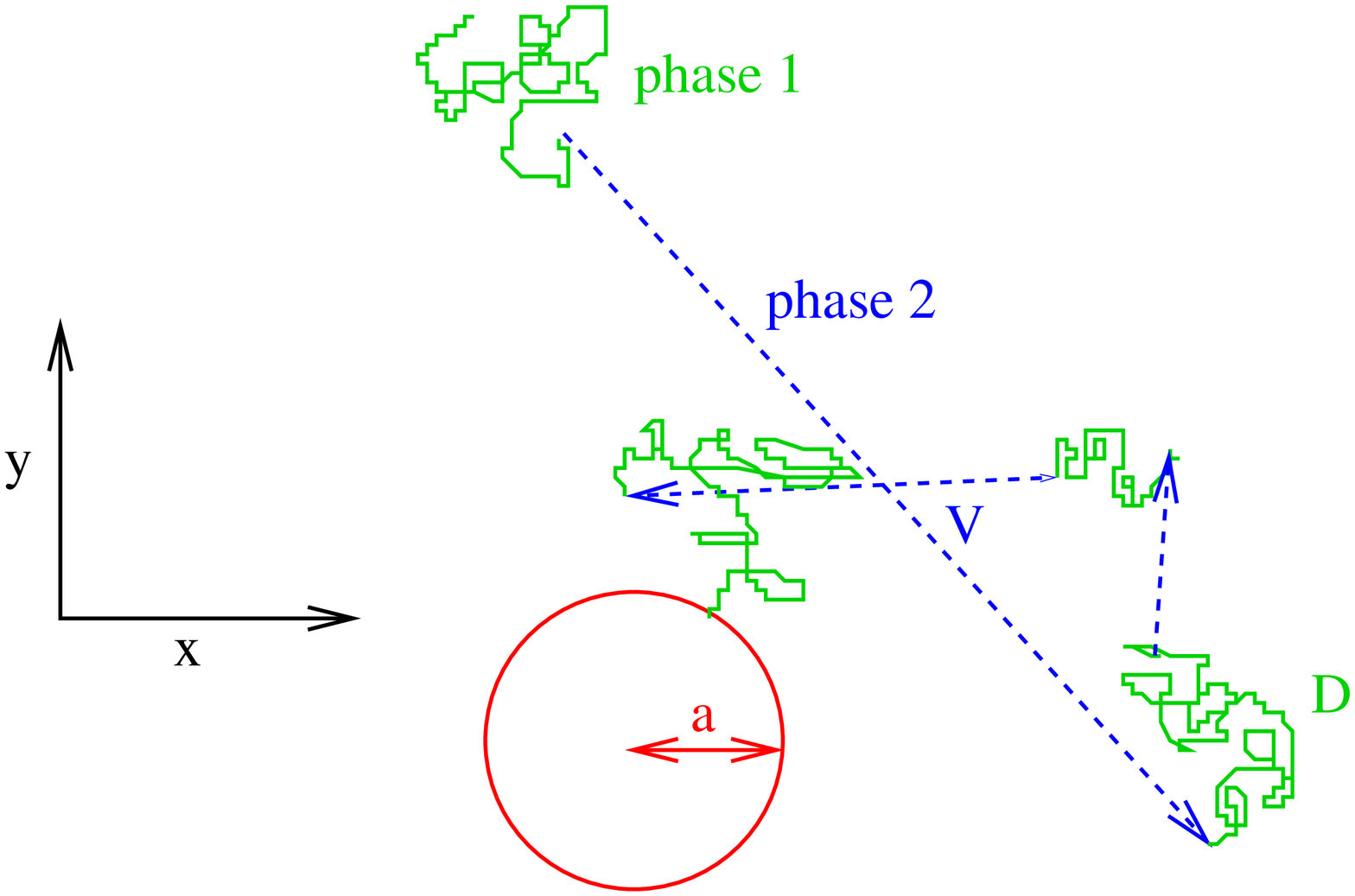}{Diffusive mode in dimension 2}{2D}{5}

We now assume that the searcher  diffuses during the detection phase. 
%This problem has already been treated in~\cite{PRE2006,SpecialIssue2006}. 
%We recall here the main results.   
For this process, the mean first passage time at the  target
satisfies the following backward equation~\cite{redner}:
\begin{equation}\label{back1d}
D\nabla^2_{\bf r}t_1(\overrightarrow{r})+\frac{1}{2\pi\tau_1}\int_{0}^{2\pi}(t_2(\overrightarrow{r})-t_1(\overrightarrow{r}))d\theta_{\bf V}=-1
\end{equation}
\begin{equation}\label{back2d}
\overrightarrow{V}\cdot\nabla_{\bf r}t_2(\overrightarrow{r})-\frac{1}{\tau_2}(t_2(\overrightarrow{r})-t_1(\overrightarrow{r}))=-1
\end{equation}
with $t_1(\overrightarrow{r})=0$ inside the target ($r\le a$). 
We use the same decoupling assumption than for the static case \refm{deck}. 
It eventually leads to the following approximation of the mean search time~: 
\begin{equation}\disp\label{tmap}
t_m =(\tau_1+\tau_2)\frac{\disp1-a^2/b^2}{\disp(\alpha^2 D\tau_1)^2}\left\{\disp a\alpha(b^2/a^2-1)\frac{\disp M}{\disp 2L_+}-\frac{\disp L_-}{\disp L_+}-\frac{\disp \alpha^2D\tau_1}{\disp 8{\widetilde D}\tau_2}\frac{\disp(3-4\ln(b/a))b^4-4a^2b^2+a^4}{\disp b^2-a^2}\right\},
\end{equation}
\begin{equation}
{\rm with}\  L_\pm={\rm I}_0\left(\frac{a}{\sqrt{{\widetilde D}\tau_2}}\right)\left({\rm I}_1(b\alpha){\rm K}_1(a\alpha)- {\rm I}_1(a\alpha){\rm K}_1(b\alpha)  \right)\pm \alpha\sqrt{{\widetilde D}\tau_2}\;{\rm I}_1\left(\frac{a}{\sqrt{{\widetilde D}\tau_2}}\right)\left({\rm I}_1(b\alpha){\rm K}_0(a\alpha)+ {\rm I}_0(a\alpha){\rm K}_1(b\alpha)  \right)\nonumber
\end{equation}
\begin{equation}
{\rm and }\  M={\rm I}_0\left(\frac{a}{\sqrt{{\widetilde D}\tau_2}}\right)\left({\rm I}_1(b\alpha){\rm K}_0(a\alpha)+ {\rm I}_0(a\alpha){\rm K}_1(b\alpha)  \right)-4\frac{a^2\sqrt{{\widetilde D}\tau_2}}{\alpha(b^2-a^2)^2}{\rm I}_1\left(\frac{a}{\sqrt{{\widetilde D}\tau_2}}\right
)\left({\rm I}_1(b\alpha){\rm K}_1(a\alpha)- {\rm I}_1(a\alpha){\rm K}_1(b\alpha)  \right)\nonumber
\end{equation}
where $\alpha=(1/(D\tau_1)+1/({\widetilde D}\tau_2))^{1/2}$ and ${\widetilde D} = v^2 \tau_2$.
We then minimize this time as a function of $\tau_1$ and $\tau_2$.

\subsubsection{$a<b\ll D/V$ : intermittence is not favorable}

In that regime, intermittence is not favorable. Indeed, the typical time required to explore the whole domain of radius $b$ is of order 
$b^2/D$ for a diffusive motion, which is shorter than the corresponding time $b/V$ for a ballistic motion. As a consequence, it is never useful to interrupt the 
diffusive phases by mere relocating ballistic phases.
We use standard methods to calculate the mean first passage time at the target in this optimal regime of diffusion only~:
\begin{equation}
 \frac{D}{r}\frac{d }{d r}\left( r \frac{d t}{d r} \right)=-1
\end{equation}
The boundary conditions $t(a)=0$ and $\frac{dt}{dr}(r=b)=0$ lead to~:
\begin{equation}\label{tdiff2D}
 t_{diff}= \frac{1}{8 b^2 D_{eff}}\left(4a^2b^2-a^4-3b^4+4b^4 \ln \frac{b}{a} \right), 
\end{equation}
and we find in the limit $b \gg a$~: 
\begin{equation}\label{tdiff2Db}
 t_{diff} = \frac{b^2}{8 D_{eff}}\left(-3+4 \ln \frac{b}{a} \right)
\end{equation}

\subsubsection{$a\ll D/V\ll b$~: first regime of intermittence}\label{r2dvd}

In this second regime, one can use the following approximate formula for the search time:
\begin{equation}\label{tapprox}
t_m =\frac{b^2}{4DV^2\alpha^2}\frac{\tau_1+\tau_2}{\tau_1\tau_2^2}
\left\{4\ln(b/a)-3-2\frac{(V\tau_2)^2}{D\tau_1}(\ln(\alpha a)+\gamma-\ln 2)\right\},
\end{equation}
$\gamma$ being the Euler constant. An approximate  criterion to determine if intermittence is useful can be obtained by expanding $t_m$ in powers of $1/\tau_1$ when $\tau_1\to\infty$ ($\tau_1\to\infty$ corresponds to the absence of intermittence), and requiring that the coefficient of the term $1/\tau_1$ is negative for all values of $\tau_2$. Using this  criterion, we find that  
intermittence is useful if 
\begin{equation}
\sqrt{2}\exp(-7/4+\gamma)Vb/D-4\ln(b/a)+3>0,
\end{equation}

In this regime, using Eq(\ref{tapprox}), the optimization of the search time  leads to 
\begin{equation}\label{intermediare}
\tau_{1}^{opt}=\frac{b^2}{D}\frac{4\ln w-5+c}{w^2(4\ln w -7+c)}, \ \tau^{opt}_{2}=\frac{b}{V}\frac{\sqrt{4\ln w -5+c}}{w}
\end{equation}
where $w$ is the solution of the implicit  equation $w=2Vbf(w)/D$ with
\begin{equation}
\frac{\sqrt{4\ln w-5+c}}{f(w)}=-8(\ln w)^2+(6+8\ln(b/a))\ln w-10\ln(b/a)+11
-c(c/2+2\ln(a/b)-3/2)
\end{equation}
and  $c=4(\gamma -\ln(2))$,  $\gamma$ being the Euler constant.
A useful approximation for   $w$ is given by
\begin{equation}
w\simeq \frac{2Vb}{D}f\left(\frac{Vb}{2D\ln(b/a)}\right).
\end{equation}
The gain for this optimal strategy reads~: 
\begin{equation}
 gain=\frac{t_{diff}}{t_m^{opt}}\simeq \frac{1}{2}\frac{4 \ln b/a -3 +4a^2/b^2 -a^4/b^4}{4\ln b/a -3+2(4\ln w)\ln (b/aw) }\left(\frac{1}{4\ln w -5}+\frac{wD}{bV}\frac{4\ln w -7}{(4\ln w -5)^{3/2}}\right)^{-1}
\end{equation}
If intermittence significantly speeds up the search in this regime (typically by a factor 2), it does not change the order of magnitude of the search time.

\subsubsection{$ D/V\ll a\ll b$~: ``universal'' regime of intermittence}

 In the last regime $D/V\ll a\ll
b$, the optimal strategy is obtained for
\begin{equation}\label{grandv}
\tau_{1}^{opt}\simeq \frac{D}{2V^2}\frac{\ln^2 (b/a)}{2\ln (b/a) -1}, \;\tau_{2}^{opt}\simeq \frac{a}{V}(\ln(b/a)-1/2)^{1/2}
\end{equation}
and the gain reads:
\begin{equation}
 gain=\frac{t_{diff}}{t_m^{opt}}\simeq \frac{\sqrt{2}aV}{8D}\left(\frac{1}{4\ln(b/a)-3}\frac{{\rm I}_0\left(2/\sqrt{2\ln(b/a)-1}\right)}
{{\rm I}_1\left(2/\sqrt{2\ln(b/a)-1}\right)}+\frac{1}{2\sqrt{2\ln(b/a)-1}}\right)^{-1}
\end{equation}
Here, the optimal strategy leads to a significant decrease of the search time which can be rendered arbitrarily smaller than the 
 search time in absence of intermittence.

\subsubsection{Summary}

We studied the case where the detection phase 1 is modeled by a diffusive mode, and obtained an approximation of the mean first passage time at the target.
We found that  intermittence is favorable (i.e. better than diffusion alone), in the regime of large system size $b\gg D/V$. 
The optimal intermittent strategy then follows two subregimes~: 
\begin{itemize}
\item if $a\ll D/V$, the best strategy is given by (\ref{intermediare}). The search is significantly reduced by intermittence but keeps the same order of magnitude as in the case of a 1-state diffusive search.
\item  if $a\gg D/V$, the best strategy is given by (\ref{grandv}), and weakly depends on $b$. In this regime, intermittence is very efficient as shown by the large gain obtained for $V$ large.
\end{itemize}

\subsection{Ballistic mode}

\imagea{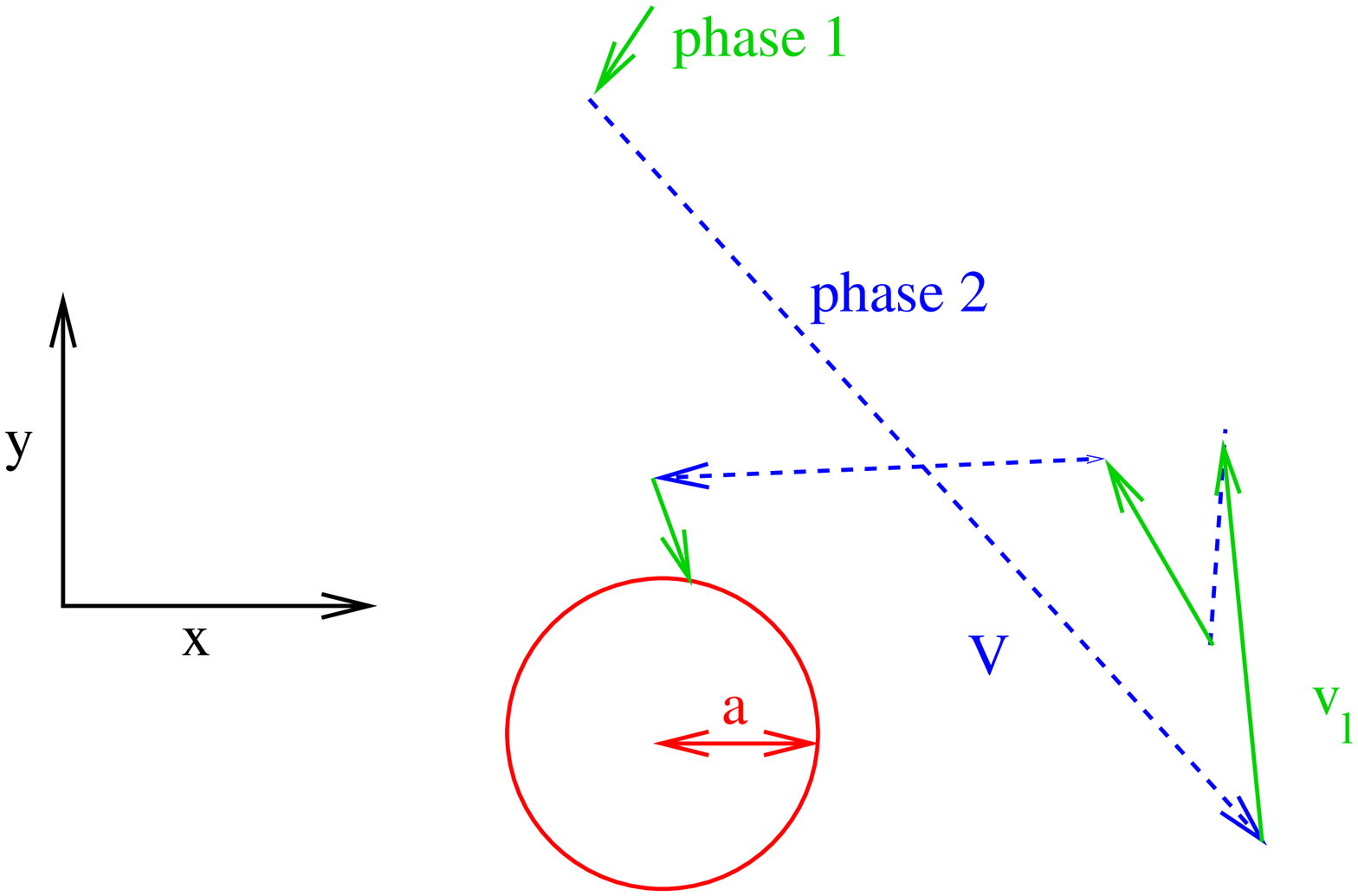}{Ballistic mode in two dimensions}{2v}{5}

In this case, the searcher has access to two different speeds:
one ($V$) is fast but prevents  target detection, 
and the other one ($v_l$) is slower but enables target detection.  %The results presented in this section are new.

\subsubsection{Simulations}

Since an explicit  expression of the mean search time is not available, a numerical study is performed. Exploring the parameter space numerically enables to identify the regimes where the mean search time is minimized. Then, for each regime, approximation schemes are developed to provide analytical expression of the mean search time.
\triplimage{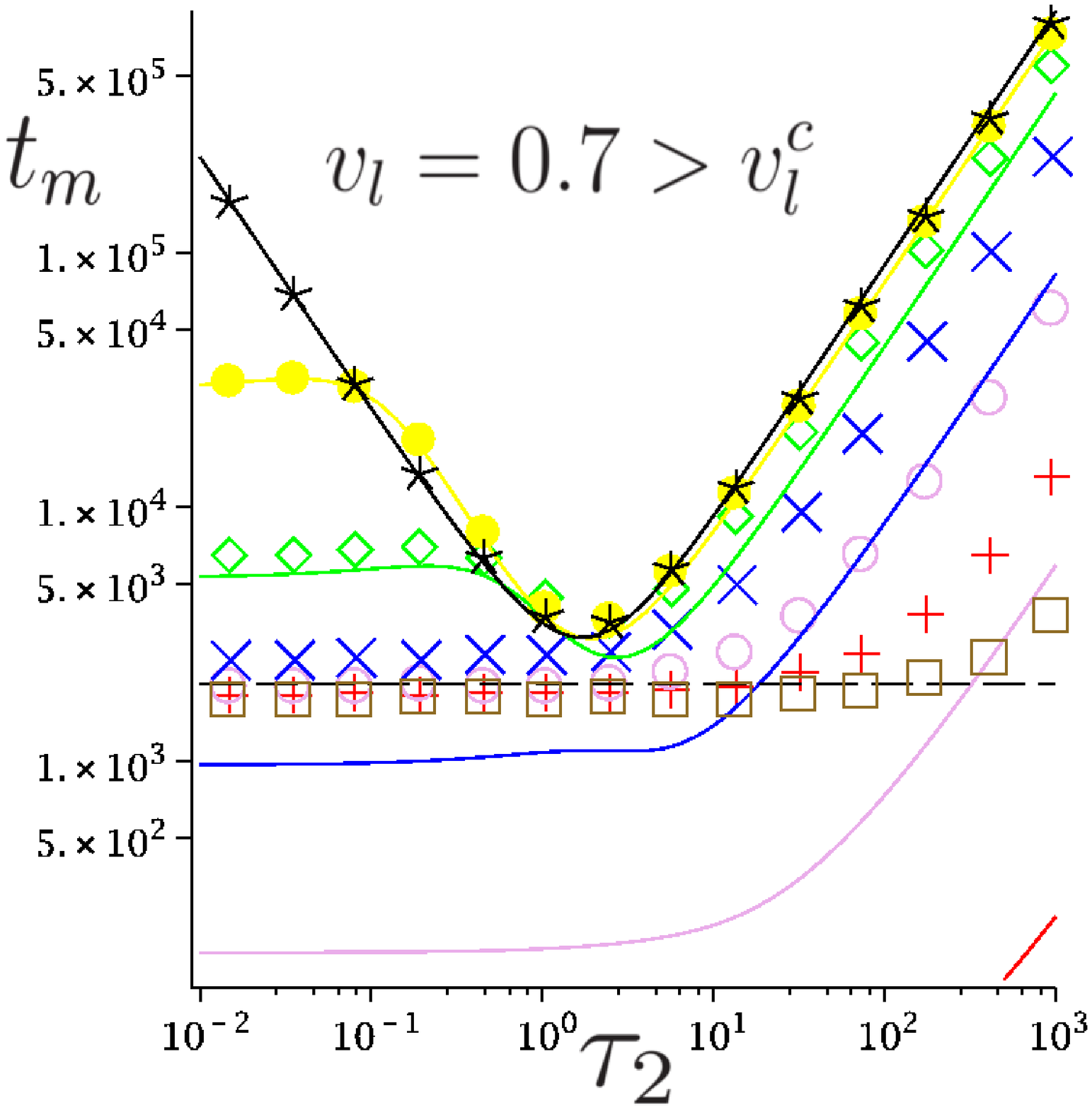}{}
{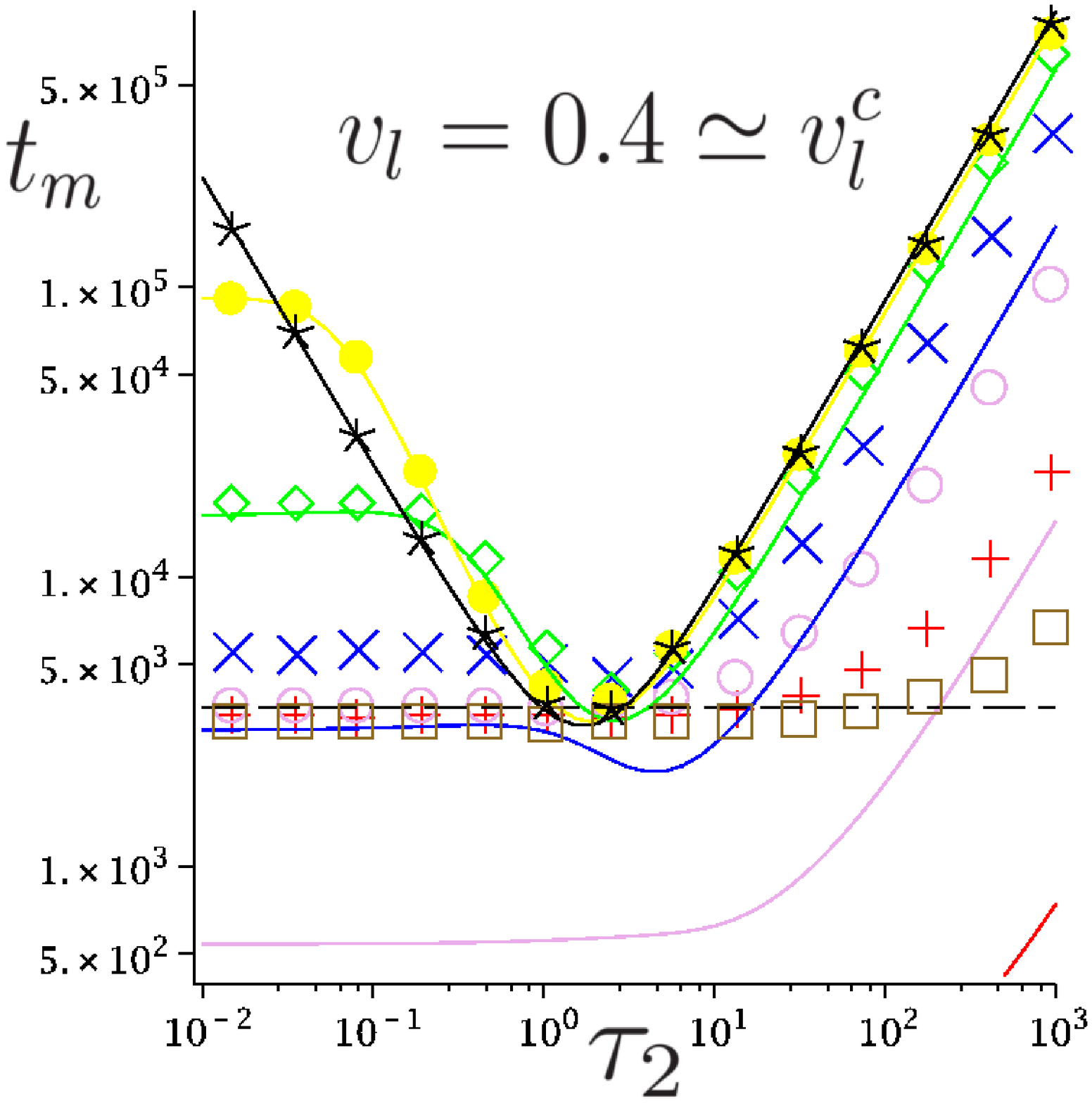}{}
{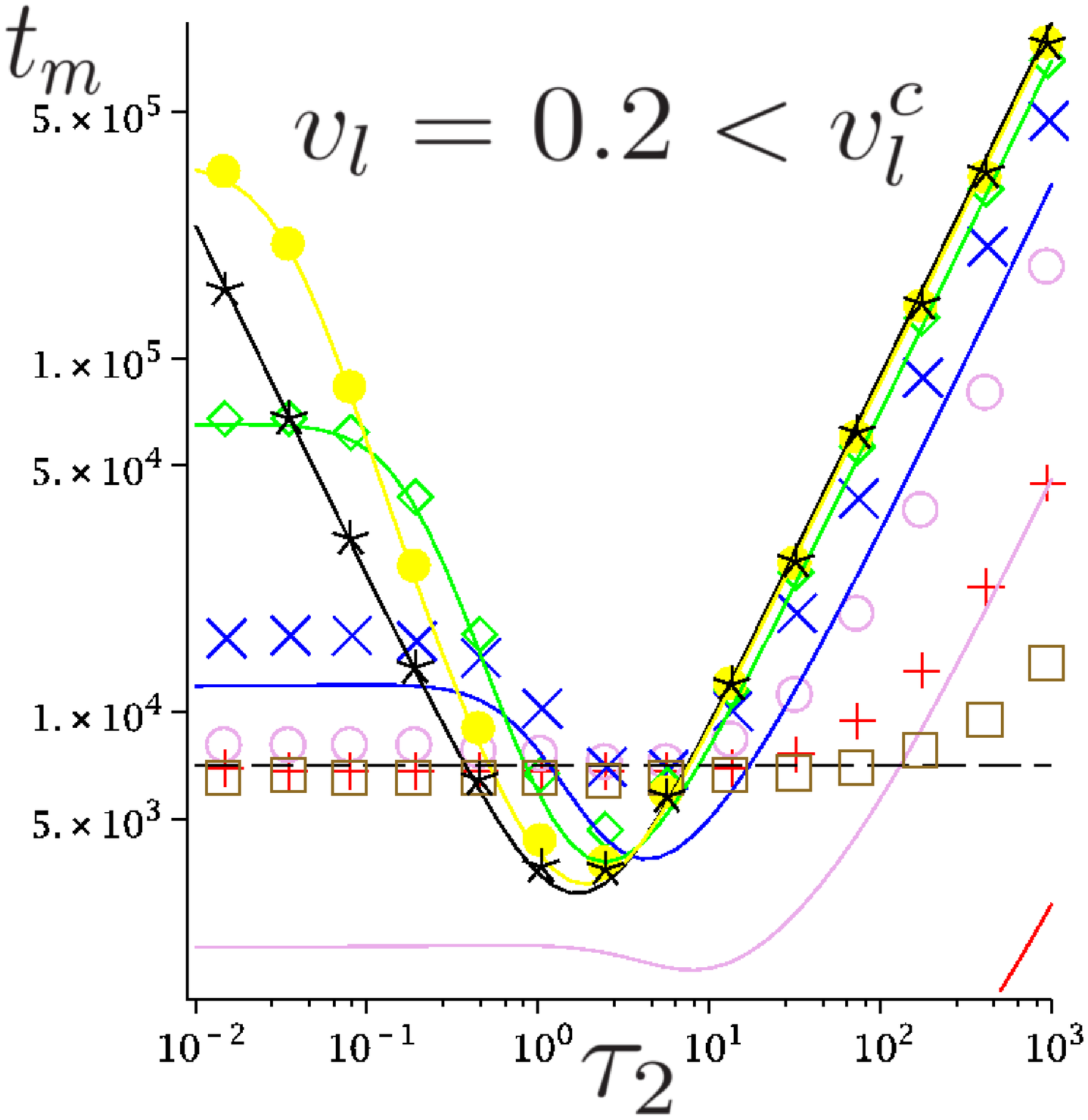}{}
{Ballistic mode in 2 dimensions. 
$\ln(t_m)$ as a function of $\ln(\tau_2)$. 
Simulations (symbols), diffusive/diffusive approximation \refm{tmap} with \refm{d2dvv} 
(colored lines), $\tau_1 \to 0$ limit \refm{Pearson2D} (black line), 
$\tau_1 \to \infty$  (no intermittence) \refm{topt2} (dotted black line). 
$b=30$, $a=1$, $V=1$. $\tau_1=0$ (black, stars), $\tau_1=0.17$ (yellow, solid circles), 
$\tau_1=0.92$ (green, diamonds), $\tau_1=5.0$ (blue, X), $\tau_1=28$ (purple, circles), 
$\tau_1=150$ (red, +), $\tau_1=820$ (brown, squares).}{2dvvb30}
The numerical results presented in  figure \ref{2dvvb30} suggest  two regimes defined according to a threshold value  $v_l^{c}$ of  $v_l$ to be determined later on ~:  
\begin{itemize}
 \item for $v_l > v_l^{c}$,  $t_m$ is minimized for $\tau_2 \to 0$
\item for $v_l < v_l^{c}$, $t_m$ is minimized for $\tau_1 \to 0$~.
\end{itemize}

\imagea{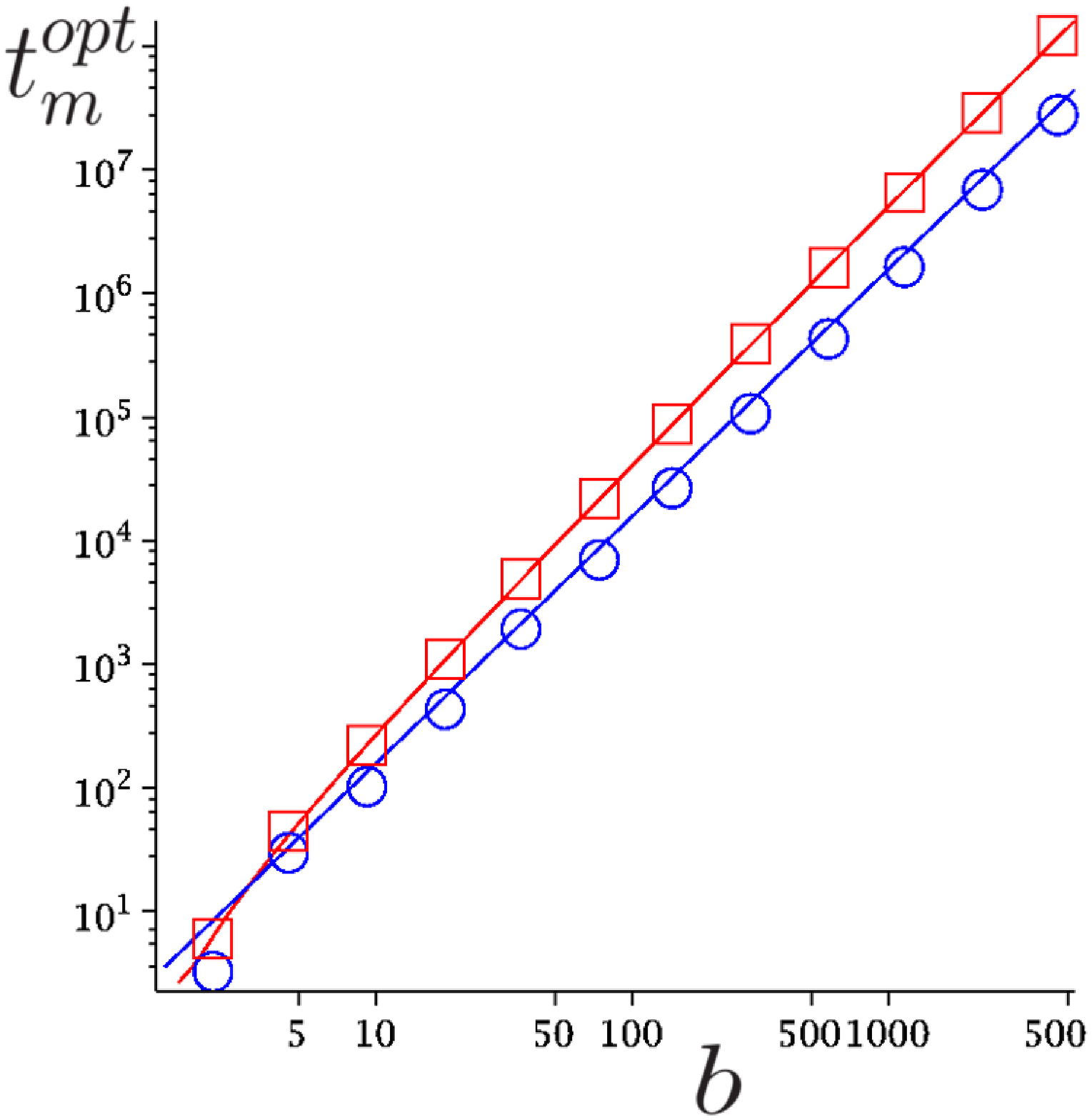}{Ballistic mode in 2 dimension. 
$t_m^{opt}$  as a function of $b$, logarithmic scale.
Regime without intermittence ($\tau_2=0$ and $\tau_1 \to \infty$, $v_l=1$), 
analytical approximation \refm{topt2} (blue line), numerical simulations (blue circles).
Regime with intermittence (with  $\tau_1=0$, $V=1$), 
analytical approximation \refm{topt1} (red line), numerical simulations (red squares). $a=1$}{optiA}{6}

\subsubsection{Regime without intermittence ($\tau_2 \to 0$, $\tau_1 \to \infty$)}

Qualitatively, it is rather intuitive that for $v_l$ large enough 
(the precise threshold value $v_l^{c}$ will be determined next),  
phase 2 is inefficient since it does not allow for target detection. 
The optimal strategy is therefore  $\tau_2 \to 0$ in this case. 
In this regime, the searcher performs a ballistic motion, 
which is randomly reoriented with frequency $1/\tau_1$. 
Along the same line as in ~\cite{viswaNat} 
(where however the times between successive reorientations are Levy distributed), 
it can be shown that the optimal strategy to find a target 
(which is assumed to disappear after the first encounter) 
is to minimize oversampling and therefore to perform a purely ballistic motion. 
In our case this means that in this regime $\tau_2 \to 0$, 
the optimal $\tau_1$ is given by  $\tau_1^{opt} \to \infty$.

In this regime, we can propose an estimate of the optimal search time  $t_{bal}$. 
The surface scanned during $\delta t$ is $2 a v_l \delta t$. 
$p(t)$ is the proportion of the total area which has not yet been scanned at $t$.
If we neglect correlations in the trajectory, one has~: 
\begin{equation}
 \frac{d p }{dt} = -\frac{2a v_l p(t)}{\pi b^2}.
\end{equation}
Then, given that  $p(t=0)=1$,  we find
\begin{equation}
 p(t)=\exp\left(-\frac{2av_l}{\pi b^2}\right),
\end{equation}
and the mean first passage time at the   target in these conditions is~: 
\begin{equation}\label{topt2}
 t_{bal} = - \int_0^{\infty} t \frac{dp}{dt} dt = \frac{\pi b^2}{2av_l} .
\end{equation}
This expression yields (\refi{optiA}) a good agreement with numerical simulations.
Note in particular that   $t_{bal} \propto \frac{1}{v_l}$.

\subsubsection{Regime with intermittence $\tau_1 \to 0$}
In this regime where $v_l < v_l^{c}$, the numerical study shows that the search time is minimized for  $\tau_1 \to 0$ (\refi{2dvvb30}). 
We here determine the optimal value of $\tau_2$ in  this regime. To proceed we  approximate the problem by  the case of a diffusive mode previously studied  \refm{tmap}, with an effective diffusion coefficient~: 
\begin{equation}\label{d2dvv}
 D = \frac{v_l^2 \tau_1}{2}.
\end{equation}
This approximation is very satisfactory in the regime  $\tau_1 \to 0$  as shown in \refi{2dvvb30}. 

We can then use the results of~\cite{PRE2006,SpecialIssue2006} in the  $\tau_1 \to 0$ regime and obtain:
\begin{equation}\label{Pearson2D}
t_m=\tau_2 \left(1-\frac{a^2}{b^2}\right) \left(1-\frac{1}{4} \frac{\left(3+4 \ln\left(\frac{a}{b}\right)\right) b^4-4 a^2 b^2+a^4}{\tau_2^2  V^2 \left(b^2-a^2\right)}+\frac{a}{V \tau_2 \sqrt{2}} \left(\frac{b^2}{a^2}-1\right) \frac{I_0\left(\frac{a \sqrt{2}}{\tau_2  V}\right)}{ I_1\left(\frac{a \sqrt{2}}{\tau_2  V}\right)}\right)
\end{equation}
The calculation of  $\tau_2^{opt}$ minimizing $t_m$ 
 then gives:
\begin{equation} \label{tau2opt}
 \tau_2^{opt} = \frac{a}{v}\sqrt{\ln\left( \frac{b}{a} \right) -\frac{1}{2}}
\end{equation}
In turn, replacing $\tau_2$ by $\tau_2^{opt}$ \refm{tau2opt} in \refm{Pearson2D}, we obtain the minimal mean time of target detection~:  
\begin{equation}\label{topt1}
t_m^{opt}= \frac{a }{u \sqrt{2}  V}  \left(1-\frac{a^2}{b^2}\right) 
\left(1-
u^2\frac{\left(3+4 \ln\left(\frac{a}{b}\right)\right) b^4-4 a^2 b^2+a^4}{a^2 2 \left(b^2-a^2\right)}
+\frac{ u\left(\frac{b^2}{a^2}-1\right) I_0\left( 2 u \right) }
{  I_1\left( 2 u\right)}\right)
\end{equation}
with $u=\left(\ln\left(\frac{b}{a}\right)-1 \right)^{-\frac{1}{2}}$.
It can be noticed that $t_m^{opt} \propto \frac{1}{ V}$.  Note that if $b \gg a$ 
this last expression can be greatly simplified: 
\begin{equation}\label{2dvvtmoptsimp}
 t_m^{opt} \simeq \frac{2b^2}{aV}\sqrt{\ln\left(\frac{b}{a}\right)}
\end{equation}
Finally the gain reads (using \refm{topt2} and \refm{2dvvtmoptsimp}):
\begin{equation}\label{2dvvgain}
gain=\frac{t_{bal}}{t_m^{opt}}\simeq \frac{\pi V}{4 v_l}\left(\ln\left(\frac{b}{a}\right)\right)^{-0.5}
\end{equation}
Numerical simulations of  \refi{optiA} shows the validity of these approximations.

\subsubsection{Determination of $v_l^c$}
It is straightforward than $v_l^c< V$. 
Indeed, if $v_l= V$, 
phase 2 is useless, since it prevents target detection. Actually, an estimate of $v_l^c$ can be obtained 
from (\ref{2dvvgain}) as   the value of $v_l$ for which $gain=1$~: 
\begin{equation}\label{2dvvvlc}
v_l^c \simeq \frac{\pi V}{4}\left(\ln\left(\frac{b}{a}\right)\right)^{-0.5} \propto \frac{V}{\sqrt{\ln(b/a)}}.
\end{equation}
\imagea{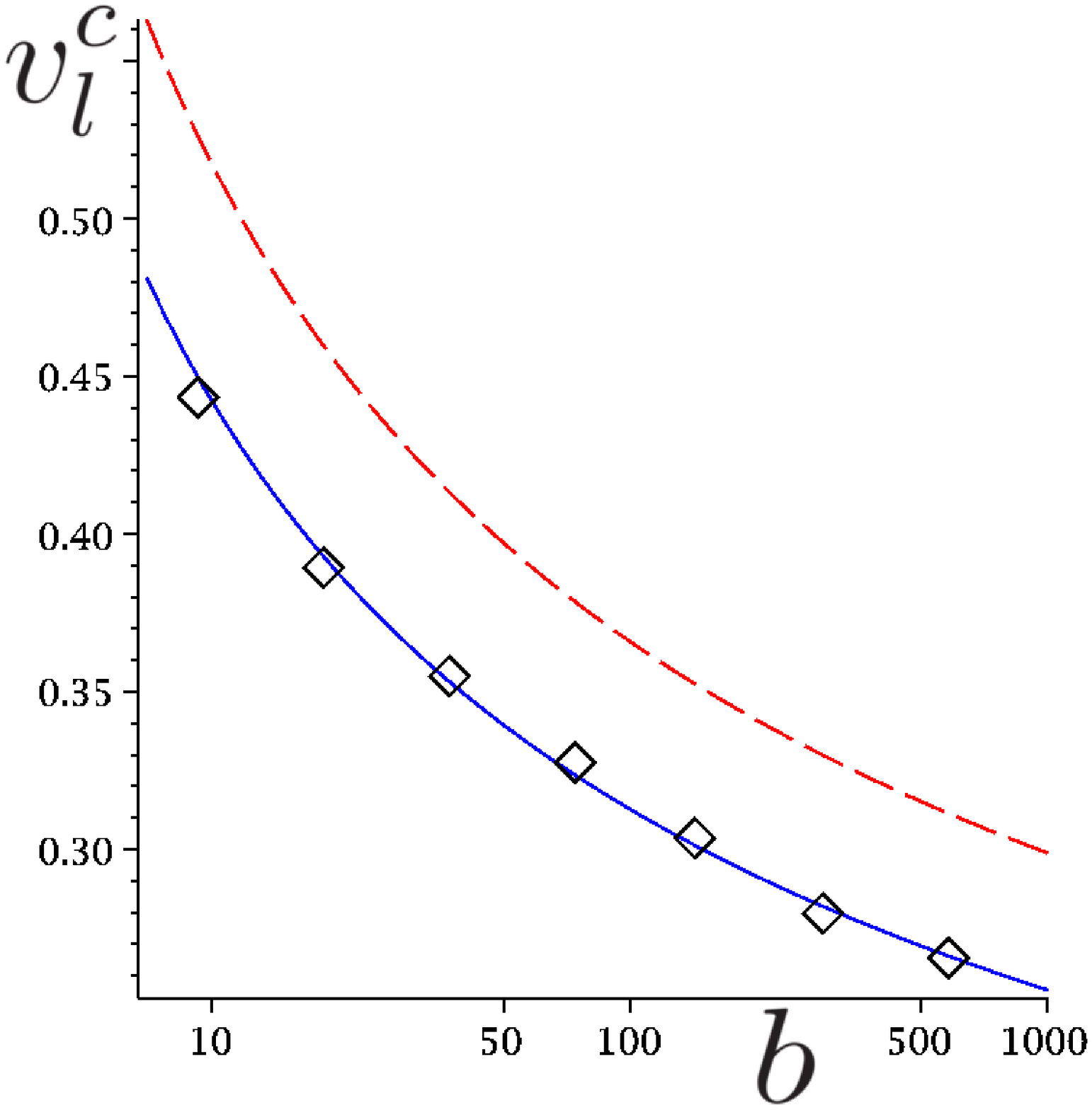}{Ballistic mode in two dimension. $v_l^c$ as a function of $\ln(b)$
by simulations (symbols), predicted expression \refm{2dvvvlc} (red dotted line),
predicted expression multiplied with a fitted numerical constant (blue line). $V=1$, $a=1$.
}{2dvvvlcfig}{5}
We note that this expression (\refi{2dvvvlcfig}) gives the correct dependence 
on $b$, but it however departs from the value obtained by numerical simulations. 
This is due to fact that the expression of $t_m^{opt}$ with intermittence \refm{2dvvtmoptsimp}
is an under estimate, 
while  $t_{bal}$ given in  \refm{topt2} is an upper estimate. 
It is noteworthy that intermittence is less favorable with increasing $b$. 
This effect is similar to the 1 dimensional case, even though it is less important here. 
It can be understood as follows: 
at very large scales the intermittent trajectory is reoriented many times and 
therefore scales as diffusion, which is less favorable than the non intermittent ballistic motion.

\subsubsection{Summary}

We studied the case of ballistic mode in the detection phase 1 in dimension 2. 
When $v_l > v_l^c$, the optimal strategy is to remain in  phase 1 and  to explore the 
domain in a purely ballistic way. 
Therefore, $\tau_2^{opt} \to 0$ ,  $\tau_1^{opt} \to \infty$.
When $v_l < v_l^c$, we find on the contrary $\tau_1^{opt} \to 0$
and $\tau_2^{opt} = \frac{a}{V}\sqrt{\ln\left( \frac{b}{a} \right) -\frac{1}{2}}$. 
The threshold value is given by $v_l^c \propto \frac{ V}{\sqrt{\ln(b/a)}}$ and shows  
that when the target density decreases, 
intermittence is less favorable.

\subsection{Conclusion of the 2-dimensional problem}

Remarkably, for the   three different modes of detection
(static, diffusive and ballistic), we find a regime where intermittence permits to 
 minimize the search time for one and the same $\tau_2^{opt}$, given by 
 $\tau_2^{opt}= \frac{a}{V}\sqrt{\ln\left( \frac{b}{a} \right) -\frac{1}{2}}$. 
As in dimension 1, this indicates that  optimal intermittent strategies are robust and widely  
independent of the details of the description of the detection mechanism.

\section{Dimension 3}

The 3 dimensional case is also  relevant to biology. 
At the microscopic scale, it corresponds for example to 
intracellular traffic in the bulk cytoplasm of cells, 
or at larger scales to animals living in 3 dimensions, such as 
 plankton~\cite{BartumeusPlancton}, or \textit{C.elegans} in its natural habitat ~\cite{CelHabitat}. 
As it was the case in dimension 2, 
different assumptions have to be made to obtain analytical expressions of the search time. 
We checked the validity of our assumptions 
with numerical simulations using the same algorithms as in dimension 2.

\subsection{Static mode}

\imagea{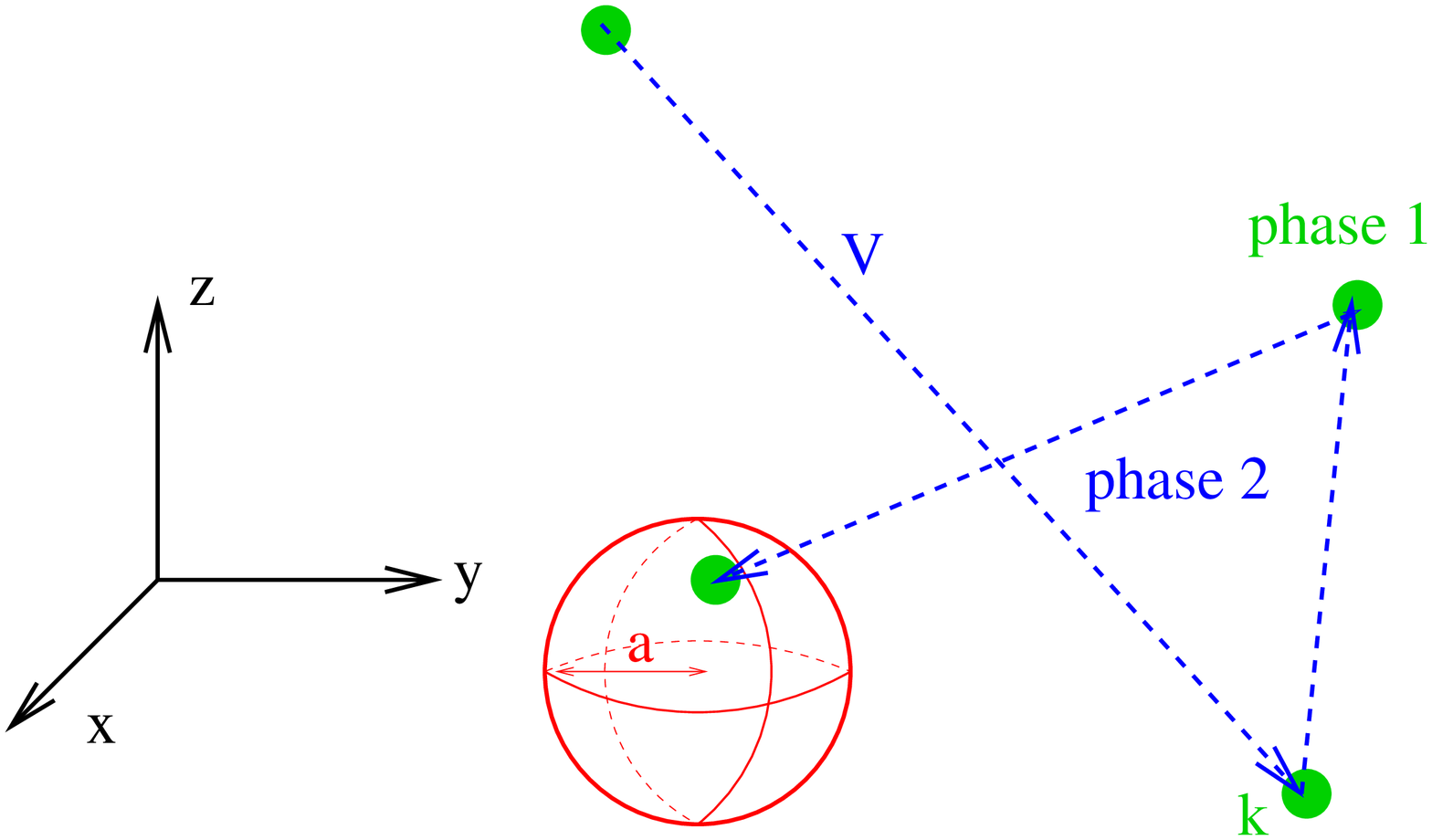}{Static mode in three dimensions}{3k}{5}

We study in this section the case where the detection phase is modeled by the static mode, for which  
the searcher does not move during the detection phase and has 
a finite reaction rate with the target if it is within a detection radius $a$.

\subsubsection{Equations}

Denoting $t_1 (r)$ the mean first passage time at the target starting from a distance $r$ from the target  in phase 1 (detection phase),
and $t_{2,\theta,\phi}(r)$ the mean first passage time at the target starting from a distance $r$ from the target in phase 2 (relocation phase) 
with a direction characterized by $\theta$ and $\phi$, we get~: 
\begin{equation}
 \overrightarrow{V}.\overrightarrow{\bigtriangledown}t_{2,\theta,\phi} +\frac{1}{\tau_2}\left( t_1- t_{2,\theta,\phi}\right)=-1.
\end{equation}
Then one has outside the target  ($r>a$)~:
\begin{equation}
\frac{1}{\tau_1}\left( \frac{1}{4\pi}\int_0^{\pi} d\theta sin\theta\int_0^{2\pi}d\phi t_{2,\theta,\phi} - t_1 \right) = -1, 
\end{equation}
and inside the target ($r\le a$)~: 
\begin{equation}
\frac{1}{\tau_1}\frac{1}{4\pi}\int_0^{\pi} d\theta sin\theta\int_0^{2\pi}d\phi t_{2,\theta,\phi} - \left(\frac{1}{\tau_1} + k \right) t_1  = -1. 
\end{equation}
With $t_2=\frac{1}{4\pi}\int_0^{\pi} d\theta sin\theta\int_0^{2\pi}d\phi t_{2,\theta,\phi}$, we obtain outside the target ($r>a$)~:  
\begin{equation}
\frac{1}{\tau_1}\left(t_2 - t_1 \right) = -1 
\end{equation}
and inside the target ($r<a$)~: 
\begin{equation}
\frac{1}{\tau_1} t_2 - \left(\frac{1}{\tau_1} + k \right) t_1  = -1. 
\end{equation}
Making a similar decoupling approximation as  in dimension 2, we finally get~: 
\begin{equation}
\frac{ V^2 \tau_2}{3} \bigtriangleup t_2  -\frac{1}{ \tau_2}(t_1-t_2)=-1.
\end{equation}
We solve these equations for inside and outside the target, using the following  boundary conditions~:
\begin{equation}
 \left.\frac{d t_2^{out}}{dr}\right|_{r=b}=0
\end{equation}
\begin{equation}
t_2^{out}(a)=t_2^{in}(a)
\end{equation}
%la condition suivante est loin d etre evidente et j ai l impression que c est aussi un eapprox
\begin{equation}
 \left.\frac{d t_2^{out}}{dr}\right|_{r=a}=\left.\frac{d t_2^{in}}{dr}\right|_{r=a}
\end{equation}
and the condition that 
$t_2^{in}(0)$ should be finite.

\subsubsection{Results}

We find an explicit expression of the mean search time
\begin{equation}\label{tm3dvk}
t_m= \left( \tau_1+\tau_2 \right)  \left( {\frac {1}{k \tau_1}}
+ \frac{1}{b^3 V^2 \tau_2^2}\left( -2\,{b}^{3} \left( {b}^{2}-{a}^{2} \right) + \left( {b}^{3}-{
a}^{3} \right)  \left( 3\,{\frac {{a}^{2}}{{\alpha}^{2}}}+ \beta \right) +
\frac{1}{5}(b^5-a^5) \right)
 \right) 
\end{equation}
with \begin{equation}
 \beta=\frac{ -
\sinh \left( \alpha \right) {a}^{3}+{\alpha\,\cosh \left( 
\alpha \right) {b}^{3}}}{a\left(  -\sinh \left( \alpha
 \right) +\alpha\,\cosh \left( \alpha \right) \right)}
\end{equation}
and $\alpha=\sqrt{3\frac{k \tau_1}{1+k \tau_1}} \frac{a}{V\tau_2}$.

In the limit $b\gg a$, this can be simplified to~:
\begin{equation}\label{tm3dvkbgga}
 t_m= \left( \tau_1+\tau_2 \right)  \left( {\frac {1}{k \tau_1}} +\frac{1}{\tau_2^2 V^2}\left(\frac{ -
\sinh \left( \alpha \right) {a}^{3}+{\alpha\,\cosh \left( 
\alpha \right) {b}^{3}}}{a\left(  -\sinh \left( \alpha
 \right) +\alpha\,\cosh \left( \alpha \right) \right)}-\frac{9}{5}b^2+\frac{3a^2}{\alpha^2} \right)
 \right).
\end{equation}
Assuming further that 
 $\alpha$ is small, we use the expansion 
$\beta \simeq \frac{b^3}{a}\left(1-\tanh(\alpha)/\alpha  \right)^{-1} \simeq 
\frac{b^3}{a} \left(\frac{3}{\alpha^2}+\frac{6}{5} \right)$ and rewrite mean search time as~: 
\begin{equation}\label{tm3dvksimp}
  t_m = \frac{b^3( \tau_2+ \tau_1)}{a} \left(\frac{(1+k  \tau_1) }{ \tau_1 k a^2}+\frac{6 }{5\tau_2^2 V^2} \right).
\end{equation}
This expression of $t_m$ can be minimized for~: 
\begin{equation}\label{tau13dvk}
 \tau_1^{opt}=\left(\frac{3}{10}\right)^{\frac{1}{4}} \sqrt{\frac{a}{Vk}}
\end{equation}
\begin{equation}\label{tau23dvk}
 \tau_2^{opt}=\sqrt{1.2}\frac{a}{V},
\end{equation}
and the minimum mean search time reads finally: 
\begin{equation}\label{topt3dvk}
   t_m^{opt}=\frac{1}{\sqrt{5}}\frac{1}{k}\frac{b^3}{a^3}\left( \sqrt{\frac{ak}{V}}24^{1/4}+5^{1/4} \right)^2.
\end{equation}

\subsubsection{Comparisons with simulations}\label{r3dvk}

\begin{figure}[h!]
   \begin{minipage}[c]{.3\linewidth}
\begin{center}
      \includegraphics[width=4cm]{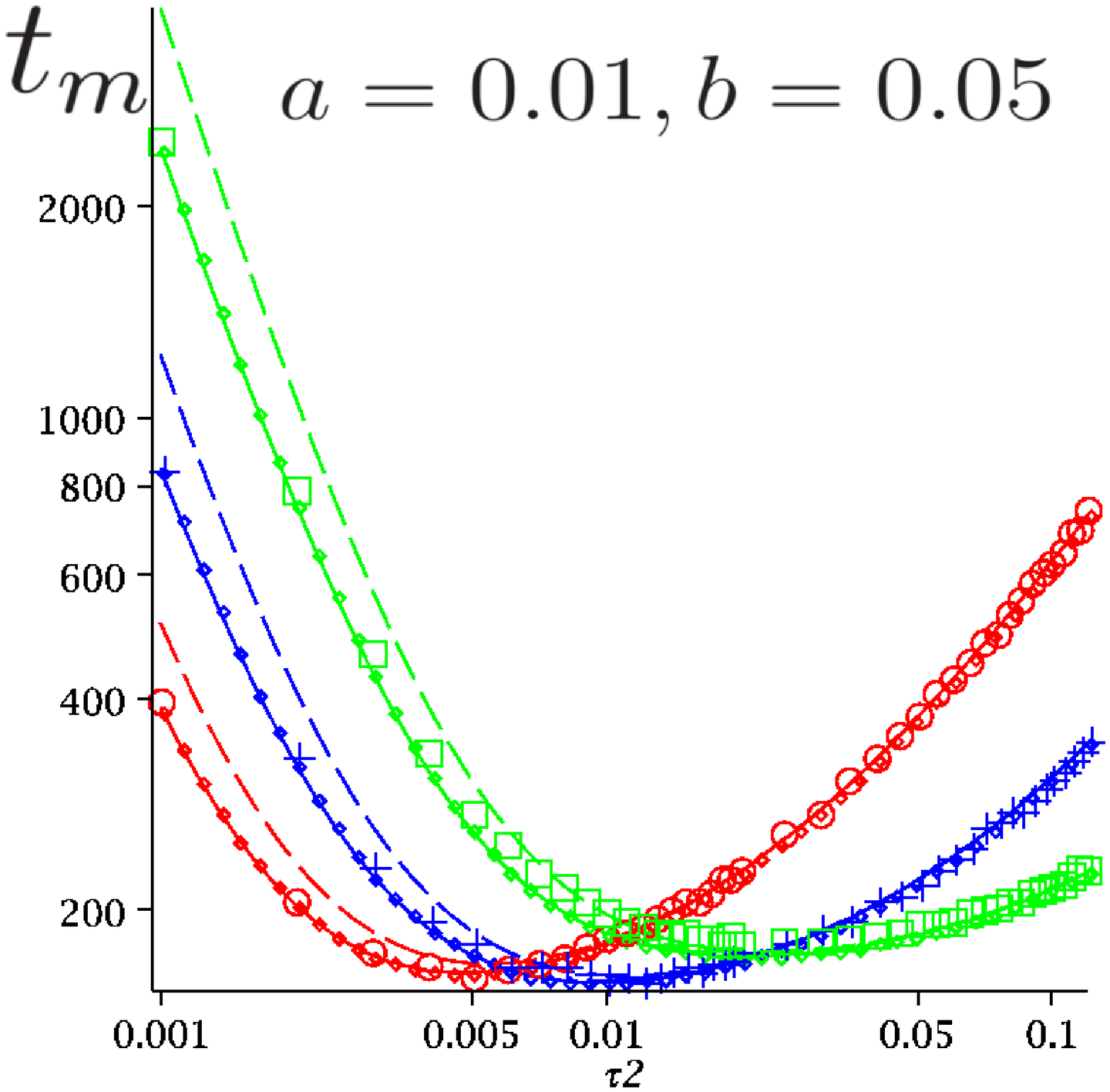}

      \includegraphics[width=4cm]{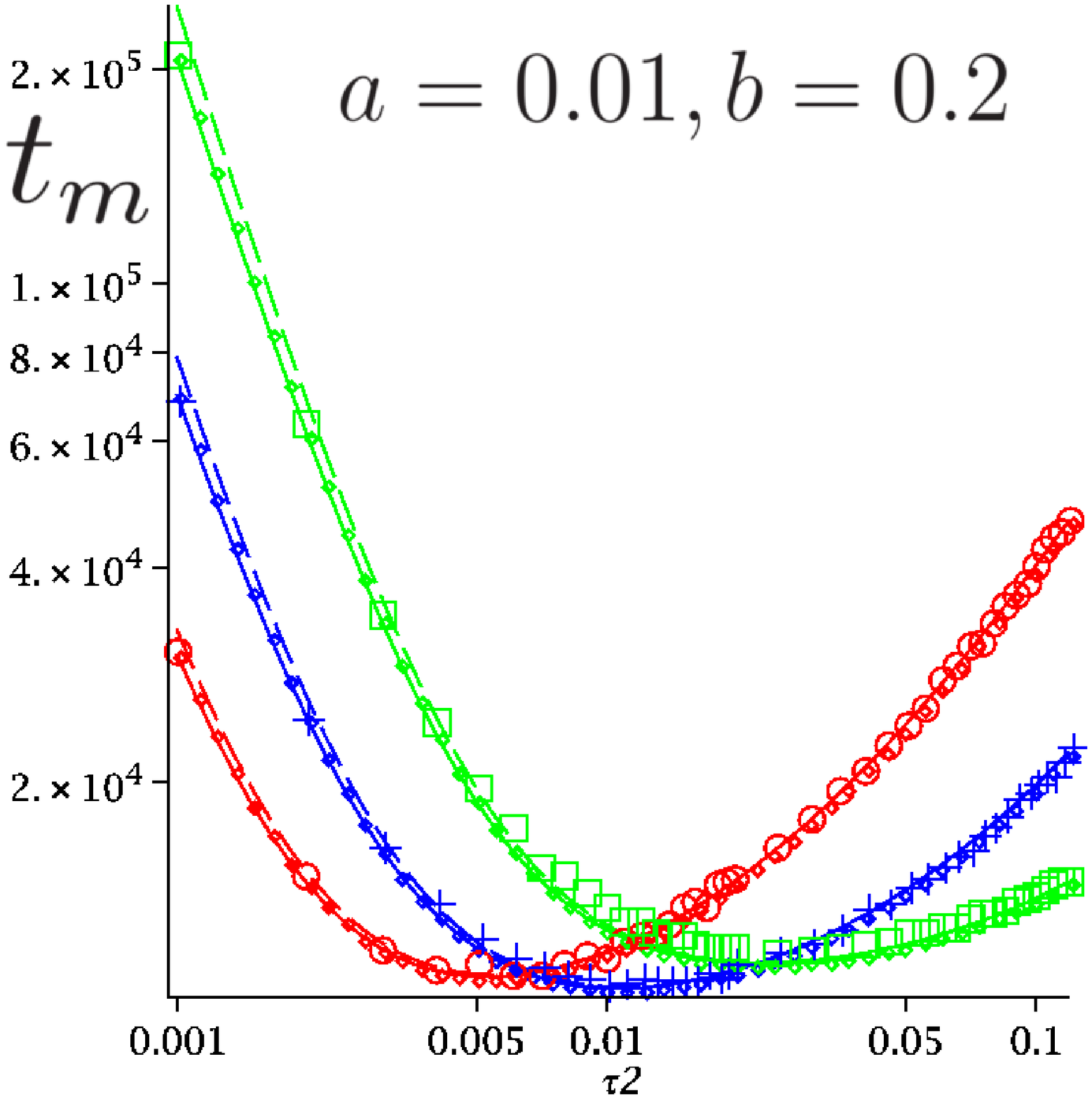}

\end{center}
   \end{minipage} \hfill
   \begin{minipage}[c]{.3\linewidth}
\begin{center}
      \includegraphics[width=4cm]{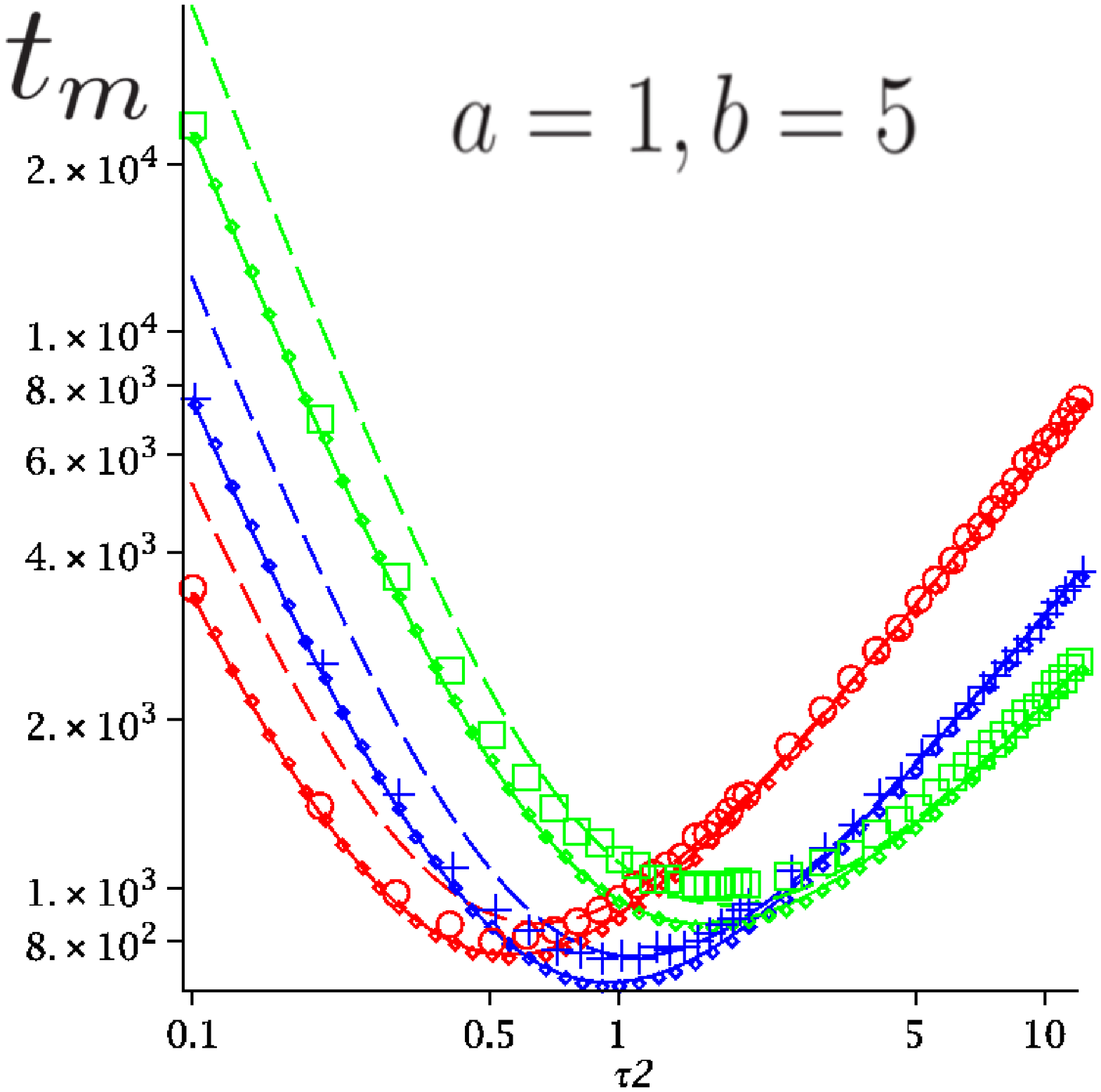}

      \includegraphics[width=4cm]{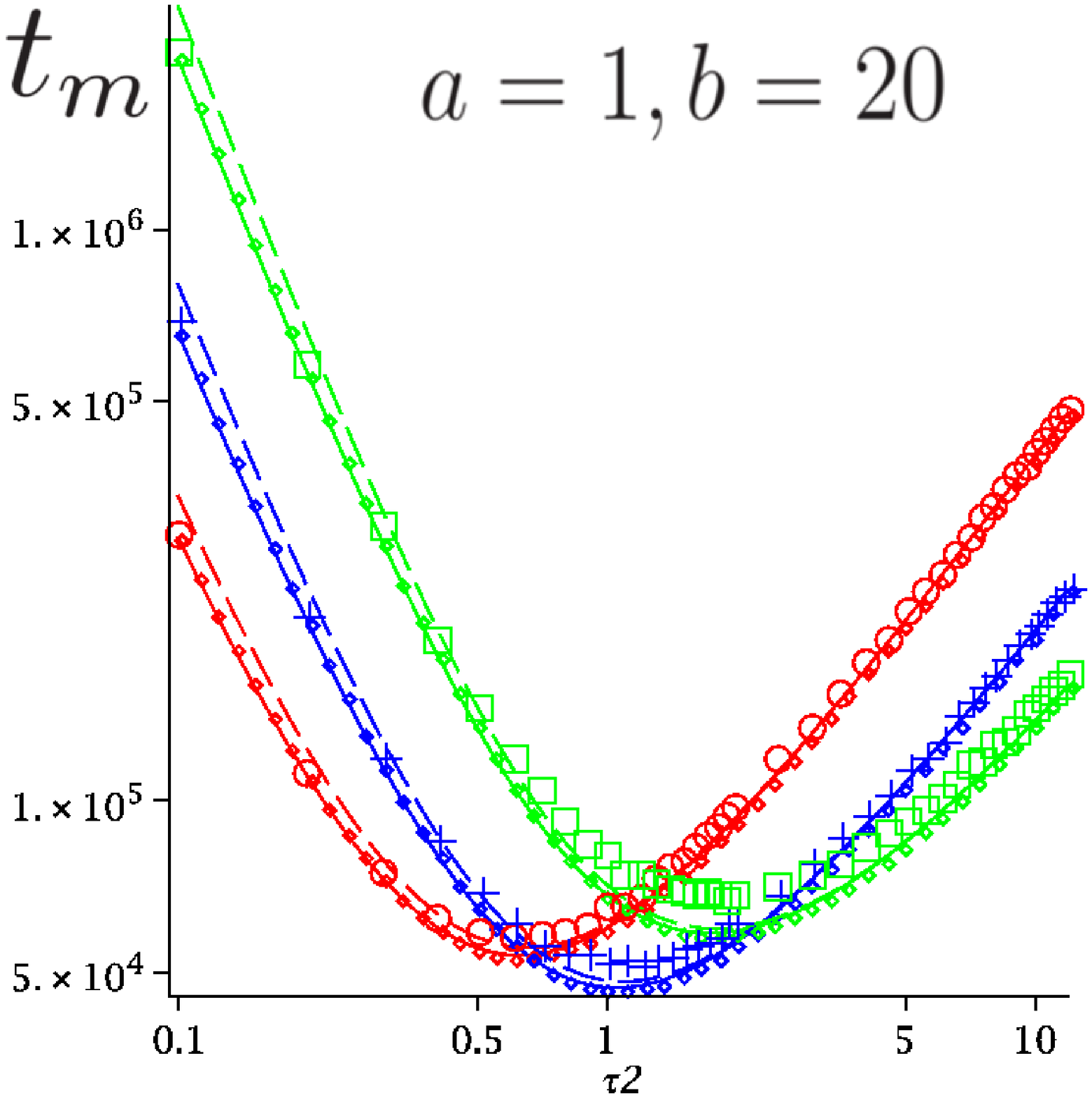}

\end{center}
   \end{minipage}\hfill
   \begin{minipage}[c]{.3\linewidth}
\begin{center}
      \includegraphics[width=4cm]{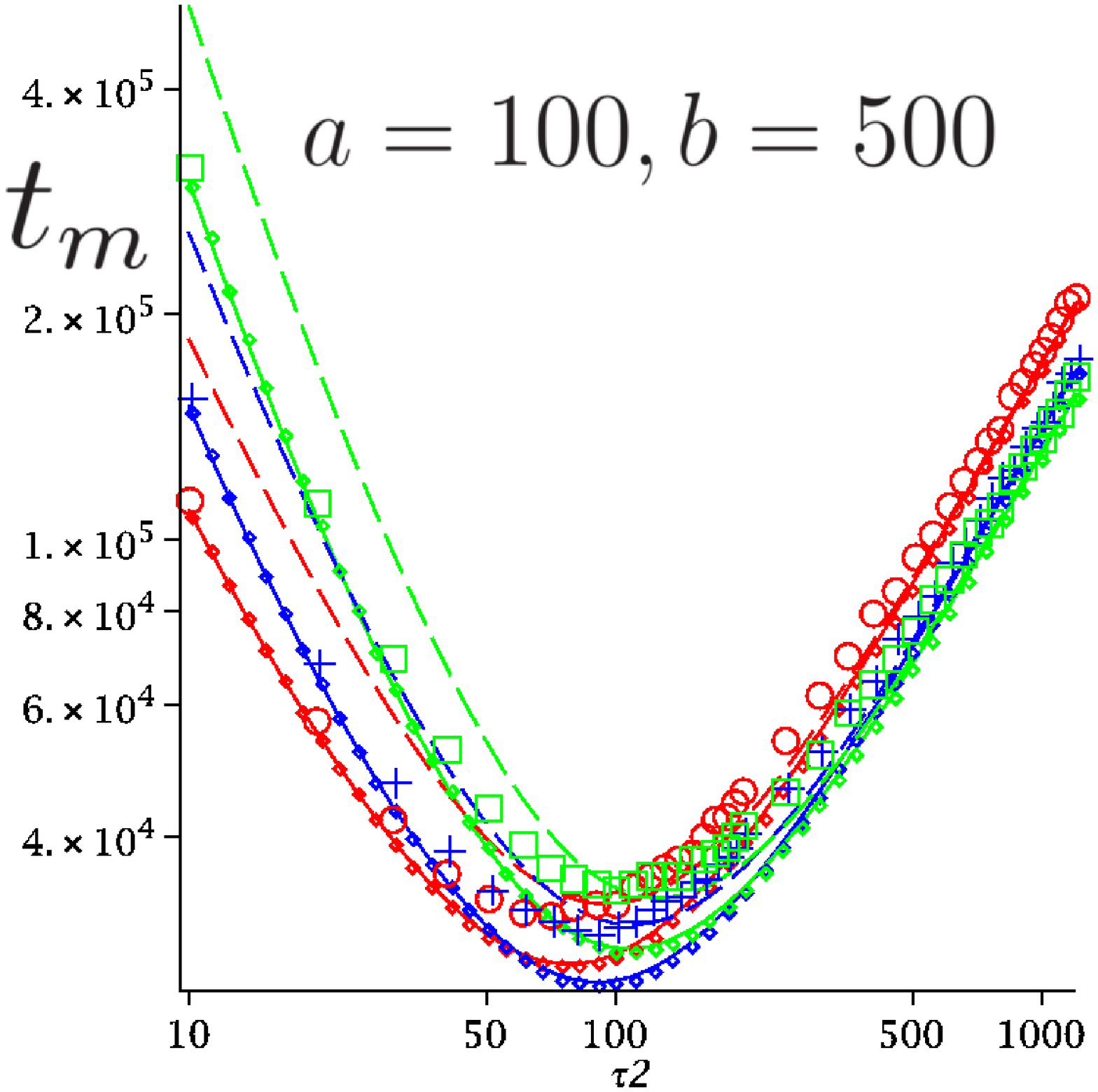} 

     \includegraphics[width=4cm]{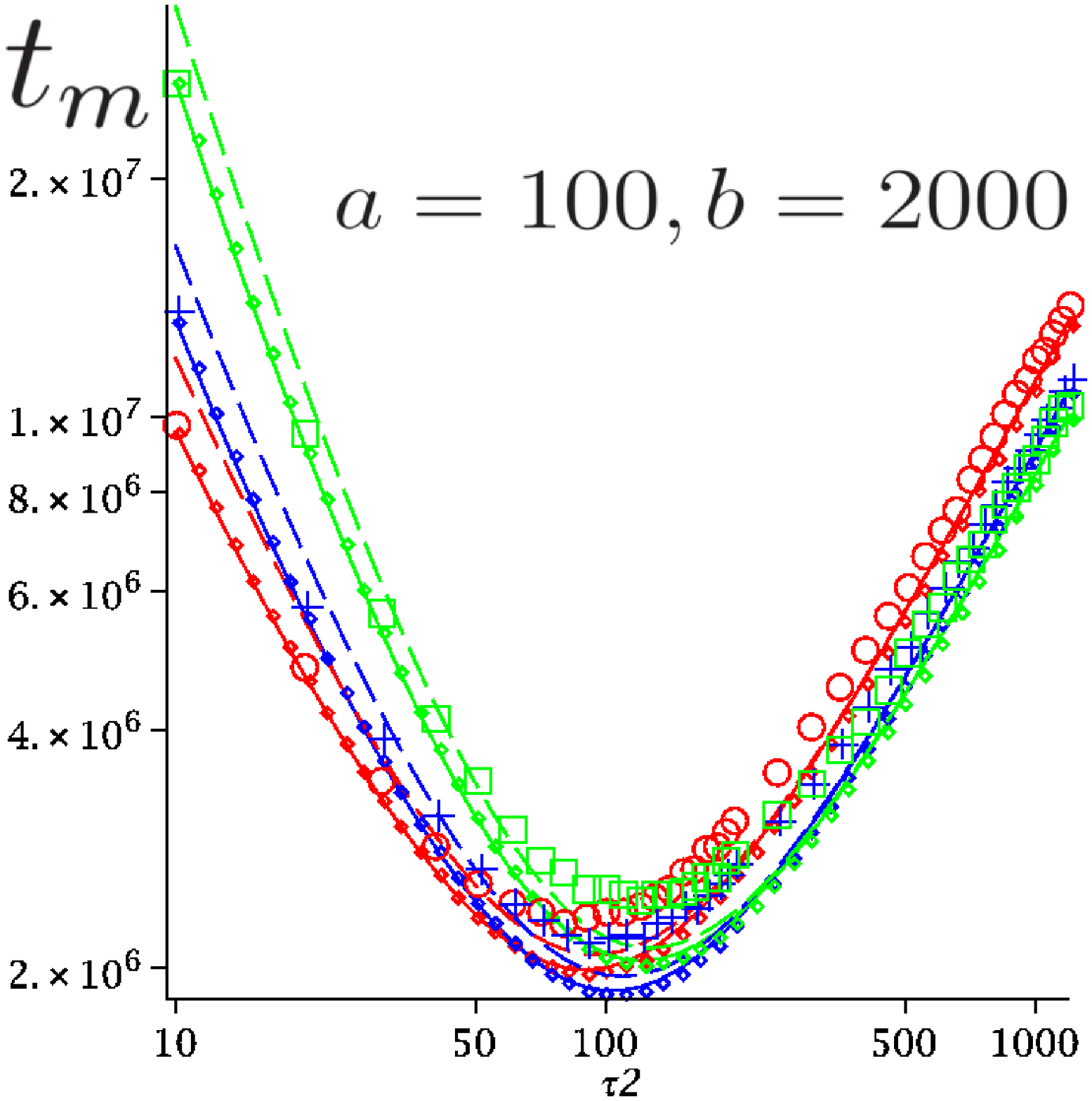}

\end{center}
   \end{minipage}

\caption{(Color online) Static mode in 3 dimensions. $\ln(t_m)$ as a function of $\ln(\tau_2)$ for different values of $\tau_1$, $a$ and $b/a$. 
Comparison between simulations (symbols), analytical expression \refm{tm3dvk} (line), expression for $b \gg a$ \refm{tm3dvkbgga} (small dots), 
and simple expression for $b\gg a$ and $\alpha$ small \refm{tm3dvksimp} (dotted line).
$\tau_1 \simeq \tau_1^{opt}  \simeq 0.74 \sqrt{\frac{aV}{k}}$ \refm{tau13dvk} (blue, crosses), 
$\tau_1 = 0.25 \sqrt{\frac{aV}{k}}$ (red, circles), $\tau_1 = 2.5 \sqrt{\frac{aV}{k}}$ (green, squares).
$V=1$, $k=1$.}\label{3Dvkcomp}
\end{figure}

Data obtained by numerical simulations (\refi{3Dvkcomp}, and additionally in  \refaa{a3dvk}) are 
in good agreement with  the analytical expression \refm{tm3dvk}.
In particular,  the position of the minimum is very well approximated, 
and the error on the value of the mean search time at  the minimum is close to 10\%. 
Note that the very simple expression \refm{tm3dvksimp} fits also rather  well the numerical data, 
except for small $\tau_2$ or small  $b$.

\subsubsection{Summary}

In the case of a  static detection mode in  dimension 3, 
we obtained  a simple approximate expression of the mean first passage time at the target 
$ t_m = \frac{b^3( \tau_2+ \tau_1)}{a} \left(\frac{(1+k  \tau_1) }{ \tau_1 k a^2}+\frac{6 }{5\tau_2^2 V^2} \right)$.
$t_m$ has a single minimum for 
$\tau_1^{opt}= \left(\frac{3}{10}\right)^{\frac{1}{4}} \sqrt{\frac{a}{Vk}}$ and 
$\tau_2^{opt}=\sqrt{1.2}\frac{a}{V}$, and 
the minimal mean search time is 
$\frac{1}{\sqrt{5}}\frac{1}{k}\frac{b^3}{a^3}\left( \sqrt{\frac{ak}{V}}24^{1/4}+5^{1/4} \right)^2$.
With the static  detection mode, intermittence is always favorable and leads 
to a single  optimal intermittent strategy. 
As in dimension 1 and 2, the optimal duration of the relocation phase 
does not depend on $k$, i.e. on the description of the detection phase. 
In addition, this optimal strategy does not depend on the typical distance between targets $b$.

One can notice than for the static mode in the three cases studied (1, 2, and 3 dimensions), 
we have the relation : $\tau_1^{opt}=\sqrt{\tau_2^{opt}/(2k)}$. 
The optimal durations of the two phases are related independently of the dimension.

\subsection{Diffusive mode}

\imagea{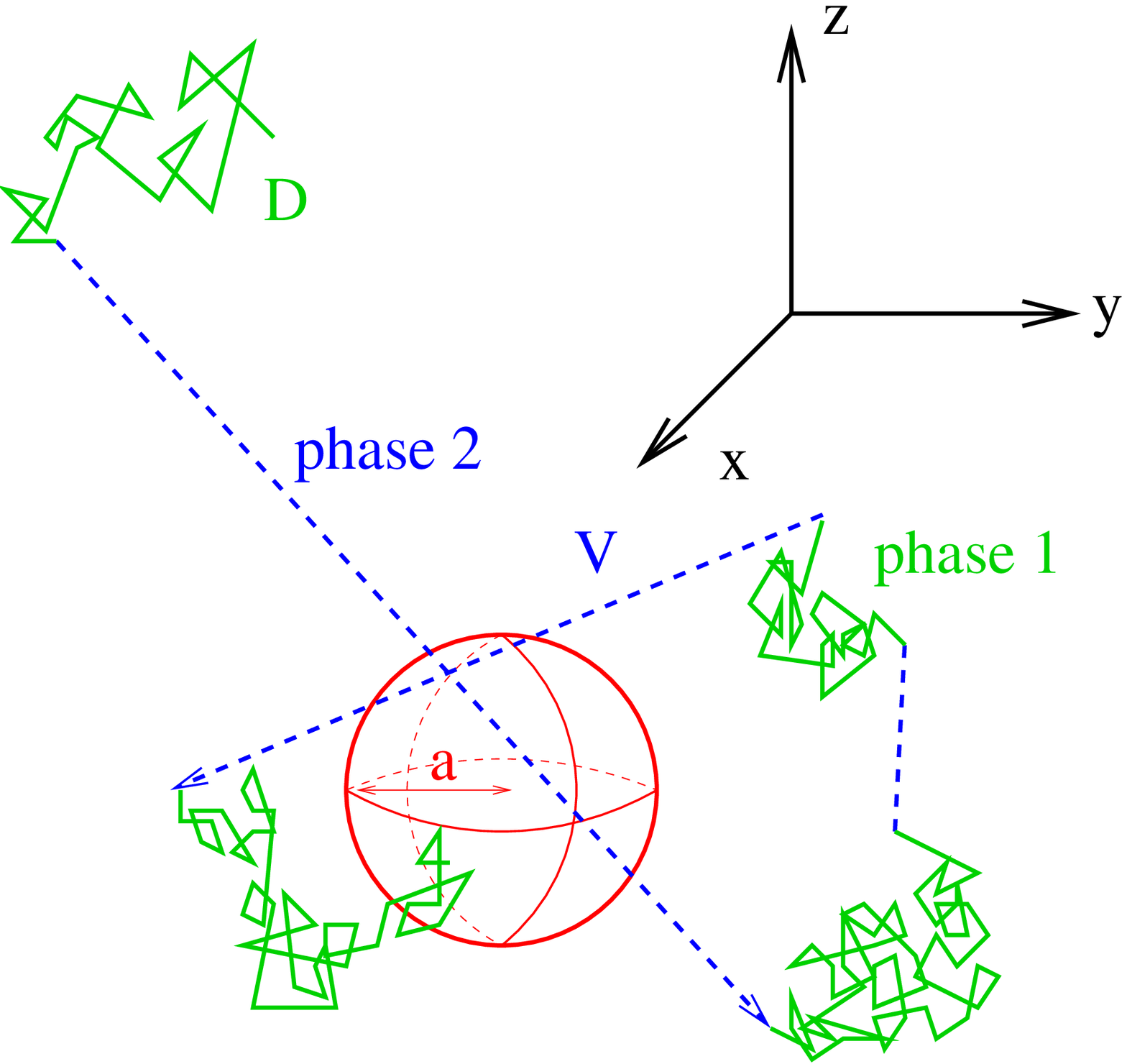}{Diffusive mode in three dimensions}{3D}{5}

We now study the case where the detection phase is modeled by a diffusive mode.
During the detection phase, 
the searcher diffuses and detects  
the target as soon as their respective
distance is less than $a$.

\subsubsection{Equations}

%With $t_1 (r)$ the mean first passage time on the target starting from a distance $r$ of the center and in phase 1 (detection phase),
%and $t_{2,\theta,\phi}(r)$ the mean first passage time on the target starting from a distance $r$ from the center and in phase 2 (relocation phase) 
%with a direction characterized by $\theta$ and $\phi$, outside the target we can write~: 

One has outside the target ($r>a$)~:
\begin{equation}
 \overrightarrow{V}.\overrightarrow{\bigtriangledown}t_{2,\theta,\phi} +\frac{1}{\tau_2}\left( t_1- t_{2,\theta,\phi}\right)=-1
\end{equation}
\begin{equation}
D \bigtriangleup t_1  +\frac{1}{\tau_1}\left( \frac{1}{4\pi}\int_0^{\pi} d\theta sin\theta\int_0^{2\pi}d\phi t_{2,\theta,\phi} - t_1 \right) = -1 
\end{equation}
and inside the target ($r\le a$)~: 
\begin{equation}
 \overrightarrow{V}.\overrightarrow{\bigtriangledown}t_{2,\theta,\phi} -\frac{1}{\tau_2} t_{2,\theta,\phi}=-1
\end{equation}
\begin{equation}
t_1=0.
\end{equation}
With $t_2=\frac{1}{4\pi}\int_0^{\pi} d\theta sin\theta\int_0^{2\pi}d\phi t_{2,\theta,\phi}$, we get outside the target ($r>a$)~:  
\begin{equation}
D \bigtriangleup t_1^{out}+\frac{1}{\tau_1}\left(t_2^{out} - t_1^{out} \right) = -1 
\end{equation}
The decoupling approximation described in previous sections then yields outside the target~:  
\begin{equation}
\frac{ V^2 \tau_2}{3} \bigtriangleup t_2^{out}  + \frac{1}{ \tau_2}(t_1^{out}-t_2^{out})=-1
\end{equation}
and inside the target ($r\le a$)~: 
\begin{equation}
\frac{ V^2 \tau_2}{3} \bigtriangleup t_2^{int}  -\frac{1}{ \tau_2}t_2^{int}=-1.
\end{equation}
These equations are completed by the following boundary conditions~: 
\begin{equation}
 \left.\frac{d t_2^{out}}{dr}\right|_{r=b}=0
\end{equation}
\begin{equation}
t_2^{out}(a)=t_2^{int}(a)
\end{equation}
\begin{equation}
 \left.\frac{d t_2^{out}}{dr}\right|_{r=a}=\left.\frac{d t_2^{int}}{dr}\right|_{r=a}.
\end{equation}

\subsubsection{Results in the general case}\label{r3dvd1}

Through standard but lengthy calculations we can solve the above system and get  
an analytical approximation of $t_m$ \refa{a3dvd1}.
In the regime  $b\gg a$, we use  the assumption $\sqrt{(\tau_1 D)^{-1}+3(\tau_2v)^{-2}} \ll b$ and obtain~:
\begin{equation}\label{tmb3dvd}
 t_m = \frac{b^3 \kappa_2^4(\tau_1+\tau_2)}{\kappa_1}\frac{\tanh(\kappa_2a)+\frac{\kappa_1}{\kappa_2}}{\kappa_1\kappa_2^2\tau_1Da\left(\tanh(\kappa_2a)+\frac{\kappa_1}{\kappa_2}  \right)-\tanh(\kappa_2 a)}
\end{equation}
with  $ \kappa_1=\frac{\sqrt{\tau2^2V^2+3\tau_1D}}{\tau_2V\sqrt{D\tau_1}}$ and 
$\kappa_2=\frac{\sqrt{3}}{V \tau_2 }$.
As shown in  Fig.\ref{comp} left or in the additional \refi{tau1effet} in 
appendix \ref{a3dvd_tau1effet}, $t_m$ only weakly depends on $\tau_1$, which indicates that  
this variable will be less important than $\tau_2$ in the minimization of the search time.
The  relevant order of magnitude for $\tau_1^{opt}$ can be evaluated by comparing the  
typical diffusion length $ L_{diff} = \sqrt{6 D t}$ and the typical ballistic length $ L_{bal}=V t$. 
An  estimate of the optimal time $\tau_1^{opt}$  can be given   
by the time scale for which those lengths are of same order, which gives~:
\begin{equation}\label{tau13dvd}
\tau_1^{opt} \sim  \frac{6 D}{V^2}.
\end{equation}
Note that taking $\tau_1 = 0$ does not change significantly $t_m^{opt}$ (\refi{comp} left), and permits to 
significantly simplify  $t_m$~: 
\begin{equation}\label{pearson}
 t_m=\frac{b^3\sqrt{3}}{V^3\tau_2^2}\left(\frac{\sqrt{3}a}{V\tau_2}-\tanh\left(\frac{\sqrt{3}a}{V\tau_2}\right)\right)^{-1}
\end{equation}
In turn, the minimization of this expression leads to~:
\begin{equation}\label{tau23dvd}
\tau_2^{opt} =\frac{\sqrt{3}a}{Vx}
\end{equation}
with $x$ solution of~:
\begin{equation}
 2 \tanh(x)-2 x+x \tanh(x)^2=0.
\end{equation}
This finally yields~:
\begin{equation}
 \tau_2^{opt} \simeq 1.078 \frac{a}{V}.
\end{equation}
Importantly this approximate expression  is very close to the expression obtained for the static mode 
($\tau_2^{opt} = \sqrt{\frac{6}{5}}\frac{a}{V} \simeq 1.095 \frac{a}{V}$) \refm{tau23dvk}, and 
there is no dependence with the typical distance between targets $b$. 
The simplified expression of the minimal $t_m$ \refm{pearson} can then be obtained as: 
\begin{equation}\label{tmopt3dvd}
 t_m^{opt} = \frac{b^3 x^2}{\sqrt{3}a^2 V} \left(x-\tanh(x) \right)^{-1} \simeq 2.18 \frac{b^3}{a^2V}
\end{equation}
and the gain reads: 
\begin{equation}\label{3dvdgain}
 gain=\frac{t_{diff}}{t_m^{opt}} \simeq 0.15 \frac{aV}{D}.
\end{equation}

\subsubsection{Comparison between analytical approximations and numerical simulations}

\doublimagem{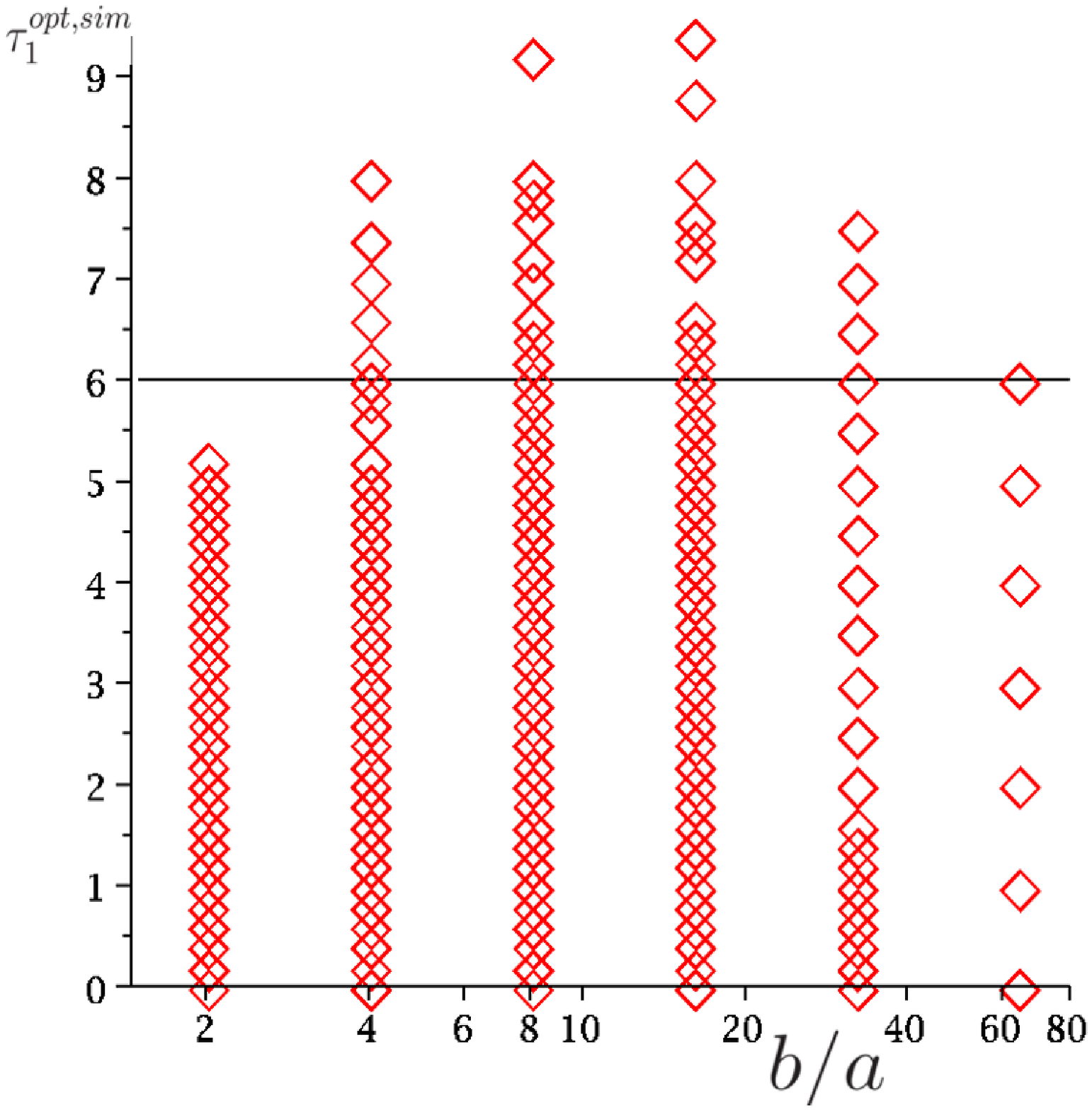}{}
{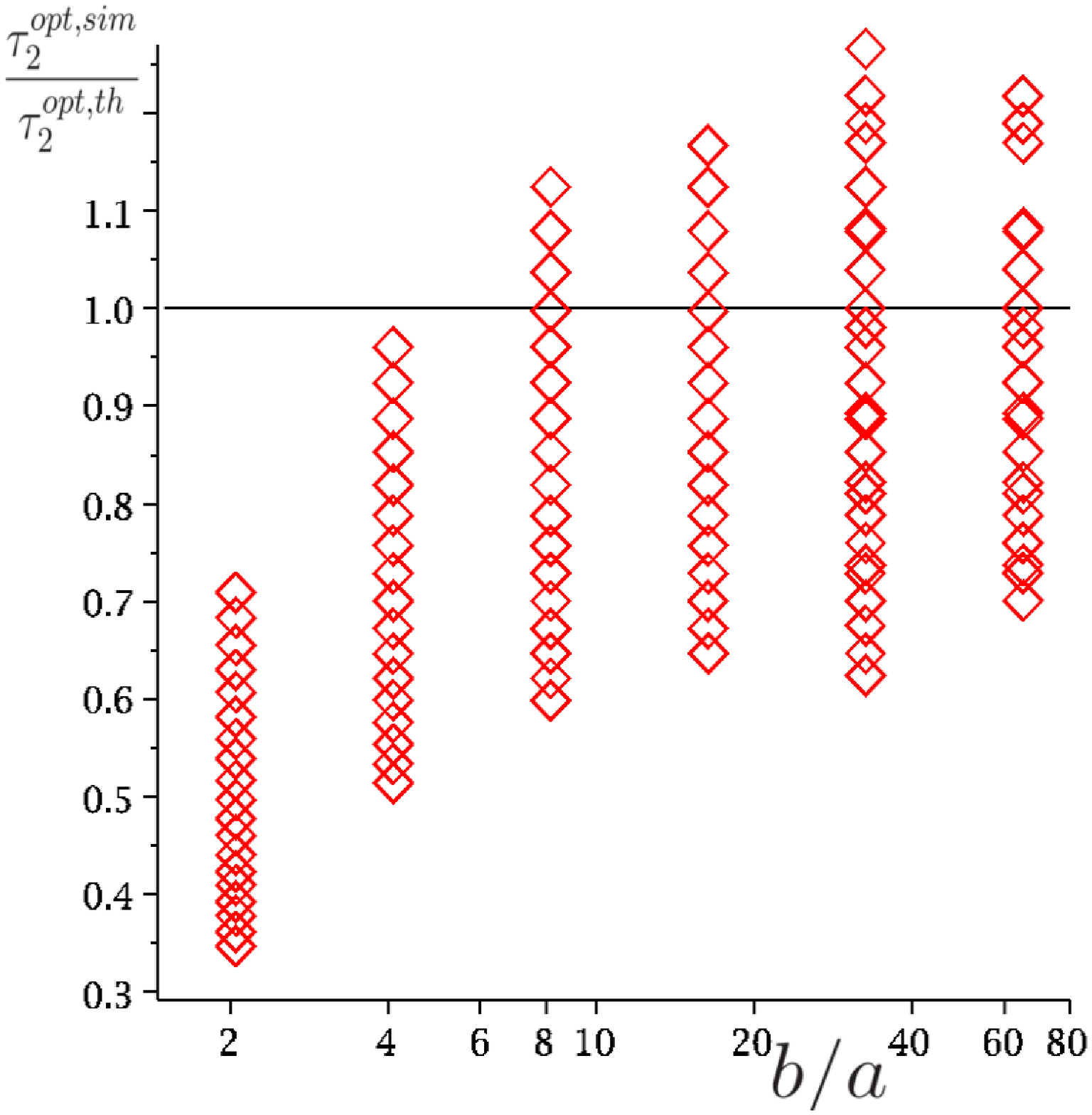}{}
{Diffusive mode in 3 dimension. Comparison between analytical approximations \refm{tau13dvd} \refm{tau23dvd} (black lines) and numerical simulations~: the symbols are the values of $\tau_1$ and $\tau_2$ for which $t_m^{simulation}<1.05t_m^{opt,simulation}$. $a=100$, $V=1$, $D=1$}{comp}

Numerical simulations reveal that the minimum  of $t_m$ with respect to  $\tau_1$ is shallow as it was expected (cf Fig.\ref{comp} left).
It approximately ranges from 0 to the theoretical estimate \refm{tau13dvd}. 
The value $\tau_2^{opt,sim}$ at the minimum is close to the expected values \refm{tau23dvd} (cf. Fig.\ref{comp} right),
except for very small $b$, which is consistent  with our assumption  $b\gg a$. 
We can then conclude than the position of the optimum in $\tau_1$ and $\tau_2$ is very well described
by the analytical approximations, even if  the value of $t_m$ at the minimum is underestimated by our analytical approximation
by about 10-20\% (\refi{3dvd_courbes}).

\doublimage{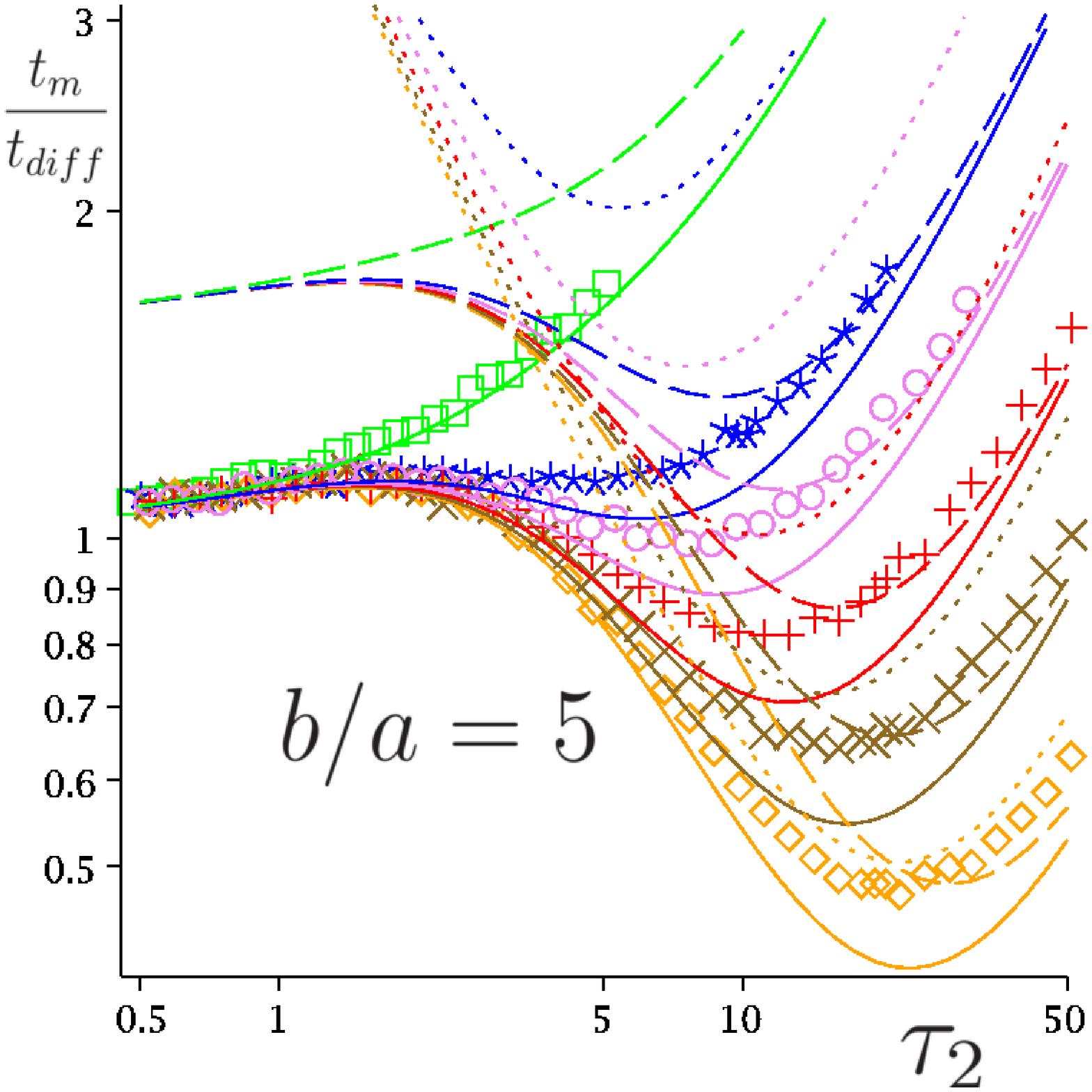}{}
{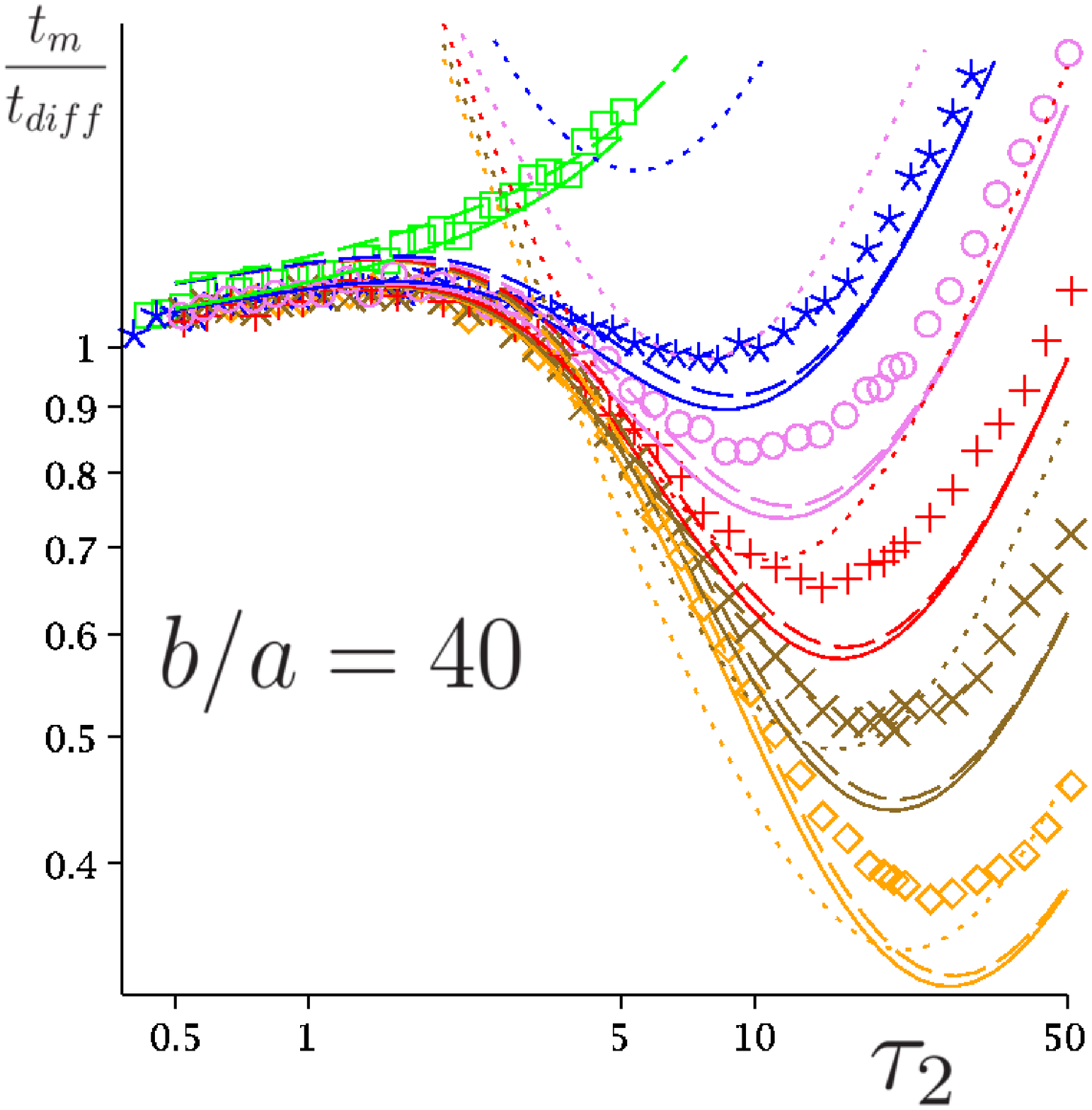}{}
{Diffusive mode in 3 dimension. $t_m /t_{diff}$ ($t_{diff}$ given by \refm{tdiff3D}) as a function of $\tau_2$ for different values of the ratio $b/a$ (logarithmic scale). 
The full analytical form \refm{tm3dvd} (plain lines) is plotted against 
the simplified expression \refm{tmb3dvd} (dotted lines), the simplified expression with $\tau_1=0$  \refm{pearson} (small dots),
 and  numerical simulations (symbols) for the following values of the parameters (arbitrary units): 
$a=1$ (green, squares), $a=5$ (blue, stars), $a=7$ (purple, circles), $a=10$ (red, +), $a=14$ (brown, X), 
$a=20$ (orange, diamonds). $\tau_1=6$ everywhere except for the small dots, $v=1$, $D=1$. $t_m /t_{diff}$ presents a minimum only for $a>a_c\simeq 4$. 
}{3dvd_courbes}

\subsubsection{Case without intermittence~: 1 state diffusive searcher }\label{r3dvd3}

If the searcher always remains in the diffusive mode, it is straightforward to obtain \refa{a3dvd3}:
\begin{equation}\label{tdiff3D}
t_{diff} = \frac{1}{15 D ab^3} \left(5b^3a^3+5b^6-9b^5a-a^6 \right), 
\end{equation}
which gives in the limit $b/a \gg 1$~:
\begin{equation}\label{tdiff3Db}
t_{diff}=\frac{b^3}{3 D a}.
\end{equation}

\subsubsection{Criterion for intermittence to be favorable}\label{r3dvd2}
 
There is a range of parameters for which intermittence is favorable, as indicated by  \refi{3dvd_courbes}.
Both the analytical expression for $t_m^{opt}$ in the regime without intermittence \refm{tdiff3D}
and with intermittence \refm{tmopt3dvd} scale as $b^3$. 
However, the dependence on $a$ is different \refa{a3dvd2}. 
In the diffusive regime, $t_m \propto a^{-1}$,
whereas in the intermittent regime $t_m \propto a^{-2}$.
This enables to define   a critical $a_c$, such that  
when $a>a_c$, intermittence is favorable: 
 $a_c \simeq 6.5 \frac{D}{V}$ is the value for which the gain \refm{3dvdgain} is 1.

\subsubsection{Summary}

We studied the case where the detection phase 1 is modeled by a diffusive mode, and  
calculated explicitly an approximation of the mean first passage time at the target.
We found that  intermittence is favorable (i.e. better than diffusion alone) 
when $a>a_c \simeq 6.5 \frac{D}{V}$~: 
\begin{itemize}
\item if $a<a_c$, the best strategy is a 1 state diffusion, without intermittence, and the 
mean first passage time at the target is  $t_m \simeq \frac{b^3}{3 D a}$. 
\item  if $a>a_c$, intermittence is favorable. The dependence on 
$\tau_1$ is not crucial, as long as it is smaller than $\frac{6 D}{V^2}$.
The value of $\tau_2$ at the optimum is $\tau_2^{opt} \simeq 1.08 \frac{a}{V}$.
The minimum search time  is then 
$t_m^{opt} \simeq 2.18 \frac{b^3}{a^2V}$. 
\end{itemize}

\subsection{Ballistic mode}

\imagea{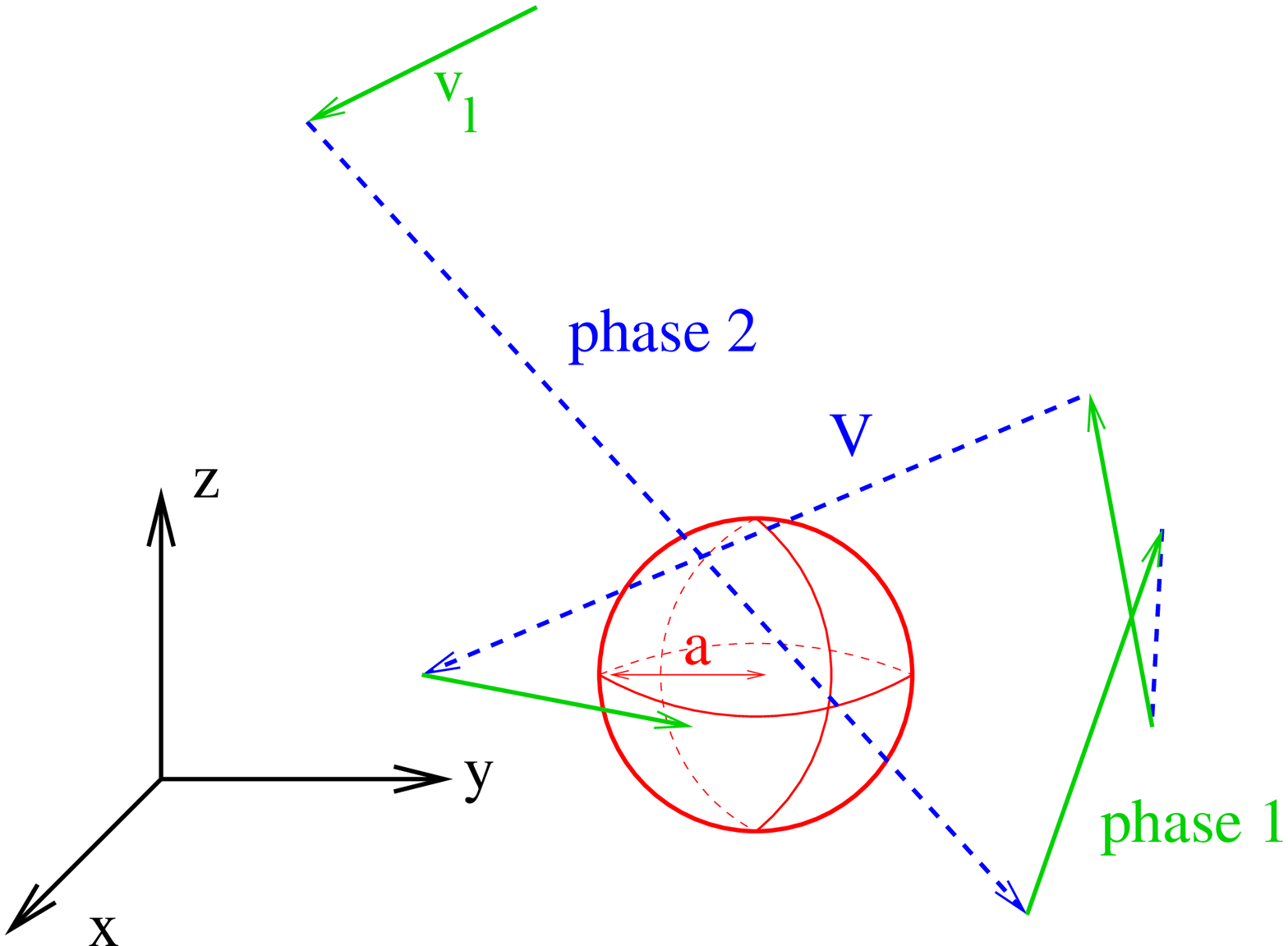}{Ballistic mode in  dimension 3}{3v}{5}

We now discuss the last case, where the detection phase 1 is modeled by a ballistic mode.% This search model  has never been studied so far. 
Since an explicit analytical determination of the search time seems out of reach, 
we proceed as in dimension 2 and  first explore numerically the parameter space to identify 
the regimes where the search time can be minimized. 
We then develop approximation schemes in each  
regime to obtain analytical expressions (more details are given in \refaa{a3dvv}).

\subsubsection{Numerical study}

\begin{figure}[h!]
   \begin{minipage}[c]{.3\linewidth}
\begin{center}
      \includegraphics[width=4cm]{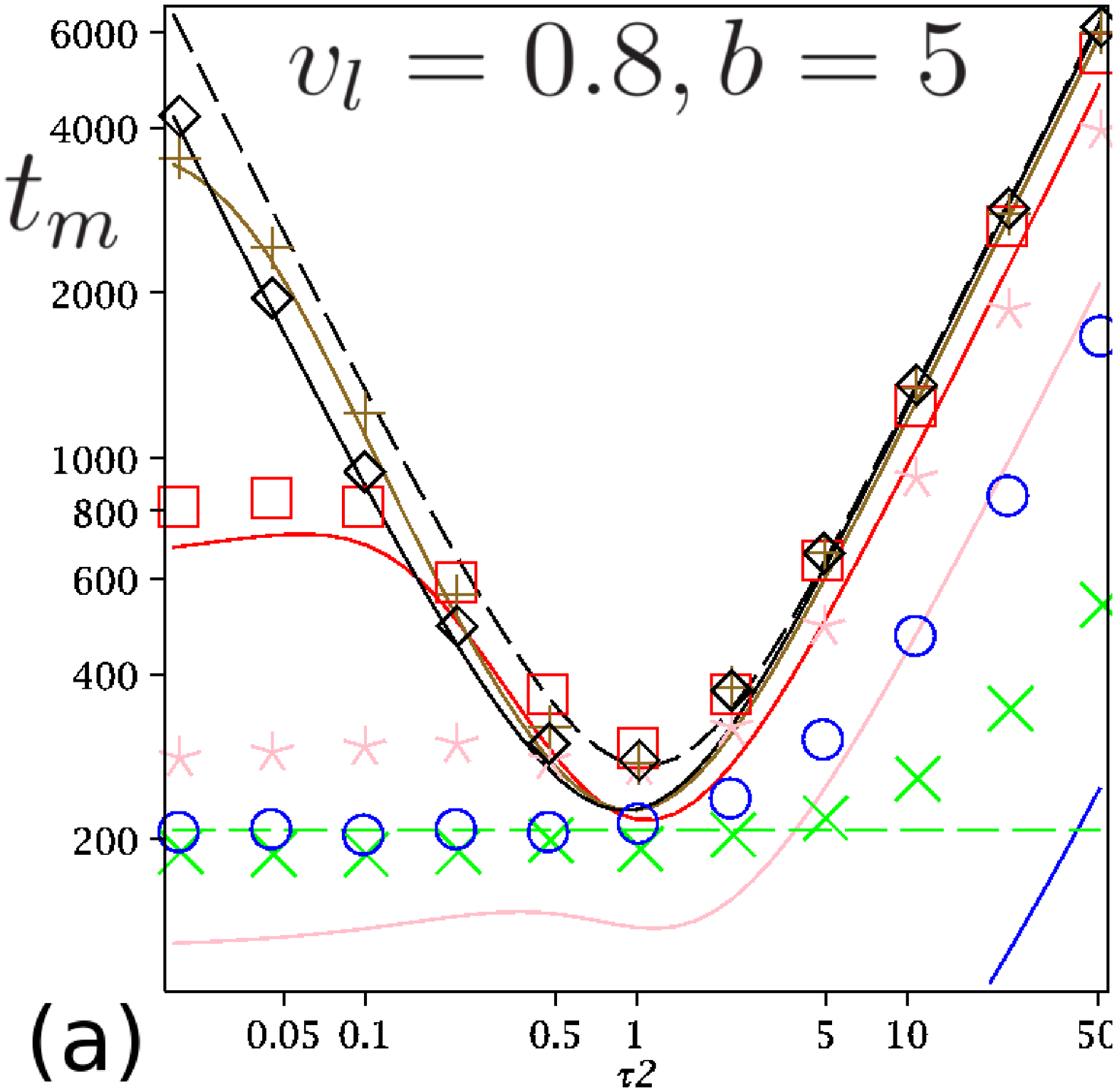}

      \includegraphics[width=4cm]{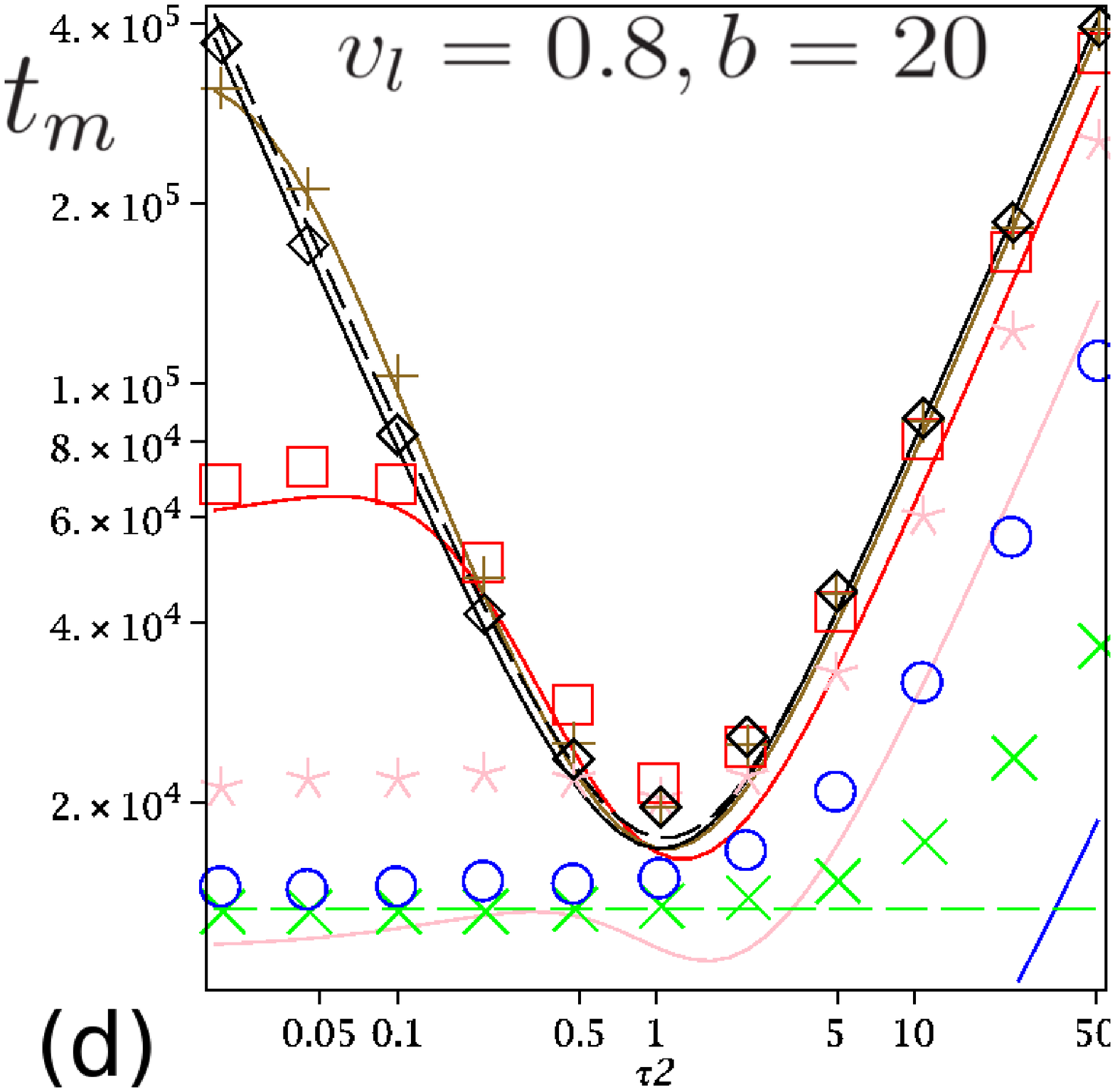}

\end{center}
   \end{minipage} \hfill
   \begin{minipage}[c]{.3\linewidth}
\begin{center}
      \includegraphics[width=4cm]{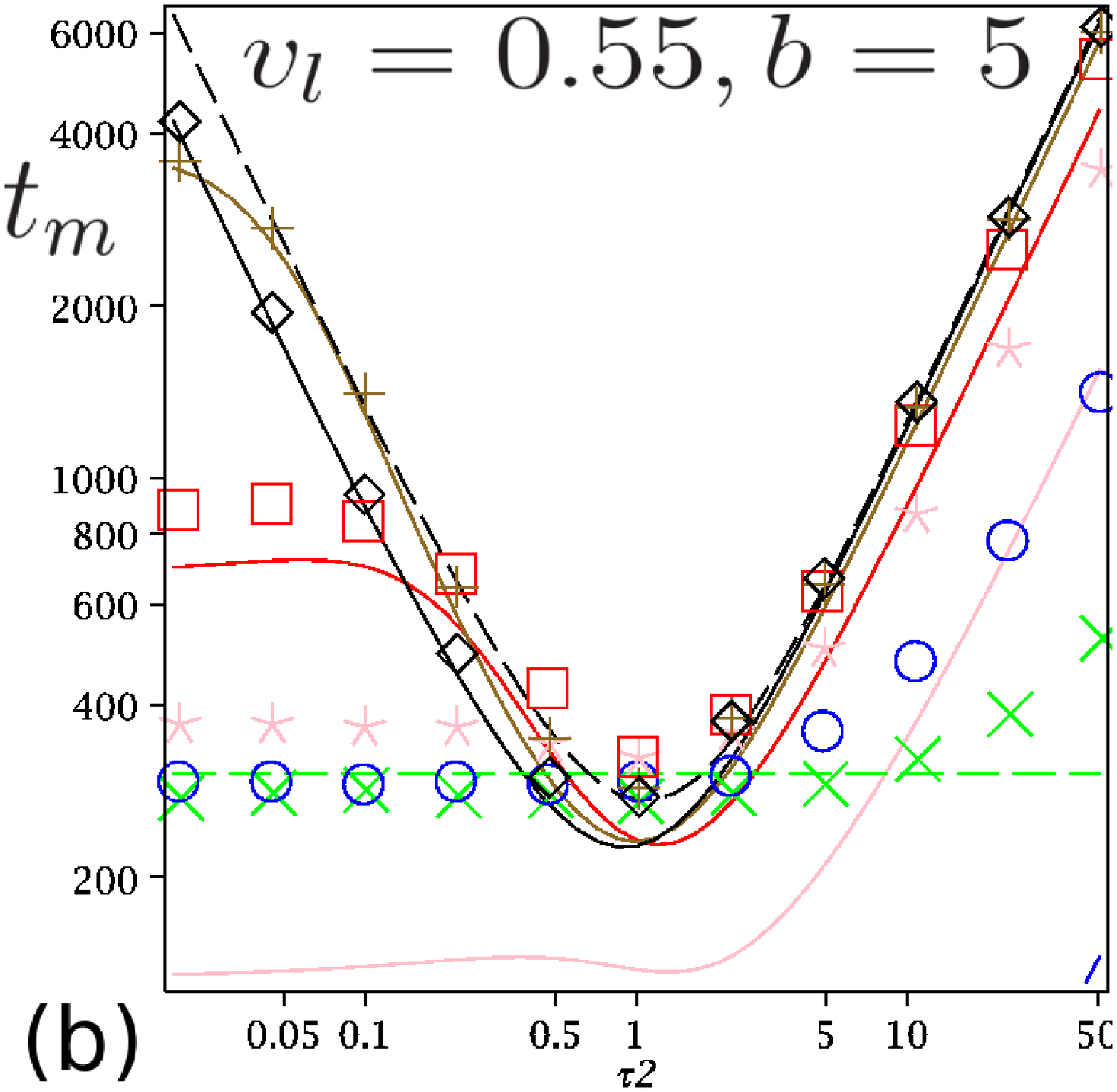}

      \includegraphics[width=4cm]{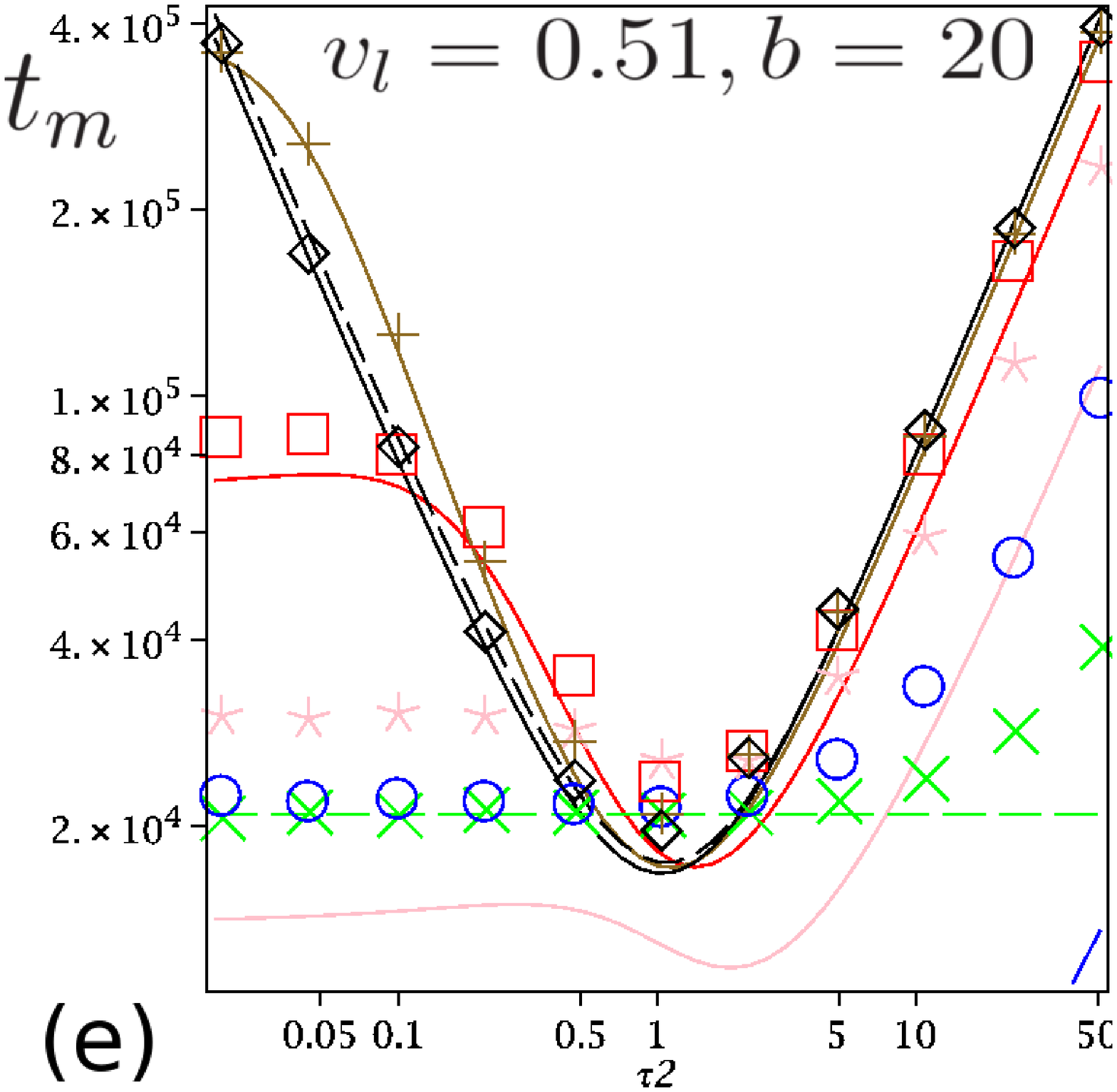}

\end{center}
   \end{minipage}\hfill
   \begin{minipage}[c]{.3\linewidth}
\begin{center}
      \includegraphics[width=4cm]{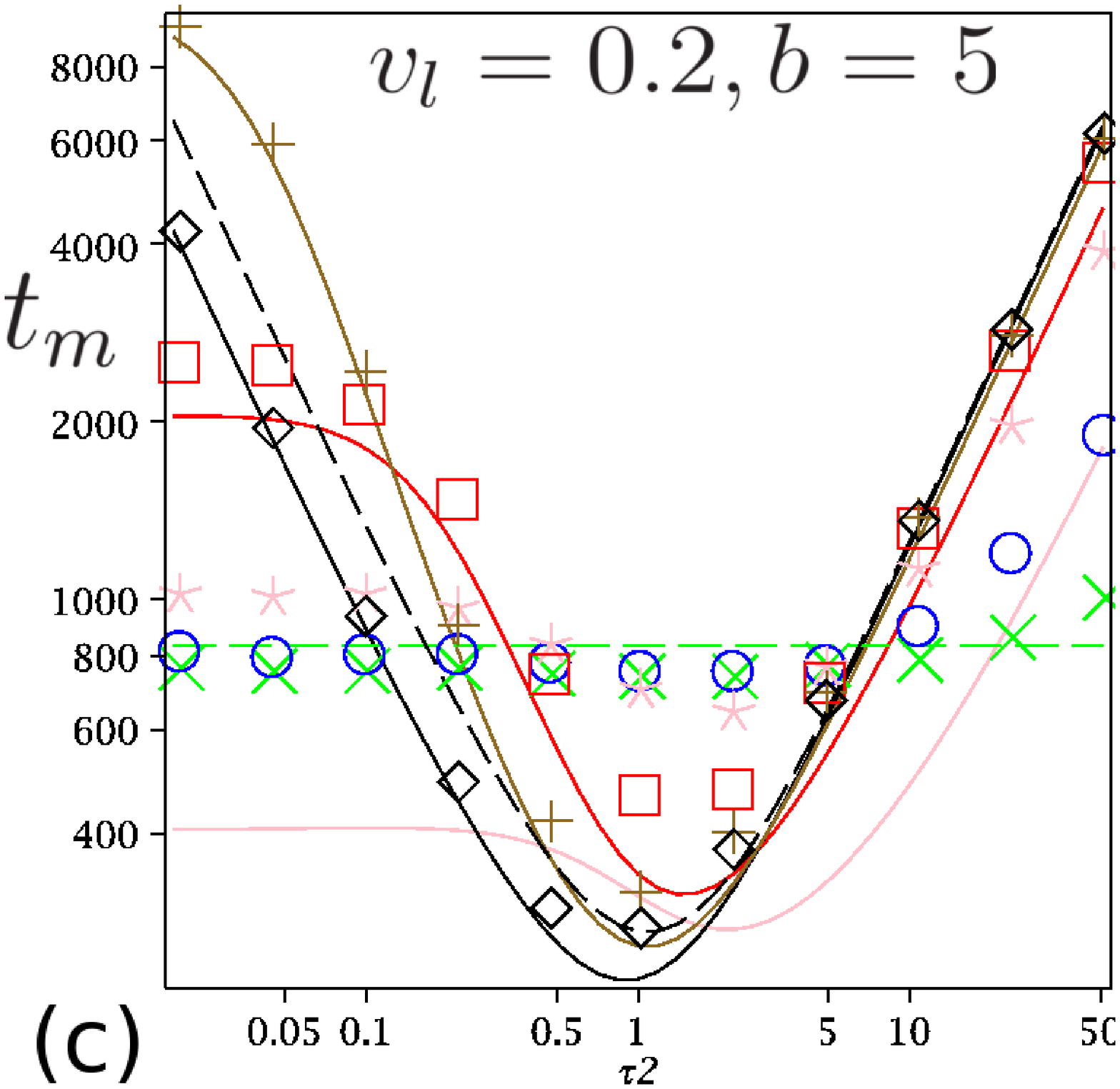} 

      \includegraphics[width=4cm]{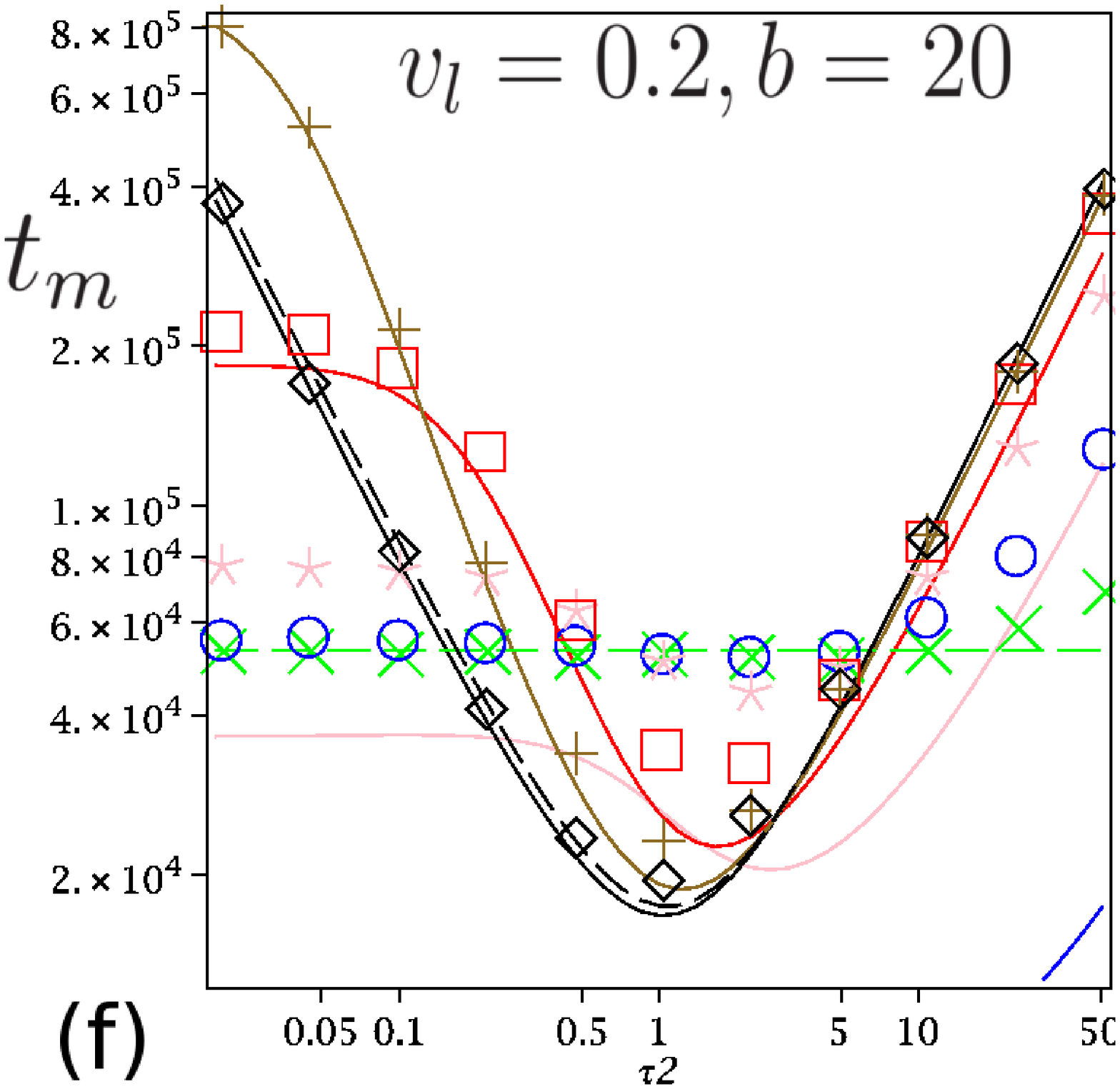}

\end{center}
   \end{minipage}
\caption{(Color online) Ballistic mode in 3 dimensions. $t_m$ as a function of $\tau_2$ in loglogscale. 
Simulations (symbols). Approximation $v_l \tau_1 \le a$ (\ref{tm3dvd} with \ref{3dvvdeff}) 
(colored lines), approximation $\tau_1=0$ \refm{pearson_complet} (black line), 
approximation $\tau_1 =0$ and  $b\gg a$ \refm{pearson} (dotted black line). 
Ballistic limit ($\tau_2 \to 0$ and $\tau_1 \to \infty$) (no intermittence) 
\refm{3dvvtmsansinter} (green dotted line). 
(a),(d)~: $v_l >v_l^c$, $\tau_{1,1}=0.04$, $\tau_{1,2}=0.2$, $\tau_{1,3}=1$, 
$\tau_{1,4}=5$, $\tau_{1,5}=25$.
(b),(e)~: $v_l \simeq v_l^c$, $\tau_{1,1}=0.08$, $\tau_{1,2}=0.4$, $\tau_{1,3}=2$, 
$\tau_{1,4}=10$, $\tau_{1,5}=50$.
(c),(f)~: $v_l <v_l^c$, $\tau_{1,1}=0.2$, $\tau_{1,2}=1$, $\tau_{1,3}=5$, 
$\tau_{1,4}=25$, $\tau_{1,5}=125$. 
 $V=1$, $a=1$. 
$\tau_1=0$ (black, diamond), $\tau_1=\tau_{1,1}$ (brown, +), $\tau_1=\tau_{1,2}$ (red, squares), 
$\tau_1=\tau_{1,3}$ (pink, stars), $\tau_1=\tau_{1,4}$ (blue, circles), $\tau_1=\tau_{1,5}$ (green, X) }\label{3dvvnouvelle}
\end{figure}

The numerical analysis puts forward  two strategies to minimize the search time, 
depending on a critical value $v_l^{c}$ to be determined (\refi{3dvvnouvelle})~:
\begin{itemize}
 \item when $v_l > v_l^{c}$, $\tau_1^{opt} \to \infty$ and $\tau_2^{opt} \to 0$. 
In this regime   intermittence is not favorable.
\item when $v_l < v_l^{c}$, $\tau_1^{opt} \to 0$,  and $\tau_2^{opt}$ finite. 
In this regime the optimal strategy is  intermittent.
\end{itemize}

\subsubsection{Regime without  intermittence (1 state ballistic searcher)~: $\tau_2 \to 0$}\label{r3dvv1}

Following the same argument as in dimension 2, without intermittence the best strategy is obtained in
the limit $\tau_1 \to \infty$ \refa{a3dvv1} in order to minimize oversampling of the search space. 
Following the derivation of \ref{topt2} (see appendix for details), 
it is found that the search time reads~: 
\begin{equation}\label{3dvvtmsansinter}
 t_{bal}=\frac{4 b^3}{3 a^2 v_l}.
\end{equation}

\subsubsection{Regime with  intermittence }

In the regime when intermittence is favorable, the numerical study suggests 
that the best strategy is realized for  $\tau_1 \to 0$ (\refi{3dvvnouvelle}).
In this  regime  $\tau_1 \to 0$, the phase 1 can be well approximated by a diffusion 
with effective diffusion coefficient $D_{eff}$ (see \refm{3dvvdeff}). 
We can then make use of  the analytical expression  $t_m$ derived in \refm{tm3dvd}.
We therefore  take  $\tau_1 = 0$  in the  expression of $t_m$  \refm{tm3dvd}, which yields~: 
\begin{equation}\label{pearson_complet}
t_m(\tau_1=0) \simeq \frac{u}{b^3aV}\left(\frac{\sqrt{3}}{5}\left( 5b^3 a^2 - 3 b^5 -2 a^5 \right)+ \frac{\left(b^3-a^3 \right)^2 u}{\sqrt{3}a(u-\tanh(u))}\right),
\end{equation}
where $u=\frac{\sqrt{3}a}{\tau_2 V}$.
In the limit  $b \gg a$, this expression can be further simplified (see  \refm{pearson}) to~: 
\begin{equation}
 t_m=\frac{b^3\sqrt{3}}{V^2\tau_2^2}\left(\frac{\sqrt{3}a}{V\tau_2}-\tanh\left(\frac{\sqrt{3}a}{V\tau_2}\right)\right)^{-1}, 
\end{equation}
and one finds straightforwardly that $\tau_2^{opt} = \frac{\sqrt{3}a}{Vx}$, 
where  $x$ is solution of $x\tanh(x)^2+2\tanh(x)-2x=0$, that is $x \simeq 1.606$.
Using this optimal  value of $\tau_2$ in the expression of $t_m$ \refm{pearson}, 
we finally get~:
\begin{equation}\label{tminter}
 t_m^{opt} = \frac{2}{\sqrt{3}}\frac{x}{\tanh(x)^2}\frac{b^3 }{a^2 V} \simeq 2.18 \frac{b^3 }{a^2 V}.
\end{equation}
These expressions show  a good agreement with numerical  simulations (\refi{3dvvnouvelle},\refi{InterApprox}).

\subsubsection{Discussion of the critical value $ v_l^{c}$}\label{r3dvv2}

\imagea{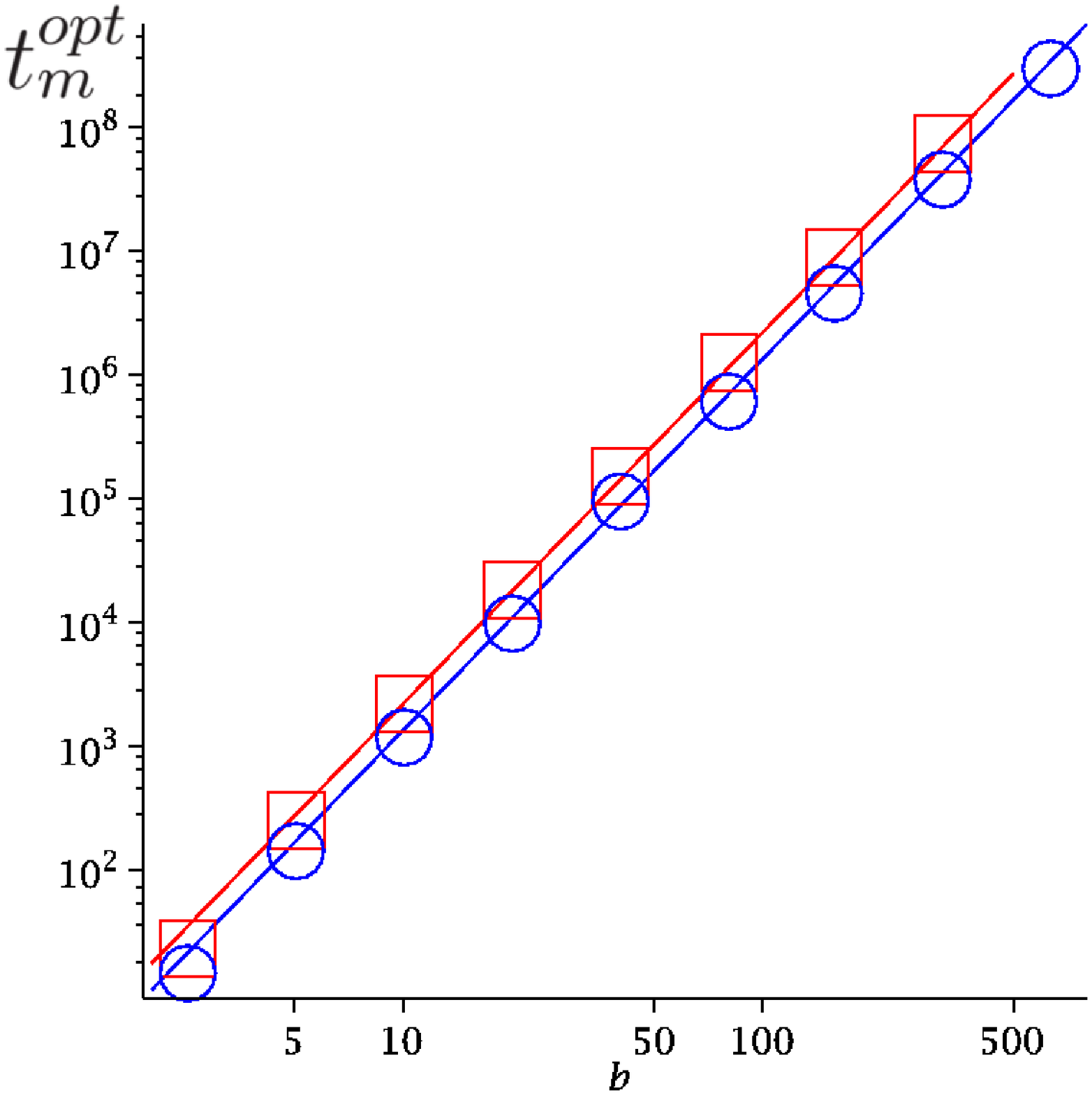}{Ballistic mode in 3 dimension. 
$t_m^{opt}$  as a function of $b$, logarithmic scale.
Regime without intermittence ($\tau_2=0$ and $\tau_1 \to \infty$, $v_l=1$), 
analytical approximation \refm{3dvvtmsansinter} (blue line), numerical simulations (blue circles).
Regime with intermittence (with  $\tau_1=0$, $V=1$), 
analytical approximation \refm{tminter} (red line), numerical simulations (red squares). $a=1$}{InterApprox}{6}

The gain is given by~: 
\begin{equation}
gain=\frac{t_{bal}}{t_m^{opt}} \simeq 0.61 \frac{V}{v_l}.
\end{equation}
As in dimension 2, it is trivial that $v_l^{c} < V$, and the critical value
$v_l^c$ can be defined as the value of $v_l$ such that  $gain=1$.  This yields
\begin{equation}
v_l^{c} \simeq 0.6 V.
\end{equation}
Importantly, $v_l^c$ neither depends on  $b$ nor $a$.
Simulations are in good agreement with this result \refa{a3dvv2}, 
except for a small numerical shift.

\subsubsection{Summary}

We studied the case where the detection phase 1 is modeled by a ballistic mode in  dimension 3. 
We have shown by numerical simulations that there are two possible optimal regimes, 
that we have then studied analytically~: 
\begin{itemize}
\item in the first regime $v_l > v_l^{c}$, the optimal strategy is a 1 state ballistic search ($\tau_1 \to \infty$, $\tau_2=0$) and  $t_m \simeq \frac{4 b^3}{3 a^2 v_l}$
\item in the second regime $v_l < v_l^{c}$, the optimal strategy is intermittent ($\tau_1=0$, $\tau_2 \simeq 1.1 \frac{a}{V}$), and $t_m \simeq 2.18 \frac{b^3 }{a^2 V}$ (in the limit $b \gg a$). 
\end{itemize}
The critical speed is obtained numerically as  $v_l^{c} \sim 0.5 V$ (analytical prediction~: $v_l^{c} \sim 0.6 V$).
It is noteworthy that when $b \gg a$, the values of $\tau_1$ and $\tau_2$ at the optimum, and the value of $v_l^c$ 
do not depend on the typical distance between targets $b$.

\subsection{Conclusion in  dimension 3}

We found that for the three possible modelings of 
the detection mode (static, diffusive and ballistic) in  dimension 3, 
there is a regime where the optimal strategy is  intermittent.  
Remarkably, and as was the case in dimension 1 and 2,  
the optimal time to spend in the fast non-reactive phase 2 is independent 
of the modeling of the detection mode and reads $\tau_2^{opt} \simeq 1.1 \frac{a}{V}$. 
Additionally, while the mean first passage time on the target scales as $b^3$, 
 the optimal values of the durations of the two phases do
not depend on the target density $a/b$.

\section{Discussion and conclusion}

The starting point of this paper was the observation that   intermittent trajectories are observed 
in various biological  examples of search behaviors, going from the microscopic scale, 
where searchers can be molecules looking for reactants, to  the macroscopic scale of foraging animals. 
We addressed the general question of determining whether such kind of intermittent trajectories could be 
favorable from a purely kinetic point of view, 
that is whether they could allow to minimize the search time for a target. % : 
% for example, at the microscopic level,
% vesicules within cells alternates 
% phases bound to ballistically moving motors with
% diffusive unbound phases; 
% at the macroscopic level, animal trajectories are often described as intermittent. 
On very general grounds, we proposed a minimal model of search strategy based on intermittence, 
where the searcher  
switches between  two phases, one slow where detection is possible, 
the other one  faster but preventing target detection. 
We studied  this minimal model in dimensions 1, 2 and 3, and under several modeling hypotheses. 
We believe that this systematic analysis can be used as a basis to study 
quantitatively various real search problems involving intermittent behaviors. 

More precisely, we calculated the mean first passage time at the target for  an intermittent searcher, 
and minimized this search time as a function of the mean duration of each of the two phases. 
The table \ref{recapgeneral} summarizes the results. 
In particular, this study shows that for certain ranges of the parameters which we determined, 
the optimal search strategy is intermittent. In other words,
there is an optimal way for the intermittent searcher to tune the mean time it spends  
in each of  the two phases. 
We found that the optimal durations of the two phases and the gain of intermittent search 
(as compared to 1 state search) 
do depend on the target density in dimension 1. 
In particular,  the gain can be very high at low target concentration. 
Interestingly, this dependence is smaller in
 dimension 2,  
and vanishes in dimension 3. 
The fact that intermittent search is more advantageous in low dimensions (1 and 2) 
can be understood as follows.  
At large scale, the intermittent searcher of our model performs effectively  a random walk, and therefore scans a space of dimension 2. 
In an environment of dimension 1 (and critically of dimension 2), 
the searcher therefore oversamples the space, and it is favorable to perform  large jumps 
to go to previously unexplored areas. On the contrary, 
 in dimension 3, 
the random walk is transient, and the searcher on average always scans previously unexplored areas, 
which makes large jumps less beneficial.

Additionally, our results show 
 that, for various modeling choices of the slow reactive phase, 
there is one and the same optimal duration of the fast non reactive phase, 
which depends only on the space  
 dimension. This further supports the robustness of optimal intermittent search strategies. 
Such robustness and efficiency -- and optimality -- could explain why intermittent 
trajectories are observed so often, and in various forms.

\begin{turnpage} 
\begin{table} 
\begin{center}
\scriptsize
\begin{tabular}{|c |c |c |c |c | c | c | }

\hline 
&Static mode&\multicolumn{3}{|c|}{Diffusive mode} &\multicolumn{2}{|c|}{Ballistic mode}\\
\hline
&always intermittence& $b < \frac{D}{V}$ & $b>\frac{D}{V}$,  $a \ll \sqrt{\frac{b}{a}} \frac{D}{V}$ & $b>\frac{D}{V}$, $a \gg \sqrt{\frac{b}{a}} \frac{D}{V}$  & $v_l>v_l^c$& $v_l<v_l^c\simeq\frac{V}{2}\sqrt{\frac{3a}{b}}$\\
1D& $\tau_1^{opt}=\sqrt{\frac{\tau_2^{opt}}{2k}}\simeq\sqrt{\frac{a}{Vk}}\left(\frac{b}{12a}\right)^{\frac{1}{4}}$ & $\tau_1^{opt} \to \infty$ &$\tau_1^{opt}\simeq\left(\frac{b^2 D}{36 V^4} \right)^{\frac{1}{3}}$ & $\tau_1^{opt}\simeq\frac{Db}{48 V^2 a}$ & $\tau_1^{opt}\to \infty$ & $\tau_1^{opt}\to 0$ \\
&\cellcolor{red}{$\tau_2^{opt}\simeq\frac{a}{V}\sqrt{\frac{b}{3a}}$} & $\tau_2^{opt}\to 0$  & $\tau_2^{opt}\simeq\left(\frac{2b^2D}{9V^4} \right)^{\frac{1}{3}}$ &\cellcolor{red}{$\tau_2^{opt}\simeq\frac{a}{V}\sqrt{\frac{b}{3a}}$}& $\tau_2^{opt} \to 0$&\cellcolor{red}{$\tau_2^{opt}\simeq\frac{a}{V}\sqrt{\frac{b}{3a}}$}\\
&$t_m^{opt}\simeq\frac{b}{ak}\sqrt{\frac{b}{3a}}\left(\sqrt{\frac{2ka}{V}}+\left(\frac{3a}{b} \right)^{\frac{1}{4}}\right)^2$ & $t_m^{opt}\simeq\frac{b^2}{3D}$ & $t_m^{opt} \simeq \left(\frac{3^5 b^4}{2^4 D V^2}  \right)^{\frac{1}{3}}$& $t_m^{opt}\simeq\frac{2b}{\sqrt{3}V}\sqrt{\frac{b}{a}}$ & $t_m^{opt} \simeq \frac{b}{v_l}$  & $t_m^{opt}\simeq\frac{2b}{V}\sqrt{\frac{b}{3a}}$\\
\hline
&always intermittence&$b < \frac{D}{V}$ & $b \gg \frac{D}{V} \gg a$ &   $b \gg a \gg \frac{D}{V} $& $v_l>v_l^c$& $v_l<v_l^c \simeq \frac{\pi V}{4}\left(\ln\left(\frac{b}{a}\right) \right)^{-\frac{1}{2}}$\\
2D& $\tau_1^{opt}=\sqrt{\frac{\tau_2^{opt}}{2k}}\simeq\sqrt{\frac{a}{2Vk}}\left(\ln\left(\frac{b}{a}\right)-\frac{1}{2} \right)^{\frac{1}{4}}$ & $\tau_1^{opt} \to \infty$ &$\tau_1^{opt}\simeq \frac{b^2}{D}\frac{4\ln w-5+c}{w^2(4\ln w -7+c)}$& $\tau_1^{opt}\simeq\frac{D}{2V^2}\frac{\left( \ln\left( \frac{b}{a}\right) \right)^2}{2\ln\left( \frac{b}{a}\right) -1 }$ & $\tau_1^{opt}\to \infty$ & $\tau_1^{opt}\to 0$\\
&\cellcolor{red}{$\tau_2^{opt}\simeq\frac{a}{V}\sqrt{\ln\left(\frac{b}{a}\right)-\frac{1}{2}}$} & $\tau_2^{opt}\to 0$  & $\tau_2^{opt}\simeq\frac{b}{V}\frac{\sqrt{4\ln w -5+c}}{w}$ &\cellcolor{red}{$\tau_2^{opt}\simeq\frac{a}{V}\sqrt{\ln\left(\frac{b}{a}\right)-\frac{1}{2}}$}& $\tau_2^{opt} \to 0$&\cellcolor{red}{$\tau_2^{opt}\simeq\frac{a}{V}\sqrt{\ln\left(\frac{b}{a}\right)-\frac{1}{2}}$}\\
&$t_m^{opt}\simeq \frac{b^2}{aV}\left(\sqrt{2}\left(\ln\left(\frac{b}{a}\right) \right)^{\frac{1}{4}}+\sqrt{\frac{V}{ak}}\right)^2$ & $t_m^{opt}\simeq\frac{b^2}{2D} \ln\left(\frac{b}{a}\right)$ &For $t_m^{opt}$, $c$ and $w$, see \ref{r2dvd}& $t_m^{opt}\simeq \frac{2b^2}{aV}\sqrt{\ln\left(\frac{b}{a}\right)}$ & $t_m^{opt} \simeq \frac{\pi b^2}{2a v_l}$  & $t_m^{opt}\simeq \frac{2b^2}{aV} \sqrt{\ln\left(\frac{b}{a}\right)}$\\
\hline
&always intermittence& $a\lesssim 6\frac{D }{V}$ &  & $b \gg a \gtrsim  6\frac{D }{V}$ & $v_l>v_l^c$& $v_l<v_l^c\simeq 0.6 V$\\
3D& $\tau_1^{opt}=\sqrt{\frac{\tau_2^{opt}}{2k}}\simeq \left( \frac{3}{10} \right)^{\frac{1}{4}} \sqrt{\frac{a}{Vk}}$ & $\tau_1^{opt} \to \infty$ & & $\tau_1^{opt}\simeq \frac{6 D}{V^2}$ & $\tau_1^{opt}\to \infty$ & $\tau_1^{opt}\to 0$\\
&\cellcolor{red}{$\tau_2^{opt}\simeq 1.1 \frac{a}{V}$}& $\tau_2^{opt}\to 0$  & &\cellcolor{red}{$\tau_2^{opt}\simeq  1.1 \frac{a}{V}$} &$\tau_2^{opt} \to 0$&\cellcolor{red}{$\tau_2^{opt}\simeq 1.1 \frac{a}{V}$}\\
&$t_m^{opt}\simeq \frac{b^3}{\sqrt{5}ka^3} \left(\sqrt{\frac{ak}{V}}24^{\frac{1}{4}}+5^{\frac{1}{4}}\right)^2$ & $t_m^{opt}\simeq \frac{b^3}{3Da}$ & & $t_m^{opt}\simeq 2.18 \frac{b^3}{Va^2}$ & $t_m^{opt} \simeq \frac{4b^3}{3a^2v_l}$  & $t_m^{opt}\simeq 2.18 \frac{b^3}{Va^2}$\\
\hline

\end{tabular}
\normalsize
\end{center}
\caption{Recapitulation of main results~: strategies minimizing the mean first passage time on the target. In each cell, validity of the regime, optimal $\tau_1$, optimal $\tau_2$, minimal $t_m$ ($t_m$ with $\tau_i=\tau_i^{opt}$). Red highlight the value of $\tau_2^{opt}$ independent from the description of the slow detection phase 1. Results are given in the limit $b \gg a$.}  \label{recapgeneral}
\end{table} 
\end{turnpage}

{\bf Acknowledgement}\\
Financial support from ANR grant Dyoptri is acknowledged.

\addcontentsline{toc}{section}{References}
%\bibliographystyle{unsrt}
%\bibliography{biblio}

\section{Appendix}

\subsection{Diffusive mode in dimension 1}

\subsubsection{Exact results \refs{r1dvd1}}\label{a1dvd1}

\begin{equation}\label{tm1DvD}
t_m=\frac{1}{3}\,\frac{ (\tau_1 + \tau_2)}{\beta^2 b } \frac{Num}{Den}
\end{equation}
With~:
\begin{equation}
 Num= \alpha_1+\alpha_2+\alpha_3+\alpha_4+\alpha_5+\alpha_6+\alpha_7
\end{equation}
\begin{equation}
 Den= \gamma_1+\gamma_2+\gamma_3+\gamma_4
\end{equation}
\begin{equation}
 \alpha_1={L_2}^{3}\left( \left( 3 {L_2}^{2} \left(L_1^2-L_2^2 \right) +2{h}^{2}\beta \right) h\sqrt {\beta} S +3L_1L_2 \left({L_2}^{4}- 2{h}^{2}\beta\right) C  \right)
\end{equation}
\begin{equation}
 \alpha_2=-L_1h L_2^{5} \left( 2\beta+3L_2^2 \right)  R C 
\end{equation}
\begin{equation}
 \alpha_3= L_1 \left(2{h}^{4}{\beta}^{2}- 3L_2^8 \right) B C
\end{equation}
\begin{equation}
 \alpha_4={h}^{2}\sqrt {\beta}\left( 6L_2^6+{h}^{2}\beta\left( \beta+L_1^2 \right)  \right) R S 
\end{equation}
\begin{equation}
 \alpha_5=\sqrt {\beta}hL_2^3 \left( 4{h}^{2}\beta+3L_2^2 \left(L_2^2 -L_1^2\right)  \right) B S
\end{equation}
\begin{equation}
 \alpha_6=  L_1 L_2^3 \left(3\left( 2h^2L_2\beta+L_2^5 \right) B +h \left( 3 L_2^2 \left( \beta+ L_2^2 \right) +2{h}^{2}\beta \right)R  \right) 
\end{equation}
\begin{equation}
 \alpha_7=-  L_1 \left( 3 L_2^8+2{h}^{4}{\beta}^{2} \right)
\end{equation}
\begin{equation}
 \gamma_1=L_2^3 L_1 R  \left( C -1 \right)
\end{equation}
\begin{equation}
 \gamma_2=\sqrt{\beta}h \left(2L_1^2+L_2^2\right) R S
\end{equation}
\begin{equation}
 \gamma_3=\sqrt{\beta} L_2^3 \left( B -1 \right) S 
\end{equation}
\begin{equation}
 \gamma_4=2h\beta L_1 \left( B C -1 \right)
\end{equation}
\begin{equation}
 B= \cosh \left({\frac {2 a}{L_2}} \right)
\end{equation}
\begin{equation}
 C=\cosh \left(2h\sqrt{L_1^{-2}+L_2^{-2}}  \right)
\end{equation}
\begin{equation}
 R= \sinh \left( {\frac {2 a}{L_2}} \right)
\end{equation}
\begin{equation}
 S=\sinh \left( 2h\sqrt{L_1^{-2}+L_2^{-2}} \right)
\end{equation}
\begin{equation}\label{1dvdbeta}
 \beta=L_1^2+L_2^2
\end{equation}
\begin{equation}
 L_1=\sqrt{D\tau_1}
\end{equation}
\begin{equation}
 L_2=V\tau_2
\end{equation}
\begin{equation}
 h=b-a
\end{equation}

\subsubsection{Numerical study \refs{r1dvd1}}\label{a1dvd2}

\begin{table}[h!]
\begin{center}
\tiny
 \begin{tabular}{|c c|c|c|c|c|c|c|c|}
\hline
\multicolumn{2}{|c|}{$b/a$} &$100$&$10^3$&$10^4$&$10^5$&$10^6$&$10^7$\\
\hline
      &$gain^{th,1}$&$0.085$&$0.39$&$1.8$&$8.5$&$39$&$180$\\
\cline{2-8}
      &$gain$&\cellcolor{red}{$1$}&\cellcolor{red}{$1$}&\cellcolor{green}{$2.1$}&\cellcolor{green}{$8.7$}&\cellcolor{green}{$40$}&\cellcolor{green}{$180$}\\
\cline{2-8}
      &$gain^{th,2}$&$0.014$&$0.046$&$0.14$&$0.46$&$1.4$&$4.6$\\
\cline{2-8}
 &$\tau_1^{th,1}$&$0.19$&$0.89$&$4.1$&$19$&$89$&$410$\\
\cline{2-8}
a= &$\tau_1^{opt}$&\cellcolor{red}{$\infty$}&\cellcolor{red}{$\infty$}&\cellcolor{green}{$6.1$}&\cellcolor{green}{$21$}&\cellcolor{green}{$90$}&\cellcolor{green}{$410$}\\
\cline{2-8}
0.005&$\tau_1^{th,2}$&$2.1$&$21$&$210$&$2100$&$21000$&$2.1.10^5$\\
\cline{2-8}
     &$\tau_2^{th,1}$&$0.38$&$1.8$&$8.2$&$38$&$180$&$820$\\
\cline{2-8}
&$\tau_2^{opt}$&\cellcolor{red}{$0$}&\cellcolor{red}{$0$}&\cellcolor{green}{$8.4$}&\cellcolor{green}{$38$}&\cellcolor{green}{$180$}&\cellcolor{green}{$820$}\\
\cline{2-8}
     &$\tau_2^{th,2}$&$0.029$&$0.091$&$0.29$&$0.91$&$2.9$&$9.1$\\
\hline
      &$gain^{th,1}$&$1.8$&$8.5$&$39$&$180$&$850$&$4000$\\
\cline{2-8}
     &$gain$&\cellcolor{green}{$2.4$}&\cellcolor{green}{$9.4$}&\cellcolor{green}{$41$}&\cellcolor{green}{$190$}&\cellcolor{green}{$850$}&\cellcolor{green}{$4000$}\\
\cline{2-8}
      &$gain^{th,2}$&$1.4$&$4.6$&$14$&$46$&$140$&$460$\\
\cline{2-8}
&$\tau_1^{th,1}$&$4.1$&$19$&$89$&$410$&$1900$&$8900$\\
\cline{2-8}
a=&$\tau_1^{opt}$&\cellcolor{green}{$3.5$}&\cellcolor{green}{$15$}&\cellcolor{green}{$78$}&\cellcolor{green}{$390$}&\cellcolor{green}{$1900$}&\cellcolor{green}{$8800$}\\
\cline{2-8}
0.5&$\tau_1^{th,2}$&$2.1$&$21$&$210$&$2100$&$21000$&$2.1.10^5$\\
\cline{2-8}
&$\tau_2^{th,1}$&$8.2$&$38$&$180$&$820$&$3800$&$18000$\\
\cline{2-8}
 &$\tau_2^{opt}$&\cellcolor{green}{$7.6$}&\cellcolor{green}{$36$}&\cellcolor{green}{$170$}&\cellcolor{green}{$810$}&\cellcolor{green}{$3800$}&\cellcolor{green}{$18000$}\\
\cline{2-8}
    &$\tau_2^{th,2}$&$2.9$&$9.1$&$29$&$91$&$290$&$910$\\
\hline
      &$gain^{th,1}$&$39$&$180$&$850$&$3900$&$18000$&$85000$\\
\cline{2-8}
     &$gain$&\cellcolor{blue}{$150$}&\cellcolor{blue}{$470$}&$1500$&$5500$&$21000$&\cellcolor{green}{$91000$}\\
\cline{2-8}
      &$gain^{th,2}$&$140$&$460$&$1400$&$4600$&$14000$&$46000$\\
\cline{2-8}
&$\tau_1^{th,1}$&$89$&$410$&$1900$&$8900$&$41000$&$1.9.10^5$\\
\cline{2-8}
a=&$\tau_1^{opt}$&\cellcolor{blue}{$2.2$}&\cellcolor{blue}{$22$}&$230$&$2500$&$21000$&\cellcolor{green}{$1.4.10^5$}\\
\cline{2-8}
50&$\tau_1^{th,2}$&$2.1$&$21$&$210$&$2100$&$21000$&$2.1.10^5$\\
\cline{2-8}
&$\tau_2^{th,1}$&$180$&$820$&$3800$&$18000$&$82000$&$3.8.10^5$\\
\cline{2-8}
 &$\tau_2^{opt}$&\cellcolor{blue}{$290$}&\cellcolor{blue}{$980$}&$3500$&$15000$&$72000$&\cellcolor{green}{$3.6.10^5$}\\
\cline{2-8}
    &$\tau_2^{th,2}$&$290$&$910$&$2900$&$9100$&$29000$&$91000$\\
\hline
      &$gain^{th,1}$&$850$&$3900$&$18000$&$85000$&$3.9.10^5$&$1.8.10^6$\\
\cline{2-8}
     &$gain$&\cellcolor{blue}{$15000$}&\cellcolor{blue}{$46000$}&\cellcolor{blue}{$1.4.10^5$}&\cellcolor{blue}{$4.6.10^{5}$}&\cellcolor{blue}{$1.5.10^{5}$}&\cellcolor{blue}{$4.7.10^{6}$}\\
\cline{2-8}
      &$gain^{th,2}$&$14000$&$46000$&$1.4.10^5$&$4.6.10^5$&$1.4.10^6$&$4.6.10^6$\\
\cline{2-8}
&$\tau_1^{th,1}$&$1900$&$8900$&$41000$&$1.9.10^5$&$8.9.10^5$&$4.1.10^6$\\
\cline{2-8}
a=&$\tau_1^{opt}$&\cellcolor{blue}{$2.2$}&\cellcolor{blue}{$21$}&\cellcolor{blue}{$210$}&\cellcolor{blue}{$2100$}&\cellcolor{blue}{$21000$}&\cellcolor{blue}{$2.2.10^5$}\\
\cline{2-8}
5000&$\tau_1^{th,2}$&$2.1$&$21$&$210$&$2100$&$21000$&$2.1.10^5$\\
\cline{2-8}
&$\tau_2^{th,1}$&$3800$&$1800$&$82000$&$3.8.10^5$&$1.8.10^6$&$8.2.10^6$\\
\cline{2-8}
&$\tau_2^{opt}$&\cellcolor{blue}{$29000$}&\cellcolor{blue}{$91000$}&\cellcolor{blue}{$2.9.10^{5}$}&\cellcolor{blue}{$9.2.10^{5}$}&\cellcolor{blue}{$2.9.10^{6}$}&\cellcolor{blue}{$9.8.10^{6}$}\\
\cline{2-8}
     &$\tau_2^{th,2}$&$29000$&$91000$&$2.9.10^{5}$&$9.1.10^{5}$&$2.9.10^{6}$&$9.1.10^{6}$\\
\hline

\end{tabular}
\normalsize
\end{center}
\caption{Diffusive mode in 1 dimension. Optimization of $t_m$ as a function of $\tau_1$ and $\tau_2$ for different sets of parameters ($D=1$, $V=1$).
For each $(a,b)$, numerical values for the exact analytical function \refm{tm1DvD} are given 
 with the values expected in the regimes where intermittence is favorable, 
either with  $\frac{bD^2}{a^3 V^2} \gg 1$ ($th,1$), or with $\frac{bD^2}{a^3 V^2} \ll 1$ ($th,2$).
 $gain^{th,1}$ \refm{gainn}, $gain=t_m^{opt}/t_{diff}$, $gain^{2,th}$ \refm{gain1dvd2}. 
 $\tau_1^{th,1}$  \refm{l11dvd1}, $\tau_1^{opt}$,   $\tau_1^{th,2}$ \refm{l11dvd2}. 
 $\tau_2^{th,1}$  \refm{l21dvd1}, $\tau_2^{opt}$,   $\tau_2^{th,2}$ \refm{l21dvd2}. 
Colors indicate the regime~: red when intermittence is not favorable, 
green in the  $\frac{bD^2}{a^3 V^2} \gg 1$ regime, 
blue in the $\frac{bD^2}{a^3 V^2} \ll 1$ regime. }\label{tabnum}
\end{table}

We studied numerically the optimum of the exact $t_m$ expression \refm{tm1DvD} (\reft{tabnum}). 
We could distinguish 3 regimes~: one with no intermittence, and two with favorable intermittence, 
but with different scalings. Intermittence is favorable 
when $b >\frac{D}{V}$.
The demarcation line between the two intermittent regimes is
$ \frac{bD^2}{a^3 V^2}=1$.

\subsubsection{Details of the optimization of the regime where intermittence is favorable, with  $ \frac{bD^2}{a^3 V^2} \gg 1$ \refs{r1dvd3}}\label{a1dvd3}

We suppose that target density is low~: $\frac{a}{b} \ll 1$.

We are interested in the regime where intermittence is favorable. 
We have both  $2(b-a)\sqrt{L_1^{-2}+L_2^{-2}}>2\frac{b-a}{L_1}$ and  $2(b-a)\sqrt{L_1^{-2}+L_2^{-2}}>2\frac{b-a}{L_2}$.
In a regime of intermittence, one diffusion phase does not explore a significant part of the system~: $b/L_1\gg 1$. 
Alternatively, having a ballistic phase of the size of the system is a waste of time, thus close to the optimum $b/L_2 \gg 1$. 
Consequently $2(b-a)\sqrt{L_1^{-2}+L_2^{-2}}\gg 1$. 

We use the numerical results (\reft{tabnum}) to make assumptions on the dependence
of $\tau_1^{opt}$ and $\tau_2^{opt}$  with the parameters. We define $k_1$ and $k_2$~:
\begin{equation}
 \tau_1= (k_1)^{-1} \left( \frac{b^2 D}{V^4} \right)^{\frac{1}{3}}
\end{equation}
\begin{equation}
 \tau_2= (k_2)^{-1} \left( \frac{b^2 D}{V^4} \right)^{\frac{1}{3}}
\end{equation}
We make a development of $t_m$ for $b\gg a$. We suppose $k_1$ and $k_2$ 
do not depend on $b/a$~:
\begin{equation}\label{tmnn}
t_m = \frac{1}{3}\frac{D}{V^2}\left(\frac{bV}{D} \right)^{\frac{4}{3}} \frac{k_1+k_2}{k_1 k_2}\left(k_2^2+3\sqrt{k_1} \right)
\end{equation}

We checked that this expression gives a good approximation of $t_m$ 
in this regime, in particular around the optimum (\refi{tcasn}, in section \ref{r1dvd3}).

Derivatives of \refm{tmn} as a function of $k_1$ and $k_2$ must be equal to 0 at the optimum. It leads to~: 
%To simplify the obtained system, we need the assumption $ \frac{bD^2}{a^3 V^2} \gg 1$, and we get~:
\begin{equation}
 -3k_1^{\frac{3}{2}}+3k_2^3+3\sqrt{k_1}k_2 =0
\end{equation}
\begin{equation}
 3k_1^{\frac{3}{2}}-2k_2^3-k_1k_2^2 =0
\end{equation}
On four pairs of solutions, only one  is strictly positive~:
\begin{equation}
 \tau_1^{opt}=\frac{1}{2} \sqrt[3]{\frac{2b^2D}{9V^4}}
\end{equation}
\begin{equation}
 \tau_2^{opt}=\sqrt[3]{\frac{2b^2D}{9V^4}}
\end{equation}

\subsubsection{Details of the optimization of the universal intermittent regime $ \frac{bD^2}{a^3 V^2} \ll 1$ \refs{r1dvd4}}\label{a1dvd4}

We start from the exact expression of  $t_m$ \refm{tm1DvD}.
We have to make assumption on the dependency of $\tau_2^{opt}$ with $b$ and $a$.
We define $f$ by  $\tau_2=\frac{1}{f} \frac{a}{V}\sqrt{\frac{b}{3a}}$, and we suppose that 
$f$ is independent from $a/b$. 
We make a development of $a/b \to 0$.
The first two terms give~:
\begin{equation}\label{tsimp1DvDa}
 t_m \simeq \frac{b}{a}\left(\sqrt{\frac{ab}{3}}\frac{1}{Vf}+\tau_1 \right)\frac{a+af^2+\sqrt{D\tau_1}f^2}{a+\sqrt{D\tau_1}}
\end{equation}
This expression gives a very good approximation of $t_m$ 
in the $bD^2/(a^3 v^2) \ll 1$ regime, especially close to the optimum (\refi{tsimpfig}, in section \ref{r1dvd4}).

We then minimize $t_m$ \refm{tsimp1DvDa} as a function of $f$ and $\tau_1$. 
We introduce $w$ defined as~:
\begin{equation}
 w=\frac{a V }{D}\sqrt{\frac{a}{b}}
\end{equation}
We make an assumption on the dependency of $\tau_1^{opt}$ with $a/b$, 
inferred via the numerical results~:
\begin{equation}
 s=\frac{1}{\tau_1} \frac{D}{V^2}\frac{b}{a}
\end{equation}
We write the equation \refm{tsimp1DvDa} with these quantities.
Its derivatives with $f$ and $s$ should equal zero at the optimum.
It leads to~:
\begin{equation}\label{eq1}
 -\sqrt {3}{s}^{3/2}{w}^{2}+\sqrt {3}{s}^{3/2}{w}^{2}{f}^{2}+\sqrt {3}s
\,w{f}^{2}+6\,\sqrt {s}w{f}^{3}+6\,{f}^{3}=0
\end{equation}
\begin{equation}\label{eq2}
 6\,{s}^{3/2}{w}^{2}{f}^{3}+6\,{s}^{3/2}f\,{w}^{2}-{w}^{2}{s}^{2}\sqrt 
{3}+3\,ws\,f+12\,ws\,{f}^{3}+6\,\sqrt {s}{f}^{3}=0
\end{equation}

We take the equation \refm{eq1} and here we need to make the assumption than~:  $a\ll b \ll \frac{a^3 V^2}{D^2}$.
We get~:
\begin{equation}
 \sqrt{3}{s}^{3/2}{w}^{2}(f^2-1)=0
\end{equation}
Consequently $f=1$.
We incorporate this result to equation \refm{eq2}:
\begin{equation}
 12\,{s}^{3/2}{w}^{2}-{w}^{2}{s}^{2}\sqrt 
{3}+15ws+6\,\sqrt {s}=0
\end{equation}
The relevant solution is~:
\begin{equation}\label{1dvds}
 s_{sol}=\left( \frac{1}{3}\,{\frac {\sqrt [3]{u}}{w}}+
\frac {5\,\sqrt {3}+16\,w}{\sqrt [3]{u}
}+\frac{4}{\sqrt{3}} \right) ^{2}
\end{equation}
With~:
\begin{equation}
u= \left( 270\,w+27\,\sqrt {3}+192\,{w}^{
2}\sqrt {3}+9\,\sqrt {55\,\sqrt {3}w+84\,{w}^{2}+27} \right) w
 \end{equation}
When $w \to \infty$, $s_{sol}=48$. 
As we made the assumption $\frac{bD^2}{a^3V^2}=w^{-2}\ll 1$, 
difference from the asymptote will be small (\refi{w}).

\imagea{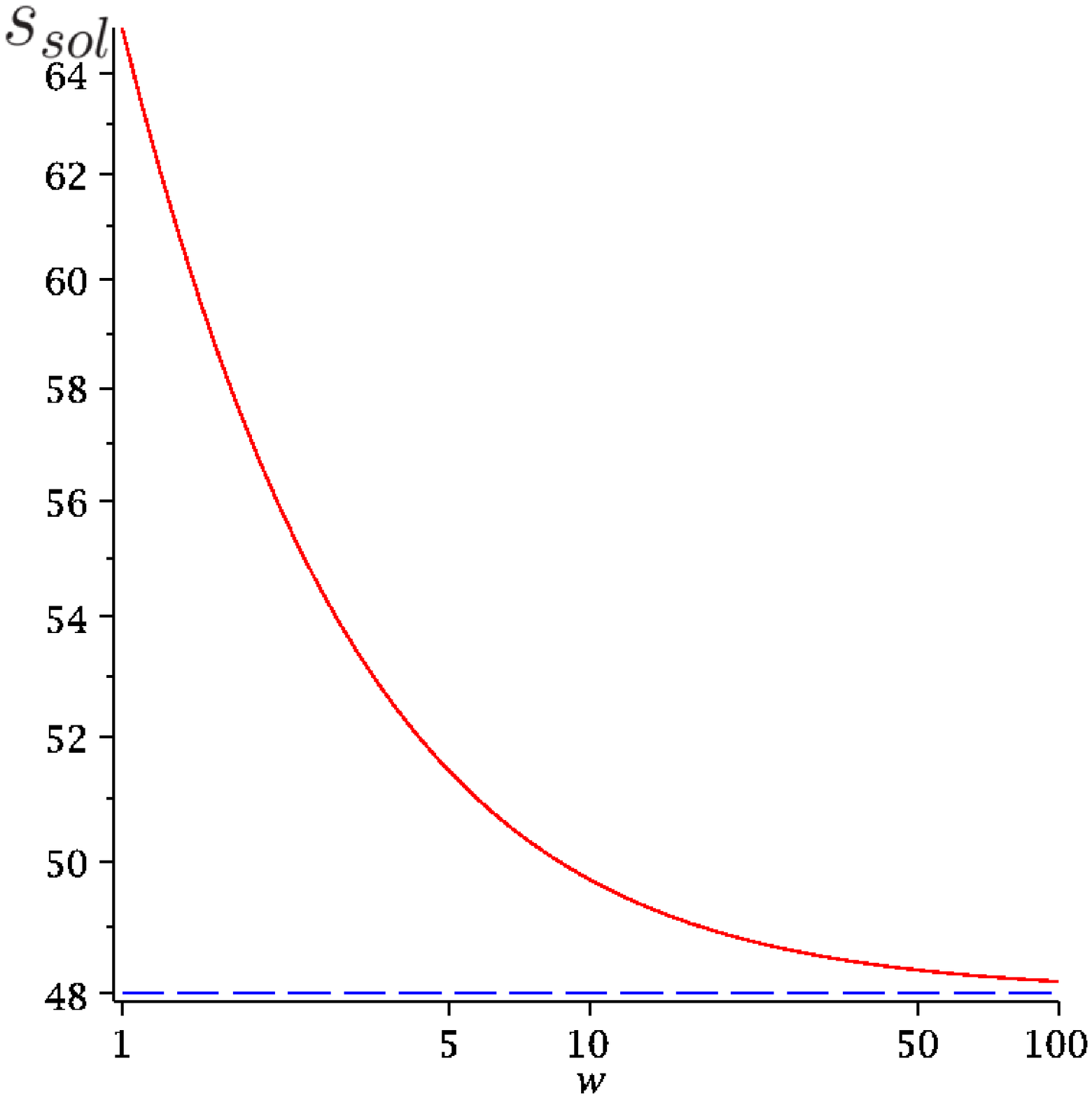}
{Diffusive mode in 1 dimension. 
$s_{sol}$ \refm{1dvds} (red line) as a function of $\ln(w)$, 
with the asymptote (blue dotted line)}{w}{6}

It leads to~:
\begin{equation}
 \tau_2^{opt}=\frac{1}{f^{opt}}  \frac{a}{V} \sqrt{\frac{b}{3a}} =  \frac{a}{V} \sqrt{\frac{b}{3a}}
\end{equation}
\begin{equation}
 \tau_1^{opt}=\frac{D b}{s^{opt}V^2 a} =\frac{D b}{48V^2 a}
\end{equation}
It corresponds to numerical results (\reft{tabnum}).

We use equations \refm{tdiff1D}, \refm{tsimp1DvD} to calculate the gain.:
\begin{equation}
  t_{m}^{opt}\simeq  \frac{2a}{V\sqrt{3}}\left(\frac{b}{a} \right)^{3/2}
\end{equation}
We get~:
\begin{equation}
 gain\simeq  \frac{1}{2\sqrt{3}}\frac{aV}{D}\sqrt{\frac{b}{a}}
\end{equation}
It is in very good agreement with numerical data (\reft{tabnum}).
Gain can be very large if density is low. 

\subsection{Ballistic mode in one dimension~: exact result \refs{r1dvv1}}\label{a1dvv1}

\begin{equation}\label{tm1v}
t_m = \frac{\tau_1+\tau_2}{b} (\gamma_1+\gamma_2+\gamma_3)
\end{equation}
\begin{equation}
 \gamma_1= {\frac {{h}^{2}  \left( h+3L_1 \right) }{3\alpha}} 
\end{equation}
\begin{equation}
\gamma_2=\frac{L_2\left( h +L_1\right)}{{\alpha}^{3/2}den} \left(g_4(-1){e^{-{\frac { 2 a }{L_2}}}}-g_4(1)\right) \left( g_3(-1) {e^{2\,{\frac {\sqrt {\alpha} a }{L_1 L_2}}}}+ g_3(1){e^{2\,{\frac {\sqrt {\alpha} b }{L_1L_2}}}}  \right)
\end{equation}
\begin{equation}
 den= g_1(1){e^{-2\,{\frac { a \, \left( L_1-\sqrt {\alpha} \right) }{L_1 L_2}}}}+g_1(-1){e^{2\,{\frac {\sqrt {\alpha} b }{L_1 L_2}}}}+g_2(1){e^{-2{\frac {a L_1-\sqrt {\alpha} b }{L_1 L_2}}}}+g_2(-1){e^{2\,{\frac {\sqrt {\alpha} a }{L_1 L_2}}}}
\end{equation}
\begin{equation}
\gamma_3= - \frac{L_2}{4   {\alpha}^{3/2}}   \frac{num_1 num_2}{den_1 den_2}
\end{equation}
\begin{equation}
 num_1 = f_1(1)+f_1(-1){e^{-{\frac {2 a }{L_2}}}}+\sigma_1+\sigma_2+\sigma_3+\sigma_4
\end{equation}
\begin{equation}
f_1(\epsilon)=2 \left( \alpha g_4(\epsilon) \left( h +L_1 \right) +L_2^4 \left(\epsilon L_2- L_1\right)  \right) 
\end{equation}
\begin{equation}
\sigma_1=\left( f_2(-1)+f_4(1,1)+f_3(1) \right) {e^{{\frac {\sqrt {\alpha}2h}{L_1 L_2}}}}
\end{equation}
\begin{equation}
\sigma_2=\left( f_2(1)+f_4(1,-1)+f_3(-1) \right) {e^{-{\frac {\sqrt {\alpha}2h}{L_1 L_2}}}}
\end{equation}
\begin{equation}
\sigma_3=\left( f_4(-1,1)+f_5(1) + f_6(1)\right) {e^{2{\frac {-a L_1+\sqrt {\alpha}h}{L_1 L_2}}}}
\end{equation}
\begin{equation}
\sigma_4= \left(f_4(-1,-1)+ f_5(-1)+ f_6(-1) \right) {e^{-2{\frac { a L_1+\sqrt {\alpha}h}{L_1 L_2}}}} 
\end{equation}
\begin{equation}
f_2 (\epsilon)= \left( \sqrt {\alpha}+ \epsilon L_2 \right) L_2 \left( L_1-L_2 \right) g_3(\epsilon) 
\end{equation}
\begin{equation}
f_3(\epsilon)=- L_2^2L_1\sqrt {\alpha} \left( h +L_1 \right)  \left( \sqrt {\alpha}+ \epsilon L_2+ \epsilon L_1 \right) 
\end{equation}
\begin{equation}
f_4(\epsilon_1,\epsilon_2)= h \alpha \left( h + L_1 \right)  \left(  \left( 2L_2+\epsilon_1L_1 \right) \left( \epsilon_2 \sqrt {\alpha}+L_2 \right)+L_1^2 \right) 
\end{equation}
\begin{equation}
f_5(\epsilon)=-\epsilon h\sqrt{\alpha} L_2 \left( L_1+L_2 \right)  \left( 2\, \left(\epsilon \sqrt{\alpha}+L_2 \right)L_2+L_1^2 \right)  
\end{equation}
\begin{equation}
f_6(\epsilon)= L_2^2 L_1 \left( L_2 \left(\epsilon \sqrt {\alpha}+L_2 \right)  \left( L_1+L_2\right) -\sqrt {\alpha} \left( h+L_1\right)   \left( \sqrt {\alpha}+ \epsilon L_2- \epsilon L_1 \right)  \right)  
\end{equation}
\begin{equation}
num_2= \varsigma_1 + \varsigma_2 -g_4(1) {e^{{\frac { 2 a }{L_2}}}} \left( f_7(1) + f_8(1) \right)-g_4(-1) {e^{-{\frac { 2 a }{L_2}}}} \left( f_7(-1) + f_8(-1) \right)
 \end{equation}
\begin{equation}
\varsigma_1=2\sqrt {\alpha} \left(  \left( L_2^2- h^{2} \right) \alpha-L_1^3 h   \right)  \left( {e^{2\,{\frac {\sqrt {\alpha} a }{L_1L_2}}}}+{e^{2\,{\frac {\sqrt {\alpha} b }{L_1 L_2}}}} \right)
\end{equation}
\begin{equation}
\varsigma_2= 2 L_2 \left(  h  \left(  h  + L_1 \right) \alpha-L_2^4 \right)  \left( {e^{2\,{\frac {\sqrt {\alpha} a }{L_1L_2}}}}-{e^{2\,{\frac {\sqrt {\alpha} b }{L_1 L_2}}}} \right) 
\end{equation}
\begin{equation}
f_7(\epsilon)= \left( \alpha+  \left(L_1+ \epsilon L_2\right)   h   \right) \sqrt {\alpha} \left( {e^{2\,{\frac {\sqrt {\alpha} a }{L_1 L_2}}}}+{e^{2\,{\frac {\sqrt {\alpha} b }{L_1 L_2}}}} \right)
\end{equation}
\begin{equation}
f_8(\epsilon)=  \left( \epsilon   h  \alpha+L_2^3+ \epsilon L_1^3 \right)  \left(- {e^{2\,{\frac {\sqrt {\alpha} a }{L_1 L_2}}}}+{e^{2\,{\frac {\sqrt {\alpha} b }{L_1 L_2}}}} \right)  
\end{equation}
\begin{equation}
den_1= \sqrt{\alpha} \left( \xi_1+\xi_2+\xi_3+\xi_4+\xi_5+\xi_6+\xi_7 \right)
\end{equation}
\begin{equation}
\xi_1=2L_1 \left(  h  +L_1\right) {\alpha}
\end{equation}
\begin{equation}
\xi_2= L_2\sqrt{\alpha} \left(  \left( \alpha+L_2^2 \right) \sinh \left( {\frac {  2h  \sqrt {\alpha}}{L_1 L_2}} \right) +2L_2\sqrt {\alpha}\cosh \left( {\frac {  2h  \sqrt {\alpha}}{L_1 L_2}} \right)  \right)  
\end{equation}
\begin{equation}
\xi_3= L_1 L_2  \left( \alpha \sinh \left(  {\frac { 2 a }{L_2}}\right) -2L_1L_2 \cosh\left(  {\frac { 2 a }{L_2}}\right)\right)
\end{equation}
\begin{equation}
\xi_4=- L_2\sqrt{\alpha} \left( \alpha+2L_1 h+L_2^2 \right) \cosh \left( {\frac { 2 a }{L_2}} \right) \sinh \left( {\frac {  2h \sqrt {\alpha}}{L_1 L_2}} \right)  
\end{equation}
\begin{equation}
\xi_5=-2\left(L_1 \left( h+L_1 \right) \alpha+L_2^4 \right) \cosh \left( {\frac { 2 a }{L_2}} \right) \cosh \left( {\frac {  2h  \sqrt {\alpha}}{L_1 L_2}} \right)  
\end{equation}
\begin{equation}
\xi_6=- L_2 {\alpha} \left( 2h +L_1 \right) \sinh \left( {\frac { 2 a }{L_2}} \right) \cosh \left( {\frac {  2h \sqrt {\alpha}}{L_1 L_2}} \right)  
\end{equation}
\begin{equation}
\xi_7=-  \sqrt{\alpha} \left( \left( 2h+L_1 \right) \alpha+L_1^{3} \right)\sinh \left( {\frac { 2a }{L_2}} \right) \sinh \left( {\frac {  2h  \sqrt {\alpha}}{L_1 L_2}} \right)  
\end{equation}
\begin{equation}
 den_2 = g_1(1){e^{-2\,{\frac { a \, \left( L_1-\sqrt {\alpha} \right) }{L_1 L_2}}}} +g_1(-1) {e^{2\,{\frac {\sqrt {\alpha} b }{L_1 L_2}}}}+g_2(1)  {e^{-2{\frac { a L_1-\sqrt {\alpha} b }{L_1 L_2}}}}+g_2(-1){e^{2\,{\frac {\sqrt {\alpha} a }{L_1 L_2}}}} 
\end{equation}
\begin{equation}
g_1(\epsilon)=L_2 \left( L_1 + L_2\right)  \left( \sqrt {\alpha}- \epsilon L_2 \right) 
\end{equation}
\begin{equation}
g_2(\epsilon)= \sqrt {\alpha} \left( 2h+L_1 \right) \left(- \epsilon \sqrt {\alpha}-L_2+ \epsilon L_1 \right) + \epsilon {\alpha}^{3/2}+L_2^3- \epsilon L_1^3 
\end{equation}
\begin{equation}
g_3(\epsilon)=  h\sqrt {\alpha} \left(-\epsilon \sqrt {\alpha}-L_2 \right) + \epsilon 2\,L_2^2 L_1 
\end{equation}
\begin{equation}
g_4(\epsilon)= \left( \left(\epsilon L_2 - L_1 \right)h+L_2^{2}\right)
\end{equation}
\begin{equation}
 h=b-a
\end{equation}
\begin{equation}
 \alpha=L_1^2+L_2^2
\end{equation}
\begin{equation}
 L_1=v_l \tau_1
\end{equation}
\begin{equation}
 L_2=V\tau_2
\end{equation}

This result have been checked by numerical simulations 
and by comparison with known limits.

\subsection{Static mode in 3 dimensions~: more comparisons between the analytical expressions and the simulations \refs{r3dvk}}\label{a3dvk}

\doublimagem{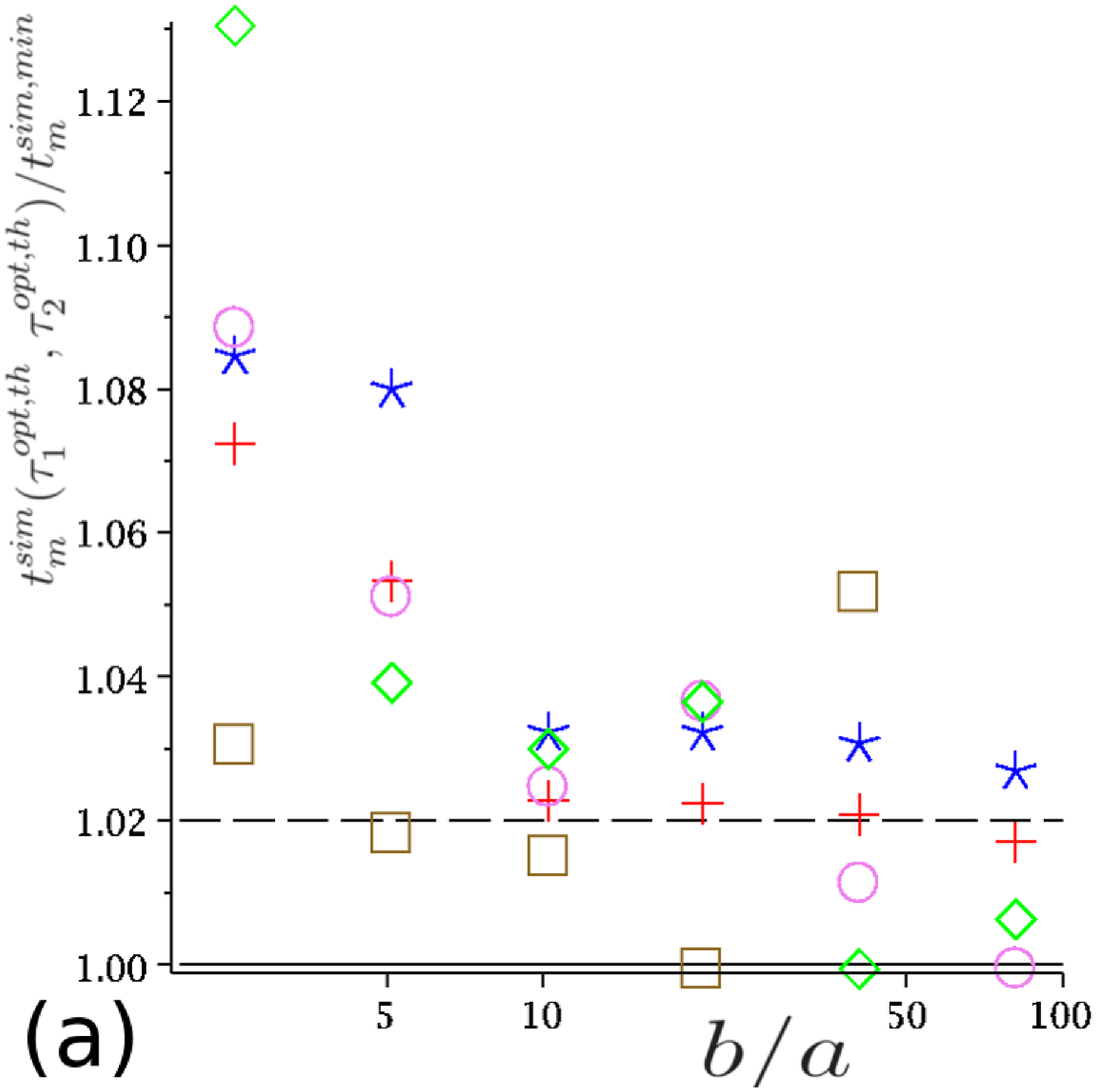}
{ }
{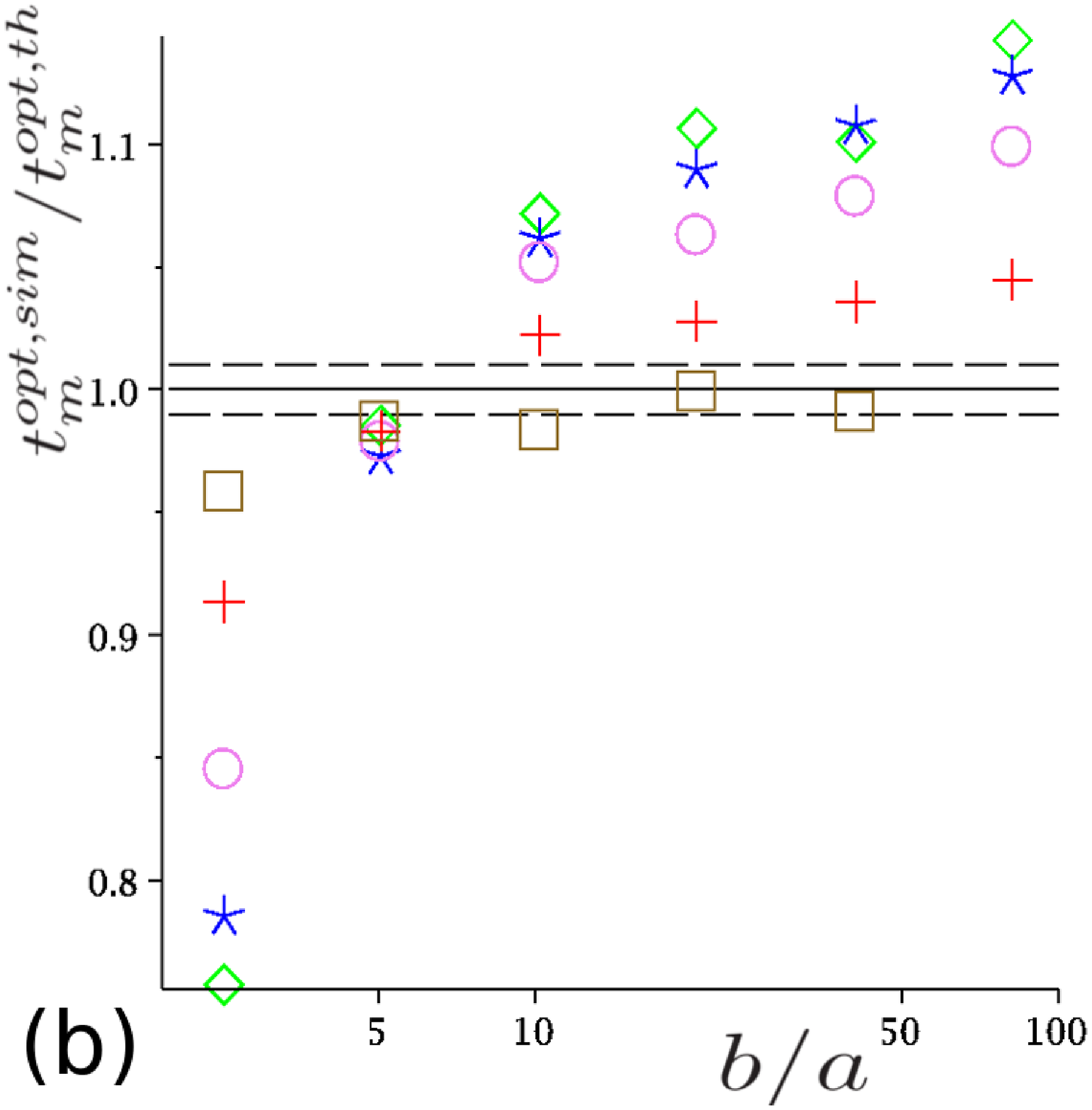}
{}
{Static mode in 3 dimensions. Study of the minimum~: 
its location in the $\tau_1,\tau_2$ space (a), 
and its value (b). $sim$ means values obtained through numerical simulations, 
$th$ means analytical values. Value expected if there was a prefect agreement between theory and simulations (black line),
 and values taking into account the simulations noise (dotted black lines) 
 (we performed 10~000 walks for each point). 
  $a=0.01$ (brown squares), $a=0.1$ (red crosses), $a=1$ (purple circles), 
$a=10$ (blue stars), $a=100$ (green diamonds).
$V=1$, $k=1$.}
{3dvkopt}

The numerical study of the minimum mean search time (\refi{3dvkopt}) 
shows that the analytical values gives the good position of the minimum 
in $\tau_1$ and $\tau_2$ as soon as $b/a$ is not too small. 
However, the value of the minimum is underestimated by 
about 10\%.

\subsection{Diffusive mode in 3 dimensions}

\subsubsection{Full analytical expression of $t_m$ \refs{r3dvd1}}\label{a3dvd1}

\begin{equation}\label{tm3dvd}
t_m=  \frac{1}{{b}^3{\alpha}^4 dp  D}
\left( X
+Y
+Z 
\right)
\end{equation}
with~:
\begin{equation}
 X = \frac{ \left( {  \tau_1^{-1}}+{\alpha}^{2}{
 dp} \right)  \left( {\frac {{\alpha}^{2} \left( {b}^{3}-{a}^{3}
 \right) }{a}}-3\,S \right)  \left( 1/3\,{\frac { \left( {b}^{3}-{a}^{
3} \right)  \left( {\alpha}^{2}{ dp}-{  \tau_2^{-1}} \right) }{a}}+{
\frac {{  \tau_2^{-1}}\, \left( \alpha\,aR+1 \right) }{{\alpha}^{2}}}+{
\frac {\alpha\,{ dp}\, \left( -1+{ TT} \right) }{{{ \alpha_2}}^
{2}}} \right)}{ \tau_1 \left(  \left( {  \tau_1^{-1}}+{\alpha}^{2}{ dp}
 \right) {  \tau_2^{-1}}\,R\alpha+{\frac { \left( -{\alpha}^{2}{ dp}+
{  \tau_2^{-1}} \right) {  \tau_1^{-1}}}{a}}+{\frac {{ TT}\,{\alpha}^{2
}{ dp}\, \left( {  \tau_1^{-1}}+{  \tau_2^{-1}} \right) }{a}} \right)}
\end{equation}
\begin{equation}
 Y=3{\frac {{  \tau_1^{-1}}\,aS}{{\alpha}^{2}}}
\end{equation}
\begin{equation}
 Z=-2/30\,{\frac {
 \left( -b+a \right) ^{3}{\alpha}^{2} \left( {a}^{3}+3\,b{a}^{2}+6\,{b
}^{2}a+5\,{b}^{3} \right)  \left( {  \tau_1^{-1}}+{\alpha}^{2}{ dp}
 \right) }{a}}
\end{equation}
\begin{equation}
 \alpha = \sqrt{(\tau_1 D)^{-1}+(\tau_2 D_2)^{-1}}
\end{equation}
\begin{equation}
 D_2=\frac{1}{3}V^2 \tau_2
\end{equation}
\begin{equation}
 dp=\frac{D D_2}{D-D_2}
\end{equation}
\begin{equation}
 \alpha_2=(\tau_2 D_2)^{-1}
\end{equation}
\begin{equation}
 R={\frac {\alpha\,b\tanh \left( \alpha\, \left( b-a \right)  \right) -1}
{\alpha\,b-\tanh \left( \alpha\, \left( b-a \right)  \right) }}
\end{equation}
\begin{equation}
S= {\frac { \left( {\alpha}^{2}ba-1 \right) \tanh \left( \alpha\, \left( 
b-a \right)  \right) +\alpha\, \left( b-a \right) }{\alpha\,b-\tanh
 \left( \alpha\, \left( b-a \right)  \right) }}
\end{equation}
\begin{equation}
 TT={\frac {{ \alpha_2}\,a}{\tanh \left( { \alpha_2}\,a \right) }}
\end{equation}

\subsubsection{Dependence of $t_m$ with $\tau_1$}\label{a3dvd_tau1effet}

\imagea{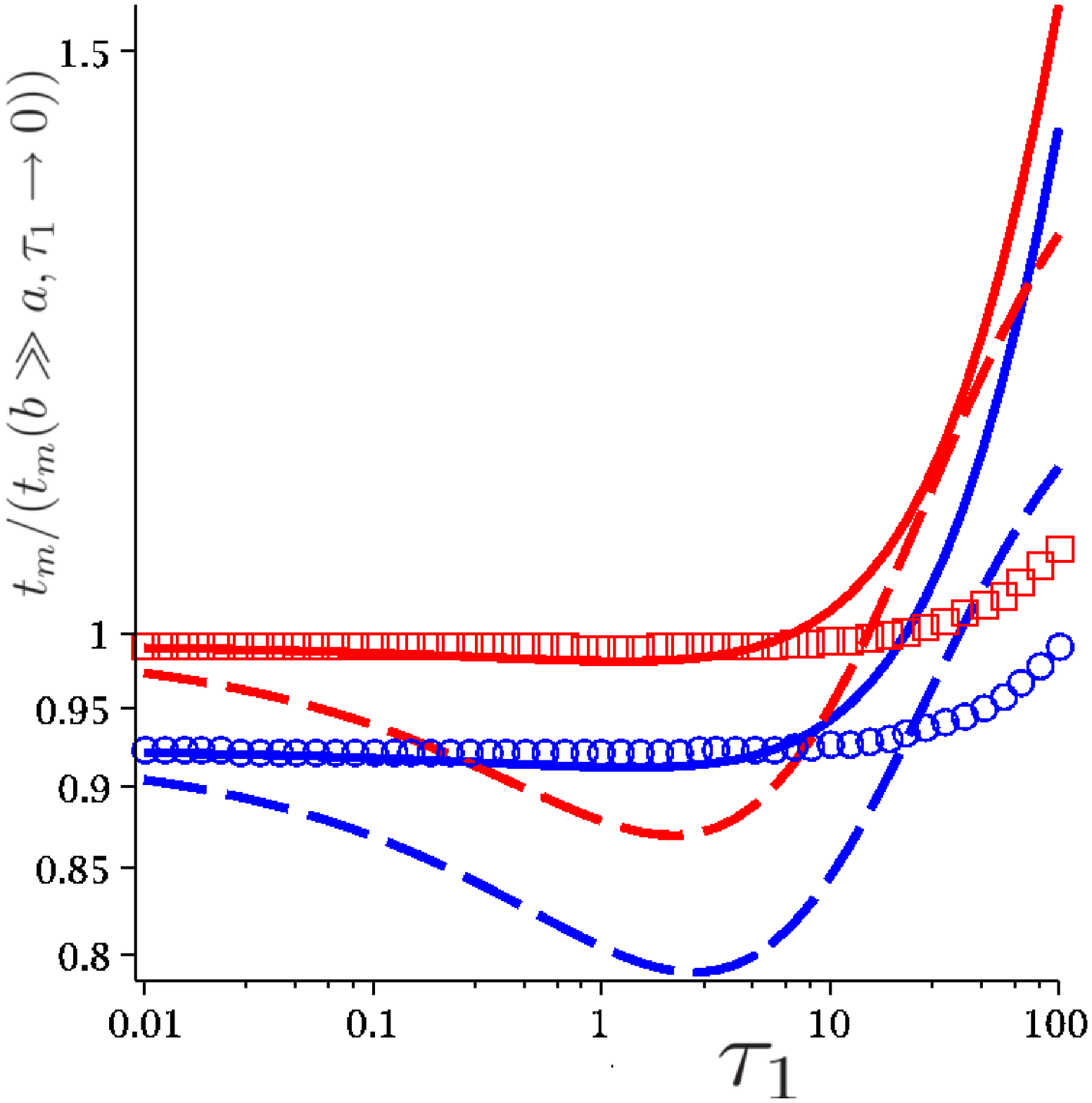}
{Diffusive mode in 3 dimension.  $t_m$ from  \refm{tm3dvd}, 
$t_m(b \gg a,\tau_1\to 0)$ from \refm{pearson}. $\tau_2=\tau_2^{opt,th}$ \refm{tau23dvd},
$D=1$, $V=1$, $a=10$ (dotted lines), $a=100$ (lines), $a=1000$ (symbols), $b/a=10$ (blue,circles), $b/a=100$ (red, squares).}
{tau1effet}{6}

The mean detection time is very weakly dependent on $\tau_1$ as long as $\tau_1 < 6D/v^2$ (\refi{tau1effet}).

\subsubsection{$t_m$ in the regime of diffusion alone \refs{r3dvd3} }\label{a3dvd3}

 We take a diffusive random walk starting from $r=r_0$ 
in a sphere with reflexive boundaries at $r=b$ and absorbing boundaries at $r=a$,
we get the following equation for $t(r_0)$ the mean time of absorption~:
\begin{equation}
 D_{eff} \frac{1}{r_0^2}\left(\frac{d}{dr_0}\left(r_0^2 \frac{d t(r_0)}{dr_0} \right) \right) = -1
\end{equation}
With the boundary conditions, the solution is~: 
\begin{equation}
 t(r_0)=\frac{1}{6 D_{eff}}\left(\frac{2b^3}{a}+a^2-r_0^2-\frac{2b^3}{r_0} \right)
\end{equation}
Than we average on $r_0$, as the searcher can start from any point of the sphere 
with the same probability~: 
\begin{equation}
t_{diff} = \frac{1}{15 D ab^3} \left(5b^3a^3+5b^6-9b^5a-a^6 \right)
\end{equation}
In the limit $b/a \gg 1$~:
\begin{equation}
t_{diff}=\frac{b^3}{3 D a}
\end{equation}

\subsubsection{Criterion for intermittence~: additional figure \refs{r3dvd2}}\label{a3dvd2}

\imagea{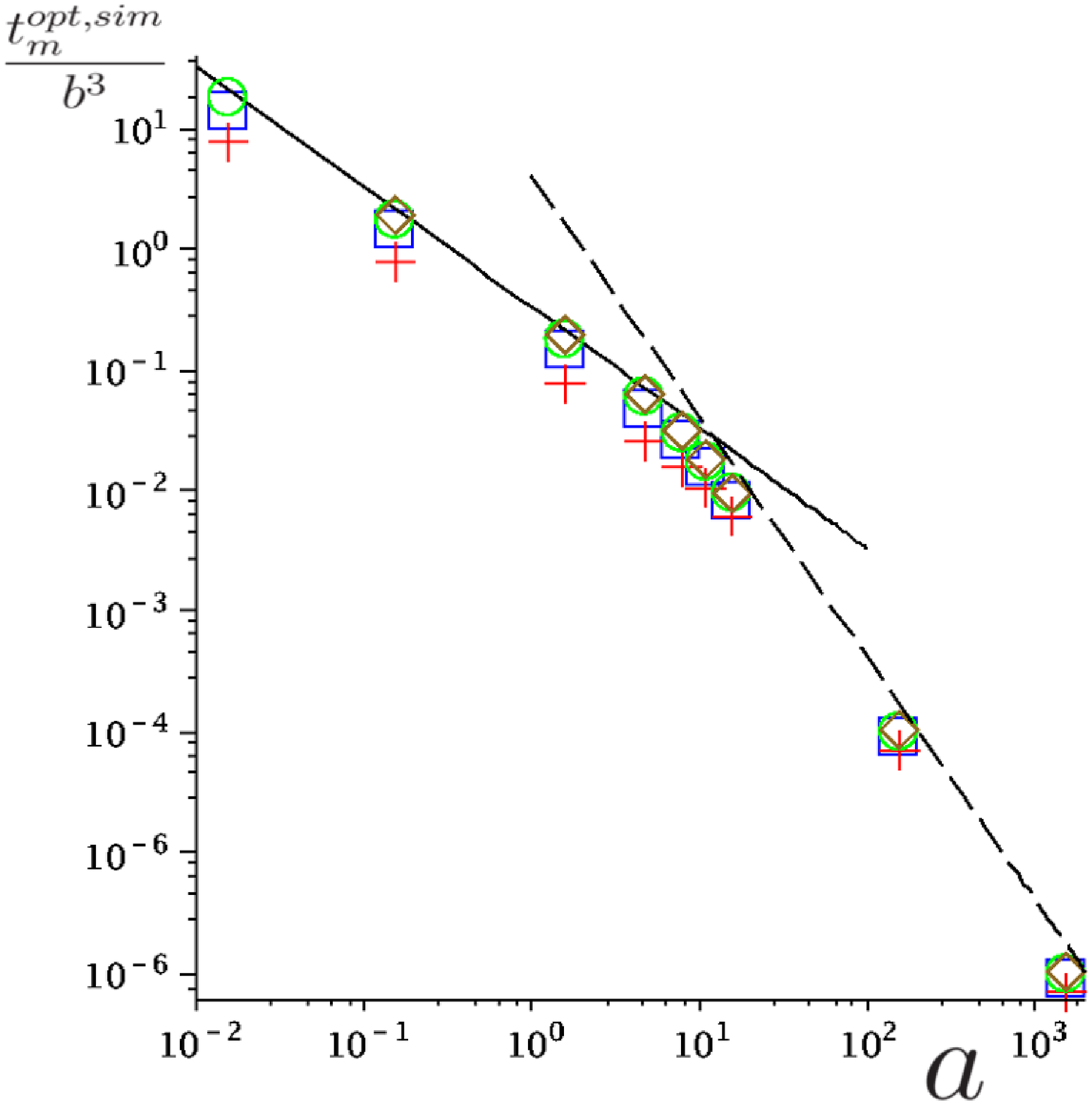}{Diffusive mode in 3 dimension.  Simulations~: 
$b/a=2.5$ (red crosses), $b/a=5$ (blue squares), $b/a=10$ (green circles), $b/a=20$ (brown diamonds).
Analytical expressions in the low target density approximation ($b/a \gg 1$)~:
$\tau_1 = 0$ \refm{pearson} (with $\tau_2=\tau_2^{opt,th}$ \refm{tau23dvd}) (dotted line),
diffusion alone \refm{tdiff3D} (continuous line). 
$V=1$, $D=1$}{aeffet}{6}

The figure \ref{aeffet} shows the dependence of $t_m^{opt}$ with $a$.

\subsection{Ballistic mode in 3 dimensions}\label{a3dvv}

\subsubsection{without intermittence  \refs{r3dvv1}}\label{a3dvv1}

%\po{Limit when $\tau_2 \to 0$ and $v_l \tau_1 \le a$}

In the regime without intermittence, $\tau_1$ is not necessarily 0. 
We calculate $t_m$ in two limits~: $\tau_1$ small or $\tau_1$ large.

% These results are in good agreement with the simulations \refi{3dvvnouvelletau2e0}.

\po{Limit $\tau_2 \to 0$,  $v_l \tau_1 \le a$}

In the limit  $v_l \tau_1 \le a$, we can consider phase 1 as diffusive, with~: 
\begin{equation}\label{3dvvdeff}
D=\frac{1}{3} v_l^2 \tau_1 
\end{equation}
We use the approached expression of $t_m$ obtained in the diffusive mode \refm{tdiff3D} with this effective diffusive coefficient. 
\begin{equation}\label{3dvvtmdiff}
t_m = \frac{1}{5 v_l^2 \tau_1 ab^3} \left(5b^3a^3+5b^6-9b^5a-a^6 \right)
\end{equation}
And in the limit $b\gg a$~:
\begin{equation}\label{3dvvtmdiffa}
  t_m =  \frac{b^3}{ v_l^2 \tau_1 a} 
\end{equation}
 %This result is in good agreement with the simulations \refi{3dvvnouvelletau2e0}

%\po{Ballistic limit $\tau_2 \to 0$, $\tau_1 \to \infty$}

\po{limit $\tau_2 \to 0$, $\tau_1 \to \infty$}

We name $V_{ol}$ the volume of the sphere. 
$g(t)$ is the volume explored by the searcher after a time $t$. 
The volume explored during $dt$ is $\pi v_l a^2 dt$.
If we consider that the probability to encounter a unexplored space is uniform,
which is wrong at short times but close to the reality at long times,
the average of first explored volume at time $t$ during $dt$ is $\frac{V_{ol}-g(t)}{V_{ol}} \pi v_l a^2 dt$.
Then in this hypothesis, $g(t)$ is solution of~: 
\begin{equation}
g(t)=\int_0^{t} \frac{V_{ol}-g(u)}{V_{ol}} \pi v_l a^2 du 
\end{equation}
This equation can be simplified taking a renormalized time $r$ as $r=\frac{\pi v_l a^2}{V_{ol}}t$,
and $f=g/{V_{ol}}$~:
\begin{equation}
f(r)=\int_0^r(1-f(w)) dw
\end{equation}
Then, as $f(0)=0$ (nothing has been explored at time 0), $f(r)=1-e^{-r}$.
The probability to encounter the target at time $t$ during $dt$ (and not before) 
is the newly explored volume at time $t$ divided by the whole volume $V_{ol}$ 
if we make the mean-field approximation. 
Then the probability $p(r)$ than the target is not yet found at time $r$ is solution of~: 
\begin{equation}
\frac{dp}{dr}=-(1-f(u))
\end{equation}
As $p(0)=1$, the result is $p(r)=e^{-r}$. 
Then the mean detection time of the target is $1$ is renormalized time, 
is to say in real time~: 
\begin{equation}
 t_{bal}=\frac{4 b^3}{3 a^2 v_l}
\end{equation}

\po{Numerical study}

\imagea{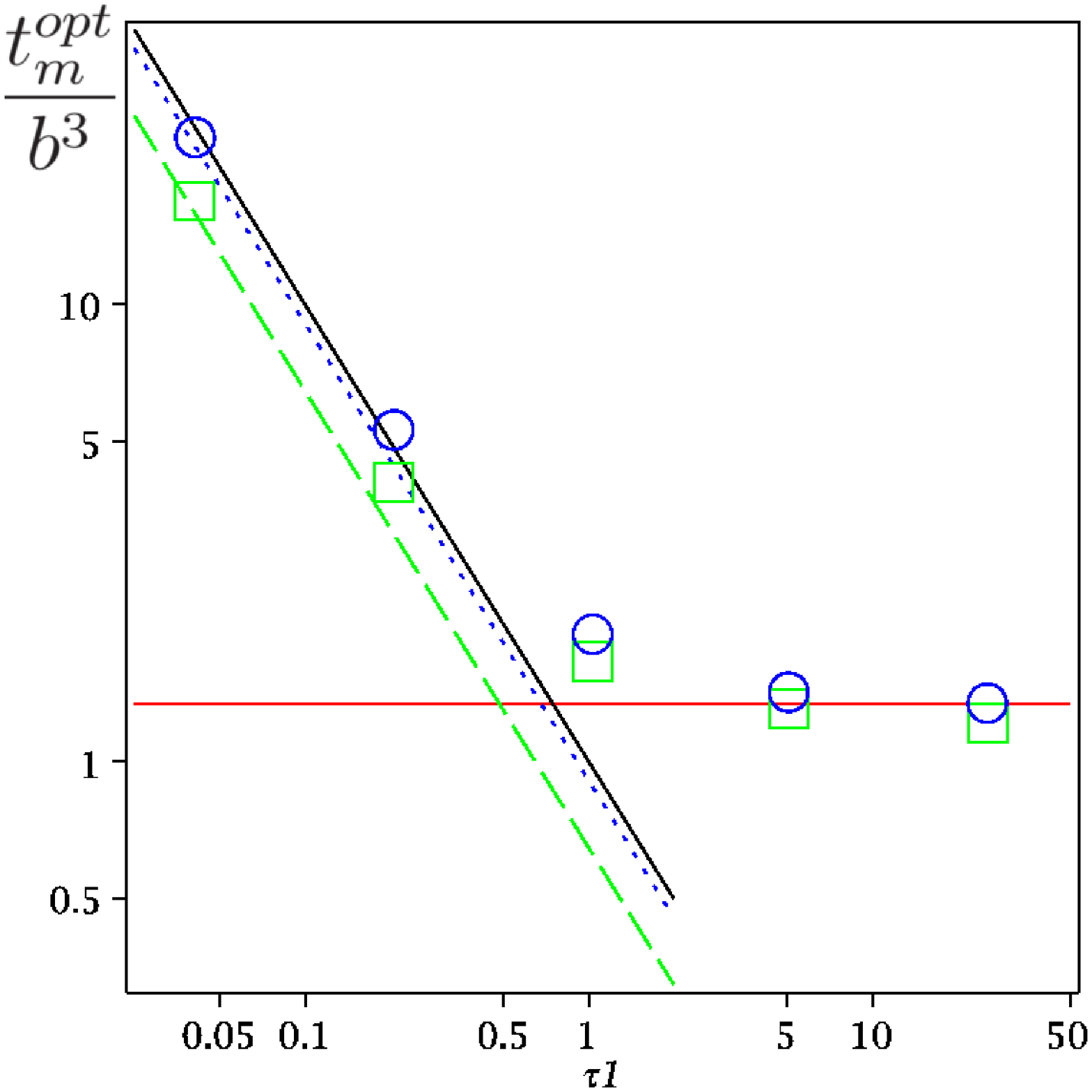}
{Ballistic mode in 3 dimensions. Regime without intermittence ($\tau_2 = 0$). 
$\ln(t_m/b^3)$ as a function of $\ln(\tau_1)$, simulations for $b=5$ (green squares) 
and $b=20$ (blue circles). 
Ballistic limit ($\tau_1 \to \infty$) (no intermittence) \refm{3dvvtmsansinter} (red horizontal line), 
Diffusive limit ($v \tau_1 < a$) \refm{3dvvtmdiff} with $b=5$ (green dotted line), $b=20$ (blue small dots),
$b \gg a$ limit \refm{3dvvtmdiffa} (black line). $a=1$, $v_l=1$.}
{3dvvnouvelletau2e0}{5}

These expressions give a very good approximation 
of the values obtained through simulations (\refi{3dvvnouvelletau2e0}, \refi{InterApprox}).
In the regime without intermittence, 
$t_m$ is minimized  for $\tau_1 \to \infty$.

\subsubsection{Numerical $v_l^c$ \refs{r3dvv2}}\label{a3dvv2}

\imagea{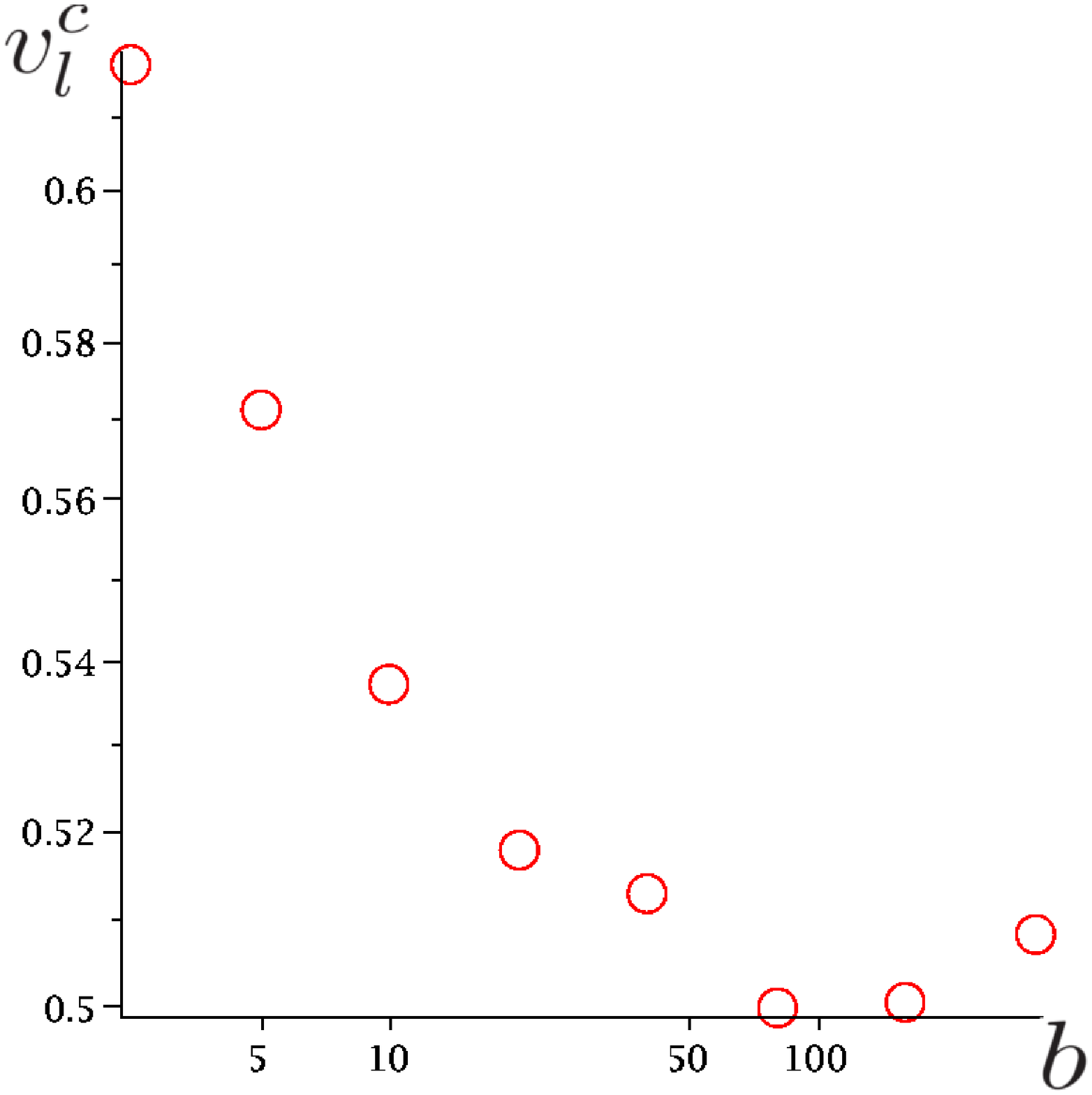}{Ballistic mode in 3 dimension. $v_l^{c}$ as a function of $\ln(b)$, through simulations. 
$a=1$, $V=1$. 
}{vc}{6}

In simulations \refi{vc}, when $b$ is small $v_l^c$ decreases,
but stabilizes for larger $b$,
which is coherent with the fact that this value is obtained
through a development in $b$. 
The value of $v_l$ for large $b$ is different (even if close) to the expected value.
The main explanation of this discrepancy is that in the intermittence regime,
the approached value of $t_m$ is about 20\% away from the value obtained through simulations.

\end{document}